\documentclass{emulateapj}
\usepackage{natbib,apjfonts}
\usepackage{placeins}
\usepackage{multirow}
\usepackage{amsmath}
\usepackage{amssymb}
\usepackage{epsfig}
\usepackage{graphicx}
\usepackage{color}
\newcommand{\kms}{\mbox{km s$^{-1}$}}

\newcommand{\Msun}{\mbox{$M_{\odot}$}}

\newcommand{\HI}{\ion{H}{1}}
\newcommand{\hii}{\ion{H}{2}}
\newcommand{\co}{\mbox{CO$(1\rightarrow0)$}}
\newcommand{\cotwo}{\mbox{CO$(2\rightarrow1)$}}
\newcommand{\cother}{\mbox{$^{13}$CO$(1\rightarrow0)$}}
\newcommand{\Kkmspc}{\mbox{K km s$^{-1}$ pc$^2$}}
\newcommand{\degrees}{\arcdeg}

\shorttitle{Giant molecular clouds in the non-grand design spiral galaxy NGC 6946}
\shortauthors{Rebolledo et al.}

\begin{document}
\submitted{Accepted for publication in The Astrophysical Journal, August 16, 2012}

\title{Giant molecular clouds and star formation in the non-grand design spiral galaxy NGC 6946}

\author{David Rebolledo$^{1}$, Tony Wong$^{1}$, Adam Leroy$^{2}$, Jin Koda$^{3}$ and Jennifer Donovan Meyer$^{3}$}

\affil{$^{1}$Astronomy Department, University of Illinois, Urbana, IL 61801, USA; dreboll3@illinois.edu\\
$^{2}$National Radio Astronomy Observatory, Charlottesville, VA 22903, USA\\
$^{3}$Department of Physics \& Astronomy, Stony Brook University, Stony Brook, NY 11794, USA}

\begin{abstract}

We present high spatial resolution observations of Giant Molecular Clouds (GMCs) in the eastern part of the nearby spiral galaxy NGC 6946 obtained with the Combined Array for Research in Millimeter-wave Astronomy (CARMA).  We have observed $\co$, $\cotwo$ and $\cother$, achieving spatial resolutions of $5\farcs4\times5\farcs0$, $2\farcs5\times2\farcs0$ and $5\farcs6\times5\farcs4$ respectively over a region of 6 $\times$ 6 kpc.  This region extends from 1.5 kpc to 8 kpc galactocentric radius, thus avoiding the intense star formation in the central kpc.  We have recovered short-spacing $u$-$v$ components by using single dish observations from the Nobeyama 45m and IRAM 30m telescopes.  Using the automated CPROPS algorithm we identified 45 CO cloud complexes in the $\co$ map and 64 GMCs in the $\cotwo$ maps.  The sizes, line widths, and luminosities of the GMCs are similar to values found in other extragalactic studies.  We have classified the clouds into on-arm and inter-arm clouds based on the stellar mass density traced by the 3.6 $\mu$m map.  Clouds located on-arm present in general higher star formation rates than clouds located in inter-arm regions.  Although the star formation efficiency shows no systematic trend with galactocentric radius, some on-arm clouds -- which are more luminous and more massive compared to inter-arm GMCs -- are also forming stars more efficiently than the rest of the identified GMCs.  We find that these structures appear to be located in two specific regions in the spiral arms.  One of them shows a strong velocity gradient, suggesting that this region of high star formation efficiency may be the result of gas flow convergence.   

\end{abstract}


\section{INTRODUCTION}

The formation and evolution of giant molecular clouds (GMCs), the sites where stars form, are thought to be influenced by the local properties of the environment and global galactic structure (\citealt{2008MNRAS.385.1893D}; \citealt{2009ApJ...700..358T}; \citealt{2011ApJ...730...11T}; \citealt{2009ApJ...699..850K}).  Although the internal physical properties of molecular clouds have been extensively studied (\citealt{1987ApJ...319..730S}; henceforth S87), a more detailed understanding of their origin and evolution in different types of environments is needed.  Maps with resolution matched to the scale of individual clouds are the key to achieve this goal; however, such maps currently exist only for a few galaxies outside the Local Group.  Due to the coupled evolution between GMCs and their host galaxies, achieving a more complete understanding of molecular cloud formation and evolution processes represents a key goal for studies of galaxy evolution.

While several mechanisms have been proposed, the predominant mode for molecular cloud formation and evolution in spiral galaxies is not yet clear.  In terms of the arm structure, spiral galaxies can be classified as {\it grand-design}, most of them consisting of two symmetrical spiral arms; {\it multi-armed}, which show several non-symmetrical arms; or {\it flocculent} galaxies with multiple shorter arms.  For grand-design galaxies, the spiral structure is generated by density waves induced by bars or companion galaxies.  The influence of the density waves in the ISM in these galaxies has been commonly modeled assuming a long lived, rigidly rotating spiral potential (\citealt{2002ApJ...570..132K}; \citealt{2004MNRAS.349..270W}; \citealt{2006MNRAS.367..873D}).  On the other hand, flocculent and multi-armed galaxies are thought to develop from local gravitational instabilities, although flocculent galaxies can present weak density waves as well (\citealt{1997ApJ...484..202T}).  In this case, numerical simulations model the galactic potential as time dependent structures, which can change on a timescale of $\sim$100 Myr (\citealt{2008MNRAS.385.1893D}; \citealt{2011ApJ...735....1W}).

The formation of massive molecular clouds in grand-design galaxies is attributed to the passing of spiral density waves.  A particular feature of this type of galaxy is the formation of spurs perpendicular to the spiral arms due to the shearing of the molecular structures as they leave the arms (\citealt{2002ApJ...570..132K}; \citealt{2004MNRAS.349..270W}).  In their study of the northeastern spiral segment of IC 342, a grand design galaxy, \citet{2010PASJ...62.1261H} showed that actively star forming GMCs are more massive, have smaller velocity dispersions and are more gravitationally bound than quiescent GMCs.  They interpreted the tendency for the star-forming GMCs to lie downstream with respect to the quiescent molecular clouds as evidence in favor of cloud a coalescence scenario, where massive molecular clouds ($\sim 10^6\ \Msun$) are formed from coagulation of smaller clouds entering the spiral arms, and the star formation is ignited once the excess kinetic energy is dissipated through inelastic collisions of the clouds (Hirota et al. 2011).  In another grand design galaxy, M51, \citet{2009ApJ...700L.132K} suggested that evolution of GMCs is driven by large scale galactic dynamics, where the most massive molecular complexes (Giant Molecular Associations, GMAs) are created by coagulation of GMCs in spiral arms.  As GMAs leave the arms, they suffer fragmentation by the strong shear motions, which would explain the spur structures observed extending from the arms into the inter-arm regions.  The remaining fragmented GMCs are not disassociated into atomic gas, and they become part of the molecular cloud population observed in inter-arm regions which survive until the passage of another spiral arm.

On the other hand, \citet{2008MNRAS.385.1893D} presented a numerical simulation of flocculent galaxies.  They find that the gaseous spiral structure essentially traces the potential minimum, in contrast to the case of grand-design spirals where an offset between the spiral shock and the potential minimum is expected.  According to the simulations provided by \citet{2008MNRAS.385.1893D} and \citet{2006MNRAS.371..530C} the most massive structures are thought to be produced in regions where collision or merging between spiral arms can occur, yielding an enhanced star formation in those overdense regions, in contrast to the situation proposed for grand-design galaxies where the molecular clouds are formed by agglomeration within spiral arms.  Conversely, the formation of inter-arm molecular gas structures are proposed to be the result of local instabilities rather than fragmentation of GMA's, and the properties of the clouds are approximately similar across the disk, with no offsets between gas and star formation expected.  High resolution observations of flocculent galaxies have provided evidence in support of this picture.  For instance, \citet{2003PASJ...55..605T} mapped $\co$ over the southern arm of the flocculent galaxy NGC 5055.  They found no obvious offset between H$\alpha$ and the molecular gas, and that on-arm and inter-arm clouds do not have significant differences in their properties. 

Resolved GMCs in the Milky Way have been shown to be in approximate virial equilibrium obeying scaling relations known as Larson's laws (\citealt{1981MNRAS.194..809L}).  These laws establish that velocity dispersion (or line width), size, and luminosity are correlated in the Milky Way GMCs.  The size-line width relation may reflect turbulent conditions in the molecular ISM.  The typical GMC temperature of 15--25 K produces thermal CO velocity dispersions of only $\sim$0.1 $\kms$, providing evidence of non-thermal supersonic turbulence within the clouds.  Nevertheless, whether the turbulence in the molecular ISM is internally or externally generated is still unclear (\citealt{2007ARA&A..45..565M}; \citealt{2011ApJ...730...11T}).  The other two Larson relations are the luminosity-line width and the luminosity-size relations.  In a recent work, \citet{2008ApJ...686..948B} (henceforth B08) present a study of resolved properties of GMCs for a sample of dwarf galaxies along with two disk galaxies of the Local Group.  They find insignificant differences between Milky Way GMCs and GMCs in dwarf galaxies, with the latter following approximately the same Larson relations as Galactic clouds.  \citet{2007prpl.conf...81B} arrived at a similar conclusion by analyzing observations of CO surveys for galaxies in the Local Group.  

Due to sensitivity and resolution limitations, observations of resolved properties of GMCs in galaxies other than Milky Way are in an early stage (\citealt{2012ApJ...744...42D}, hereafter DM12; \citealt{2009ApJ...700L.132K}).  Only a few galaxies have been mapped in CO emission across the full extent of the optical disk with sufficient resolution to probe the $\sim$100 pc size scales on which GMCs form.  High resolution CO surveys (\citealt{1999ApJS..124..403S}; \citealt{2003ApJS..145..259H}) have focused primarily on the central regions, whereas complete mapping using single-dish telescopes (\citealt{2007PASJ...59..117K}; \citealt{2009AJ....137.4670L}) has been obtained at resolutions of 11$\arcsec$ to 15$\arcsec$, corresponding to $\sim$500 pc for nearby spirals, inadequate to resolve GMCs.  Only for M33 (BIMA survey) and the Magellanic Clouds (NANTEN surveys) has it been possible to conduct a fairly complete census of GMCs (\citealt{2003ApJS..149..343E}; \citealt{2008ApJS..178...56F}).  

Using a combination of $\co$ and $\cotwo$ and pursuing a staged approach of starting with large-scale, low resolution maps and following up with high resolution imaging on smaller areas, CARMA makes it possible to study the overall distribution of GMCs, as well as the properties of individual GMCs, in galaxies outside the Local Group.  We have successfully used this observational strategy in the nearby spiral galaxy NGC 6946.  As indicated by the single-dish CO map from IRAM (\citealt{2009AJ....137.4670L}), this galaxy is extremely rich in molecular gas, with CO emission extended to a large fraction of the optical radius.  Table \ref{n6946-prop} shows the basic parameters of NGC 6946.  We have chosen NGC 6946 as our target because of its proximity, high CO surface brightness, low inclination, and the availability of high-quality datasets at a variety of wavelengths, including HI imaging from THINGS and multiband Spitzer imaging from SINGS (\citealt{2003PASP..115..928K}).  The spiral structure in NGC 6946 appears complex, with K-band images revealing four prominent, asymmetric spiral arms (\citealt{1995ApJ...452L..21R}).  While the central region of the galaxy has been well-studied in CO because of its high CO surface brightness (\citealt{2007AJ....134.1827C}; \citealt{2007PASJ...59..117K}; DM12), the eastern region has not been mapped at high resolution.

In this paper we present observations of the eastern region of NGC 6946.  Firstly, we made observations of $\co$ using CARMA covering a region of size 6$\times$6 kpc$^2$.  The angular resolution for the $\co$ map was $\sim$110 pc, which is not sufficient to resolve structures with size scales similar to GMCs (S87; B08).  Then, the second step was to make higher resolution $\cotwo$ observations ($\sim$ 50 pc) towards the brightest CO complexes to study the properties of individual GMCs.

We present our study as follows:  in Section \ref{obs} we describe our observations of NGC 6946 using CARMA, and we describe the archival data at several wavelengths that we include in our analysis.  In Section \ref{cprops} we summarize the technique used to identify GMCs and to measure their physical properties.  In Section \ref{cl-prop} we present the scaling relations of the cloud properties, and we compare them with previous studies of Galactic and extragalactic clouds.  In Section \ref{env-eff} we discuss whether the properties of the clouds differ between on-arm and inter-arm regions.  In Section \ref{discuss} we discuss the implications of our results, and in Section \ref{summary} we summarize the work presented in this paper.

\begin{table}
\caption{Properties of NGC 6946.}
\centering
\begin{tabular}{lc}
\hline\hline
Morph. \tablenotemark{a} & SABcd \\  
R.A. (J2000)$^\mathrm{a}$ & 20:34:52.3 \\
Decl. (J2000$^\mathrm{a}$ & 60:09:14 \\
Distance (Mpc)$^\mathrm{b}$ & 5.5 \\ 
Incl. (\degrees)$^\mathrm{c}$ & 33\\ 
P.A. (\degrees)$^\mathrm{c}$ & 243\\
\hline
\multicolumn{2}{l}{{\bf Notes.}} \\
\multicolumn{2}{l}{$^\mathrm{a}$ NASA/IPAC Extragalactic Database (NED).}\\
\multicolumn{2}{l}{$^\mathrm{b}$ \citet{1988JBAA...98..316T}.}\\
\multicolumn{2}{l}{$^\mathrm{c}$  \citet{2008AJ....136.2563W}.}
\end{tabular}
\label{n6946-prop}
\end{table}

\section{DATA}\label{obs}
\subsection{CARMA observations}\label{carma}

\subsubsection{3 mm}

\begin{figure*}
\centering
\begin{tabular}{c}
\epsfig{file=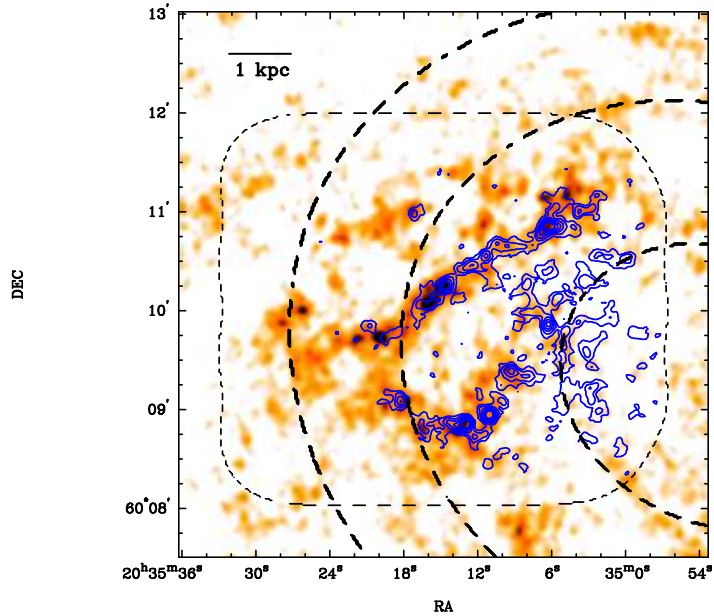,width=0.45\linewidth,angle=-90}
\end{tabular}
\caption{$\co$ integrated intensity contours overlaid on \HI\ integrated intensity map from THINGS.  Contours begin at 4$\sigma$ and are spaced by 4$\sigma$.  The noise at the center of the map is 2 K km s$^{-1}$. The dashed black line illustrates the region where full gain correction was made.  Black dashed ellipses represent, from right to left, the region dominated by molecular gas ($f_\mathrm{mol} \equiv \Sigma_\mathrm{H_2}/\Sigma_\mathrm{HI} \sim 5$; $radius = 2.8$ kpc), the region where both atomic and molecular gas are comparable ($f_\mathrm{mol} \sim 1$; $radius = 5.6$ kpc), and the region where \HI\ starts being the predominant gas phase ($f_\mathrm{mol} \sim 0.1$; $radius = 7.6$ kpc).}
\label{figure_ngc6946e}
\end{figure*}

We performed high spatial resolution observations of $\co$ and $\cother$ for the eastern part of NGC 6946 from July to August of 2009.  We used the Combined Array for Research in Millimeter-wave Astronomy (CARMA) in E and D array configurations, which have baselines of  8.5-66 meters, and 11-148 meters, respectively.  Our goals are to resolve structures close to the typical sizes of GMCs, and to make a detailed study of their properties including sizes, luminosity, velocity dispersion and star formation activity.  In order to achieve these goals, we have selected a region in the disk with a favorable observing geometry, active star formation, and where the interstellar medium transitions from being molecular gas dominated to atomic gas dominated.  We observed $\co$ and $\cother$ in a 49-point mosaic area, which covered 3.6$\times$3.6 arcmin$^2$, corresponding to a physical scale of 6$\times$6 kpc$^2$ at a distance of 5.5 Mpc (\citealt{1988JBAA...98..316T}). The correlator was set to have the $\co$ line in the upper side band (USB), and the $\cother$ in the lower side band (LSB).  By placing two overlapping 31 MHz bands with an offset of 12.9 MHz from the rest frequency line, we achieved a total velocity coverage of 150 $\kms$ and a channel width of 1.3 $\kms$.  A 500 MHz wide-band was placed outside both spectral windows for calibration.  We observed every pointing for 30 seconds, yielding a 24.5-minute observation time per cycle.  At the beginning of the track we observed 2232+117 as the passband calibrator, and at the end of every cycle, we observed 2038+513 as the gain calibrator.  Calibration, imaging and deconvolution were performed using standard procedures of the MIRIAD software package.  In order to have better angular resolution, we applied a Briggs robustness parameter of -1.5, which yields resolution for $\co$ clean map of $5\farcs2\times5\farcs0$, a $\sigma_\mathrm{rms}$ of 0.395 K, and a pixel size of 1$\arcsec$.  For the $\cother$ map, because the signal to noise is worse than $\co$, we have applied a Briggs robustness parameter of +0.5.  The resulting $\cother$ map has a resolution of $5\farcs6\times5\farcs4$, $\sigma_\mathrm{rms}$ of 0.212 K, and a pixel size of 1$\arcsec$.  Figure \ref{figure_ngc6946e} shows the $\co$ map of the region observed overlaid on a HI  image from THINGS (\citealt{2008AJ....136.2563W}).

\subsubsection{1 mm}

Our observations at 3 mm using CARMA likely do not resolve the GMCs.  Therefore, the second step was to make high resolution observations in D array towards some CO complexes to study the properties of individual GMCs using $\cotwo$.  The eleven brightest regions in $\co$ integrated intensity were selected across the area covered by 3 mm observations.  Figure \ref{figure_ngc6946e_13co} illustrates the regions targeted with $\cotwo$ observations.  The observations were taken between July and August 2010.  We set the correlator to have the $\cotwo$ line in the LSB.  In this case we use a 125 MHz band centered on the rest frequency of $\cotwo$ line, and seven 500 MHz wide bands to observe the continuum at 1 mm.  This correlator configuration yields a velocity coverage of 160 $\kms$ and a channel width of 0.5 $\kms$.  Every pointing was observed for 1.5 minutes, which yields a total cycle time of 18 minutes, including a 1.5-minute pointing on the test calibrator 1927+739.  We observed the gain calibrator 1849+670 after every cycle, and 3C454.3 was used as passband calibrator.  We have applied a +0.5 Briggs robustness parameter, which produces a map with angular resolution of 2\farcs45$\times$2\farcs03 corresponding to $\sim$ 50 pc, marginally the scale for GMC sizes.  The rms noise of the clean maps is 0.527 K.  The pixel size of the maps is 0.5$\arcsec$.  The observed regions are illustrated in Figure \ref{fig_co21maps}.

We simultaneously obtained data for  the same 1 mm regions using 14 continuum bands covering 3.5 GHz in each side band.  At $2\farcs43\times1\farcs23$ resolution, the rms sensitivity is given by $\sigma_\mathrm{rms}=3$ $\mathrm{mJy\ beam^{-1}}$, which corresponds to a dust column density $\Sigma_\mathrm{dust} \sim  7.5 \times \Msun\ $pc$^{-2}$, assuming a dust temperature of 20 K and a dust opacity at 1.2 mm of 1 cm$^2$ $g^{-1}$.  Further assuming a dust-to-gas ratio of 0.01 (\citealt{1994A&A...291..943O}), this corresponds to a gas surface density $\Sigma_\mathrm{gas} \sim  7.5 \times 10^2\ \Msun\ $pc$^{-2}$.  We have no detections over the regions observed.

\subsection{Single dish maps}\label{single-dish}

\subsubsection{NRO 45m $\co$ map}\label{nro45}
In order to recover the extended flux, we have merged our $\co$ CARMA data with a NGC 6946 single dish map from the Nobeyama 45-meter single dish telescope (NRO 45m).  NRO 45m observations of NGC 6946 are part of the CARMA and NObeyama Nearby-galaxies (CANON) CO(1-0) Survey project, which uses both CARMA and NRO 45m telescopes to perform resolved observations of GMCs in disks of nearby galaxies (Koda et al., in preparation).  DM12 recently reported observations towards the central part of NGC 6946 using the CARMA and NRO 45m telescopes, and we refer the reader to that paper for the observation strategy and details of NRO 45m map.  Here, we just summarize the most relevant information.  NGC 6946 was observed in the $\co$ transition using the Beam Array Receiver System (BEARS) instrument.  The FWHM of the 45 meter dish is 15$\arcsec$ at the $\co$ rest frequency which is degraded to 19$\farcs$7 after regridding.  The velocity resolution of the map is 2.54 $\kms$.  The rms noise of the $\co$ single dish map is 0.13 K (0.57 Jy beam$^{-1}$), which corresponds to $\sigma(\Sigma_\mathrm{H_2}) \sim 5\ \Msun\ \mathrm{pc}^{-2}$ assuming a 10 $\kms$ line width window to recover the entire flux.

\subsubsection{Heracles $\cotwo$ map}\label{heracles}
In the same way as we did for $\co$ maps, we have merged our $\cotwo$ CARMA data with a NGC 6946 single dish map from the HERACLES project (\citealt{2009AJ....137.4670L}).  HERACLES used the HERA receiver at the IRAM 30m telescope to map the $\cotwo$ transition towards 18 nearby galaxies over the full optical disk, at $13\arcsec$ angular resolution and 2.6 $\kms$ velocity resolution.  At that resolution, the $\sigma_\mathrm{rms}$ in the data cube is 25 mK, which yields an integrated intensity map with a noise level of $\sigma(\Sigma_\mathrm{H_2}) \sim 1\ \Msun\ \mathrm{pc}^{-2}$ assuming that we need to integrate over 10 $\kms$ to recover the entire flux.

\begin{figure*}
\centering
\begin{tabular}{cc}
\epsfig{file=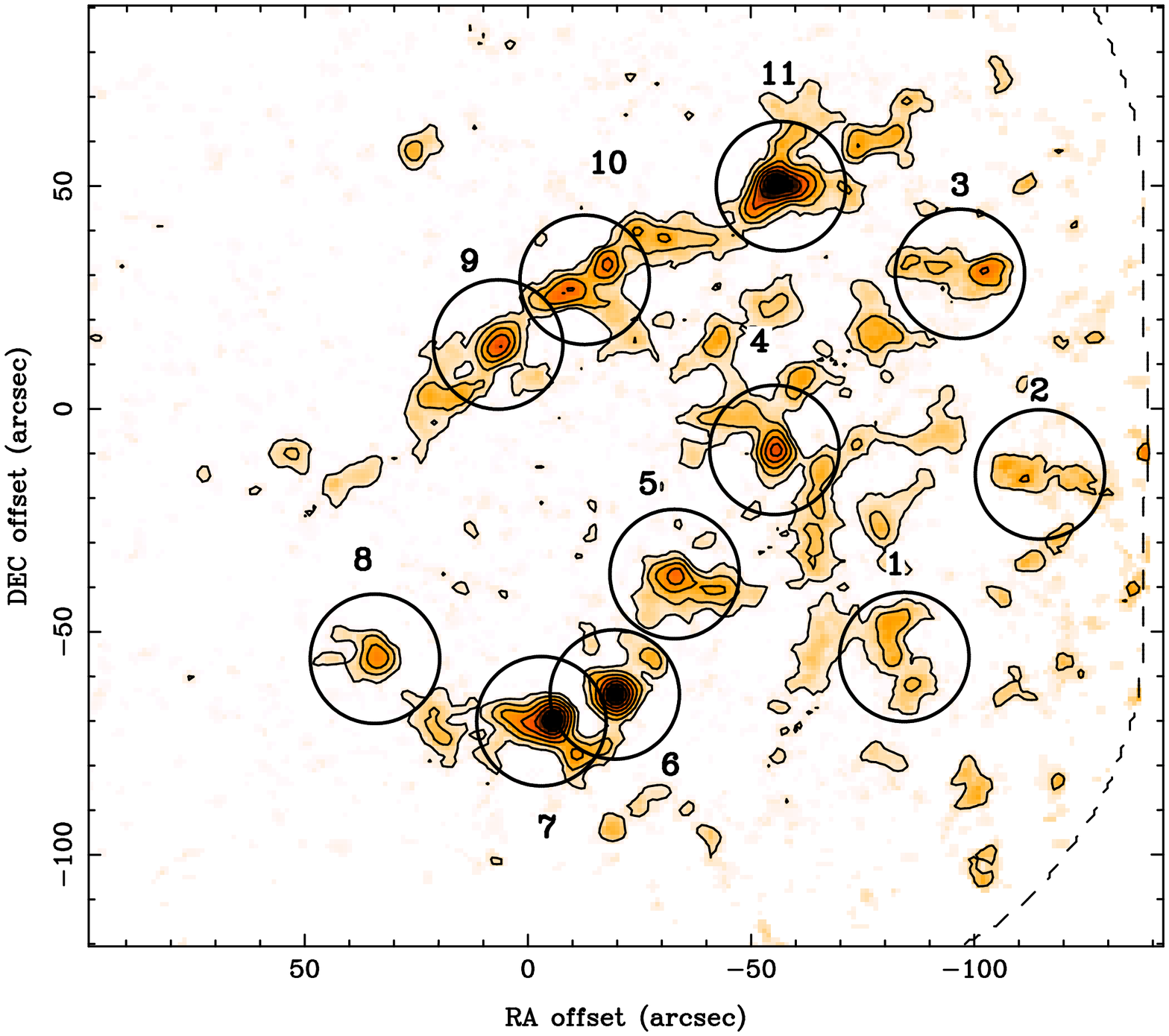,width=0.37\linewidth,angle=-90}
\epsfig{file=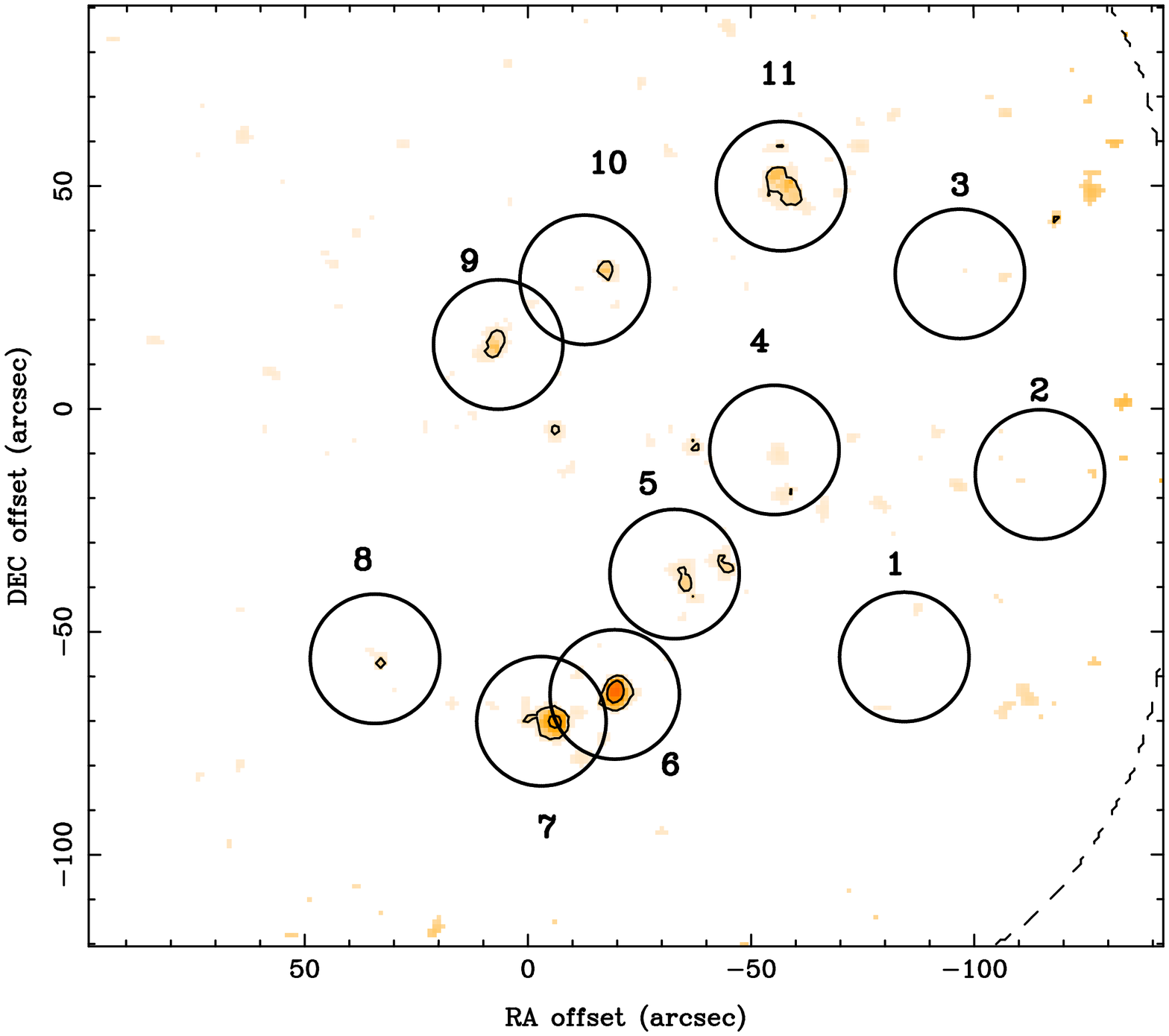,width=0.37\linewidth,angle=-90}
\end{tabular}
\caption{{\bf Left}: $\co$ integrated intensity map.  Contours are as in Figure \ref{figure_ngc6946e}.  Black circles indicate the regions where we performed $\cotwo$ observations.  {\bf Right}:  $\cother$ integrated intensity map.  Contours begin at 1.5$\sigma$ and are spaced by 1.5$\sigma$.  The noise at the center of the map is 2 K km s$^{-1}$.  Significant emission is observed in regions 5, 6, 7, 9 and 11 as is illustrated in the left panel.}
\label{figure_ngc6946e_13co}
\end{figure*}

\subsection{CARMA-Single dish merging procedure}\label{merging}

We have merged our interferometric maps from CARMA with the single dish maps from HERACLES and NRO 45m using the MIRIAD task {\it immerge}.  This task linearly merges a low resolution image, which represents the short-spacing data, with a high resolution image which represents the large spacings in the $uv$ plane.  The cubes were masked to exclude edge velocity channels, and the region where the sensitivity of the CARMA map falls below half of the maximum.  {\it Immerge} allows one to solve for the factor to scale the low resolution image by to put it on the same flux scale as the high resolution image.  Thus, this factor should be close to 1 if both data sets have correctly calibrated flux scales.  In our combination procedure, we obtained values close to 1.05.  We have calculated the flux inside the regions of significant emission provided by CPROPS (see Section \ref{cprops} for details) for both the image from CARMA observations alone, and the image resulting from merging CARMA and single dish data.  In the case of the $\co$ map, we found that most of the regions of significant emission in CARMA+NRO 45m maps present $\lesssim$ 20\% more flux than CARMA maps.  Nevertheless there are a few regions that can have up to 50\% more flux in the CARMA+NRO 45m map.  On the other hand, we found that, for all the regions observed in $\cotwo$, CARMA+IRAM 30m data sets present typically 20\% more flux inside the region of significant emission.  

\subsection{THINGS HI map}\label{things}
Atomic gas surface density is estimated from 21 cm line maps from The HI Nearby Galaxy Survey (THINGS) described in \citet{2008AJ....136.2563W}.  THINGS maps the 21 cm line in 34 nearby galaxies using the Very Large Array (VLA).  The angular resolution of the ``robust'' weighted map is $4\farcs93\times\ 4\farcs51$, the channel width is 2.6 $\kms$, and the noise of the map is $\sim$ 17 K.  Assuming a 20 $\kms$ integration velocity range to fully recover the flux, this noise corresponds to a surface density of $\Sigma_\mathrm{HI} \sim 7\ \Msun \mathrm{pc}^{-2}$.

\subsection{GALEX FUV data}
NGC 6946 FUV data are available from the {\it GALEX} Nearby Galaxies Survey (NGS; \citealt{2007ApJS..173..185G}).  The FUV band covers the wavelength range 1350-1750 \AA, and the maps have an angular resolution of 5\farcs6.  We generate a sky background subtracted FUV image, by subtracting the sky background image from the intensity map, both provided by the NGS project.  We examined the image for presence of foreground stars over the field we are studying in NGC 6946.  Following the approach presented in \citet{2008AJ....136.2782L} (henceforth, L08), we search for foreground stars by using NUV/FUV flux ratio.  We did not detect the presence of  foreground stars over the field we are analyzing, so we did not apply any masking to the FUV map.  We have corrected FUV maps for Galactic extinction using the extinction map of \citet{1998ApJ...500..525S}.  The FUV extinction estimated by using $A_\mathrm{FUV}=8.24\times\  E(B-V)$ from \citet{2007ApJS..173..293W}.

\begin{figure*}[h!]
\centering
\begin{tabular}{ccc}
\epsfig{file=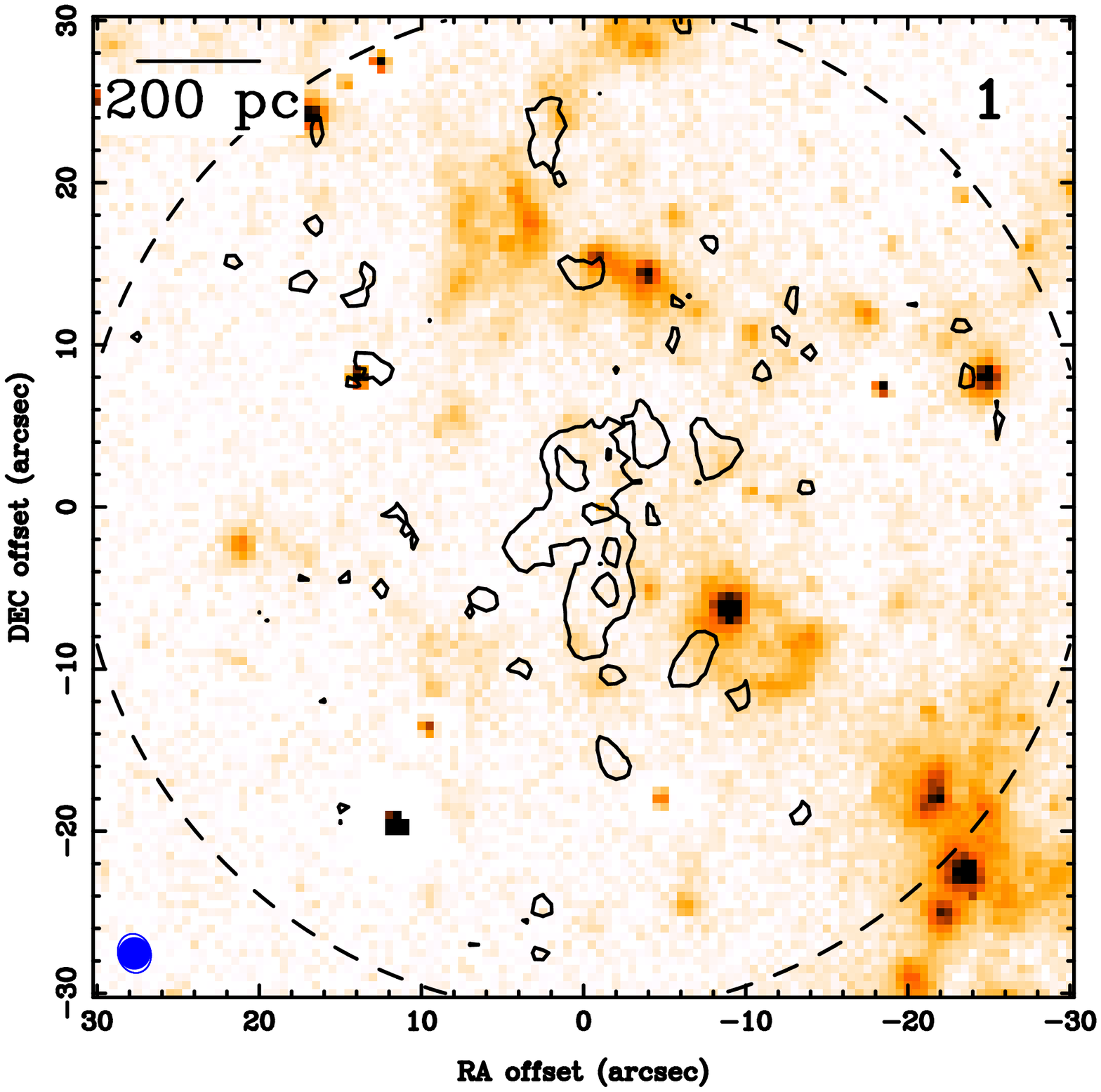,width=0.27\linewidth,angle=-90}
\epsfig{file=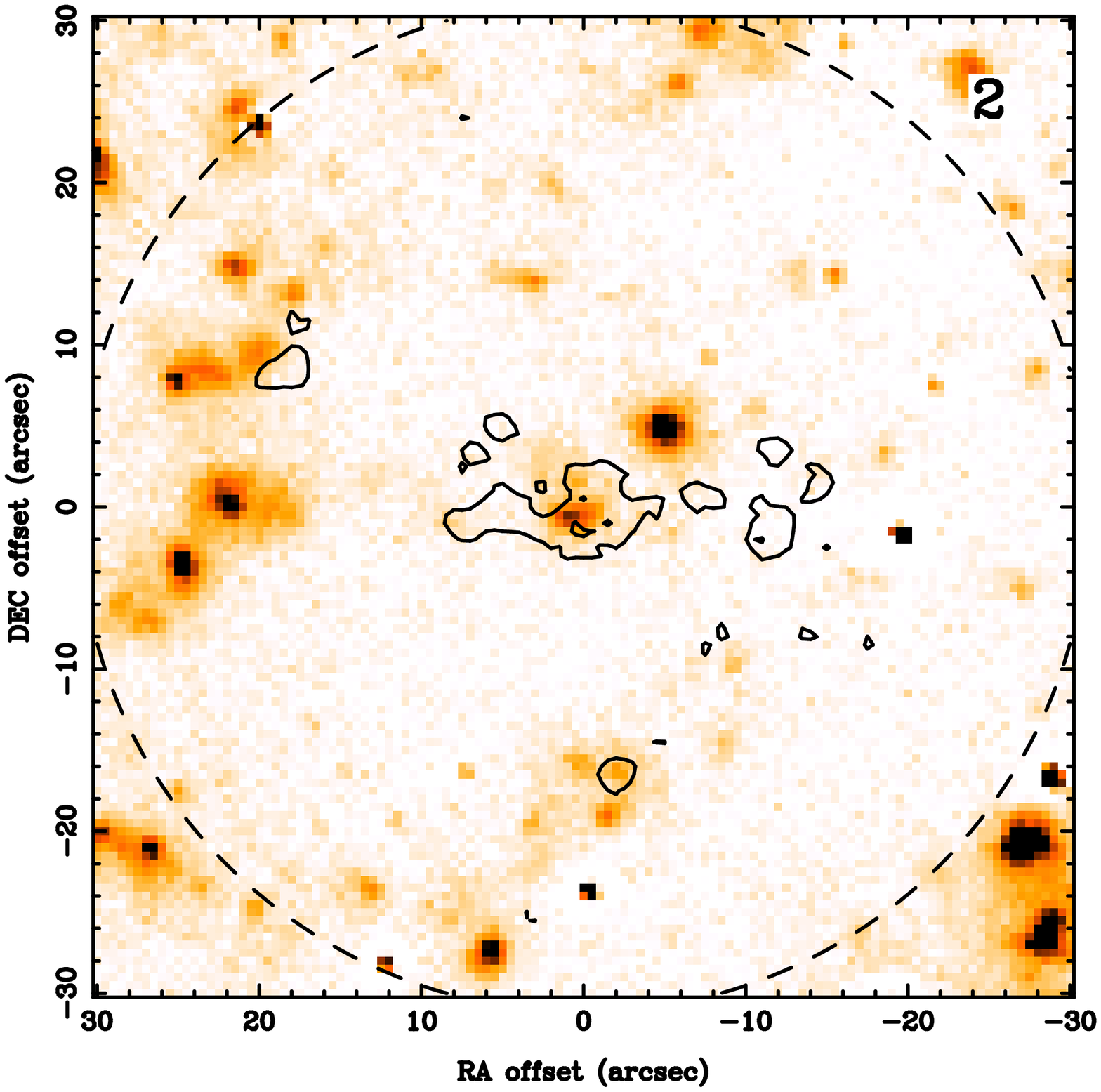,width=0.27\linewidth,angle=-90}
\epsfig{file=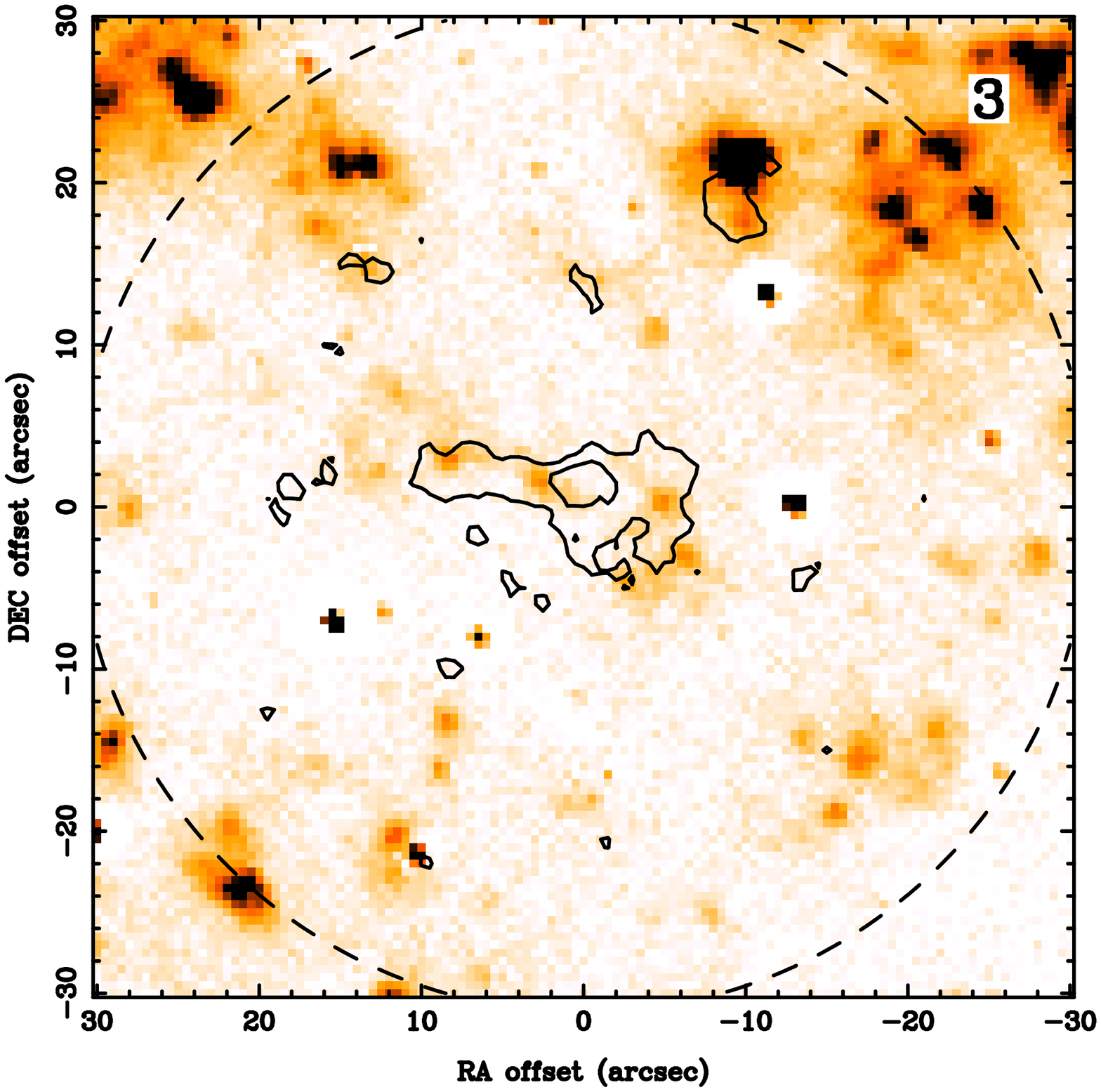,width=0.27\linewidth,angle=-90} \\
\epsfig{file=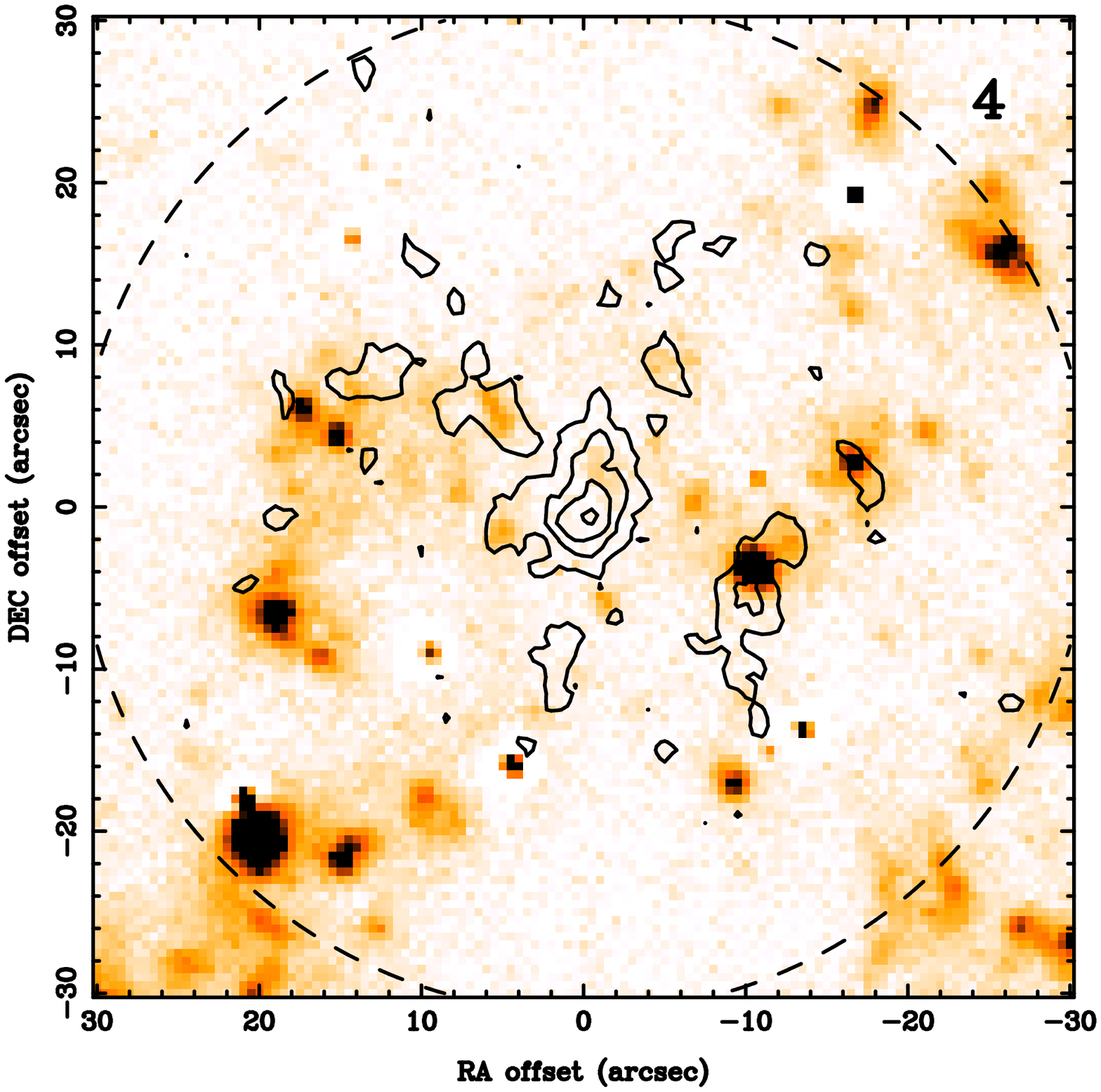,width=0.27\linewidth,angle=-90}
\epsfig{file=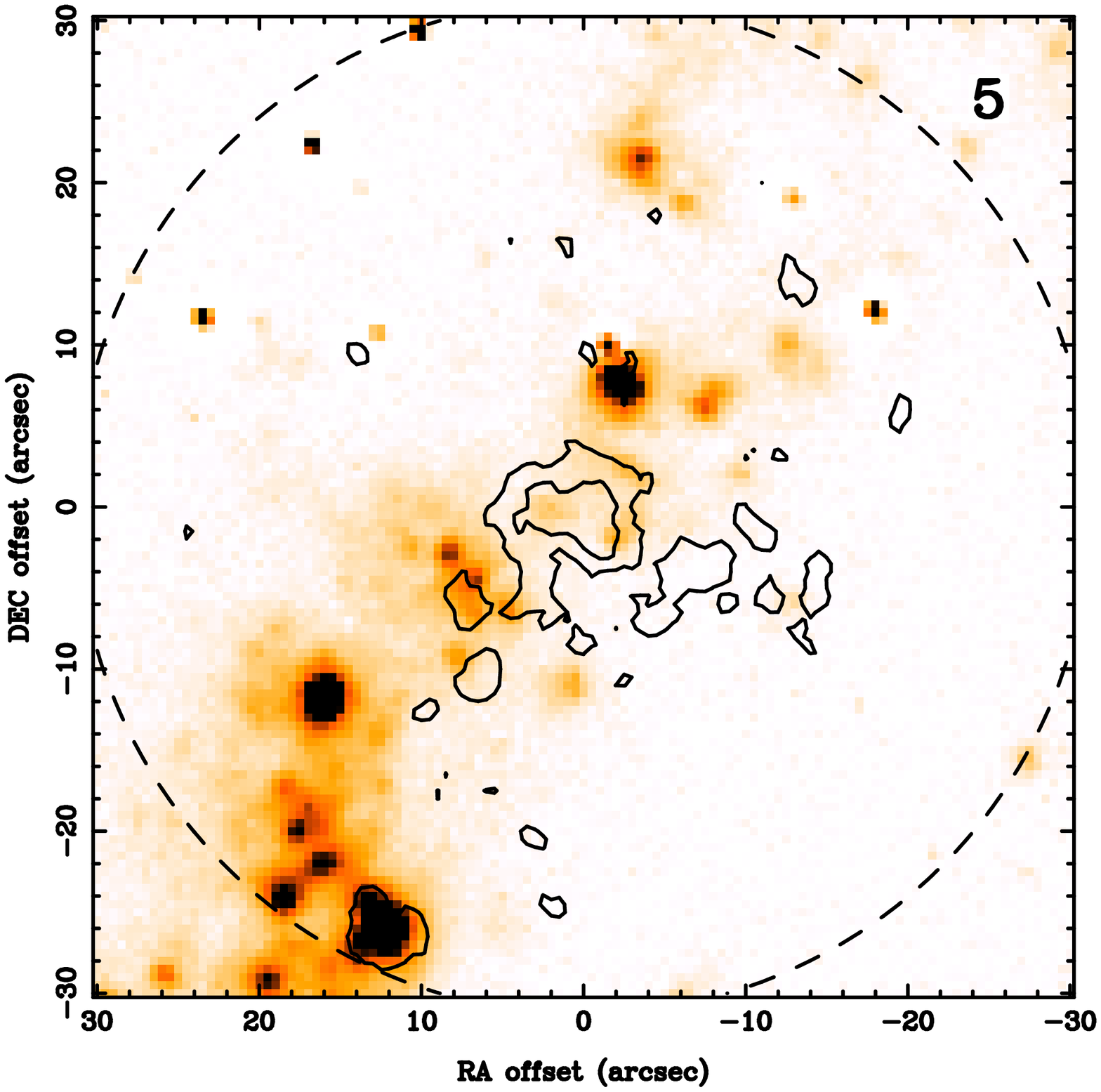,width=0.27\linewidth,angle=-90}
\epsfig{file=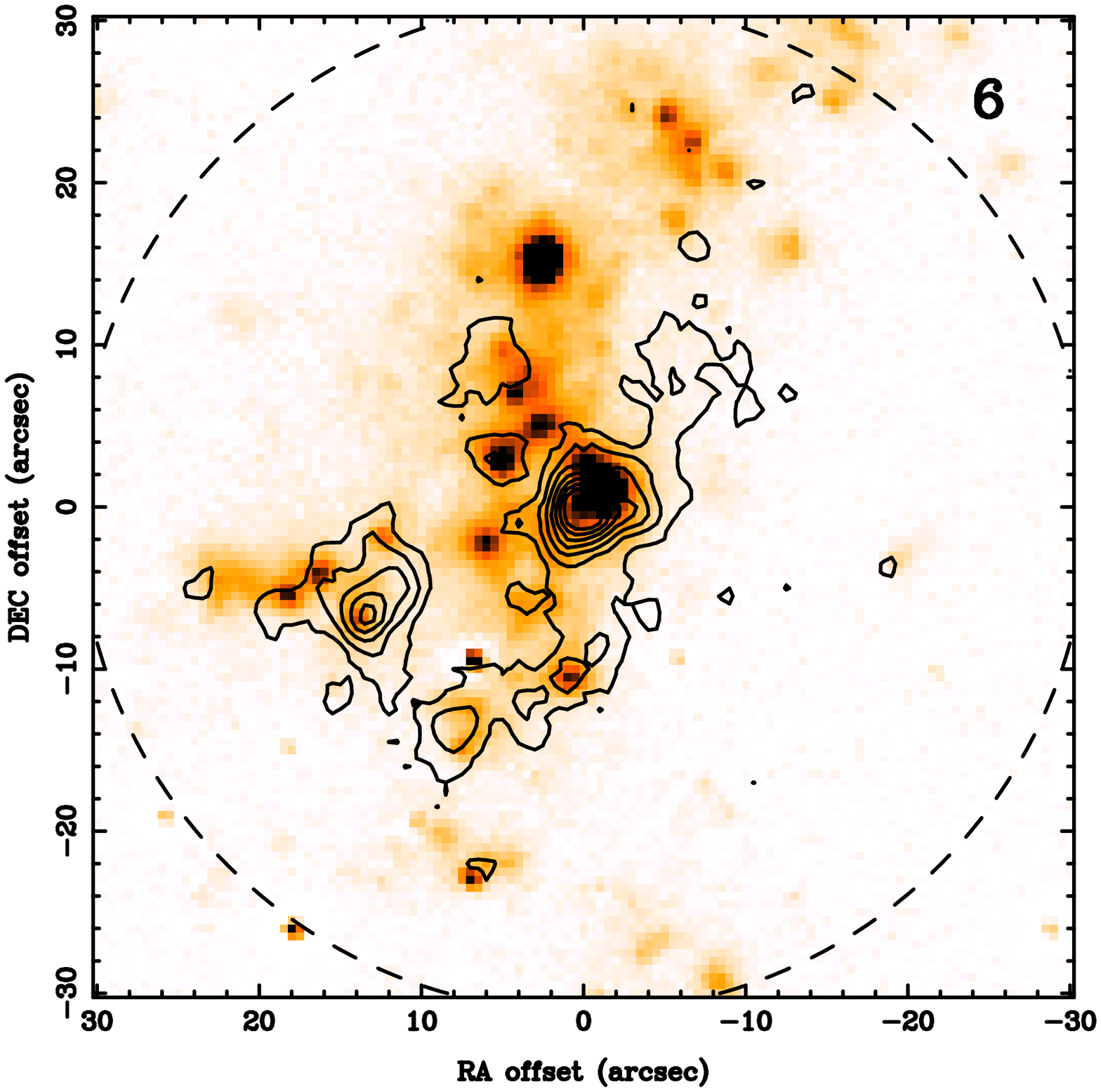,width=0.27\linewidth,angle=-90 } \\
\epsfig{file=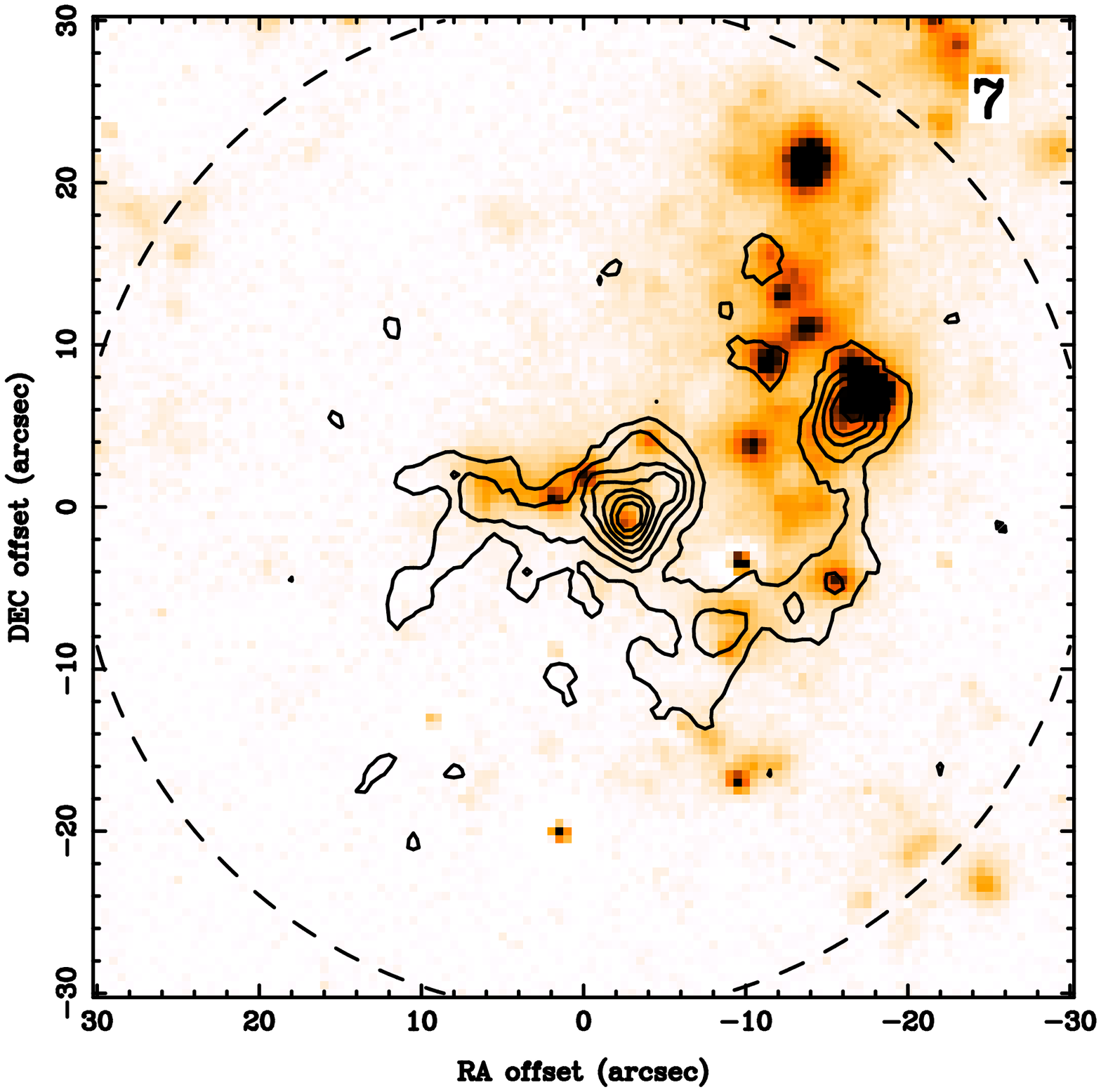,width=0.27\linewidth,angle=-90}
\epsfig{file=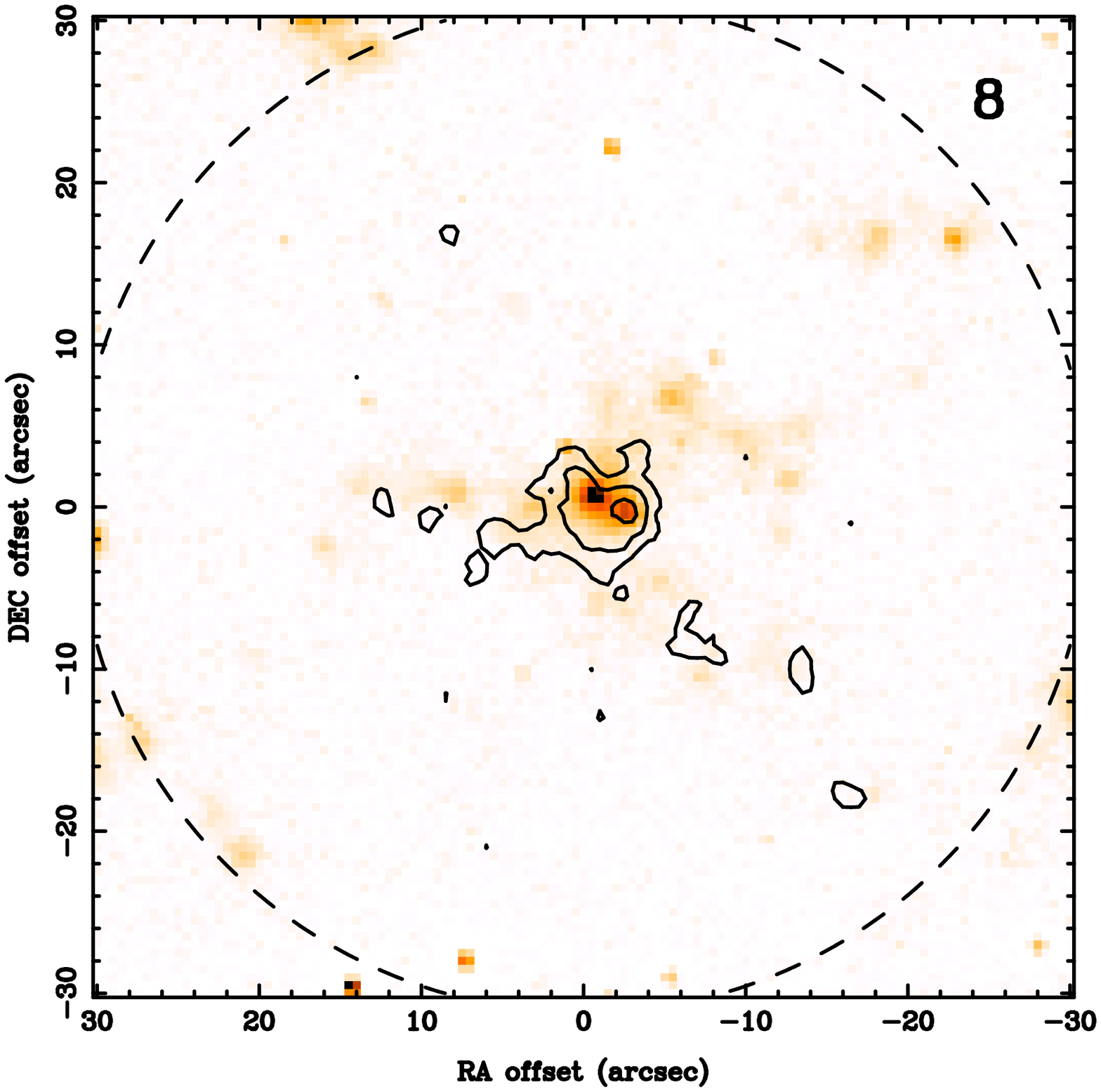,width=0.27\linewidth,angle=-90}
\epsfig{file=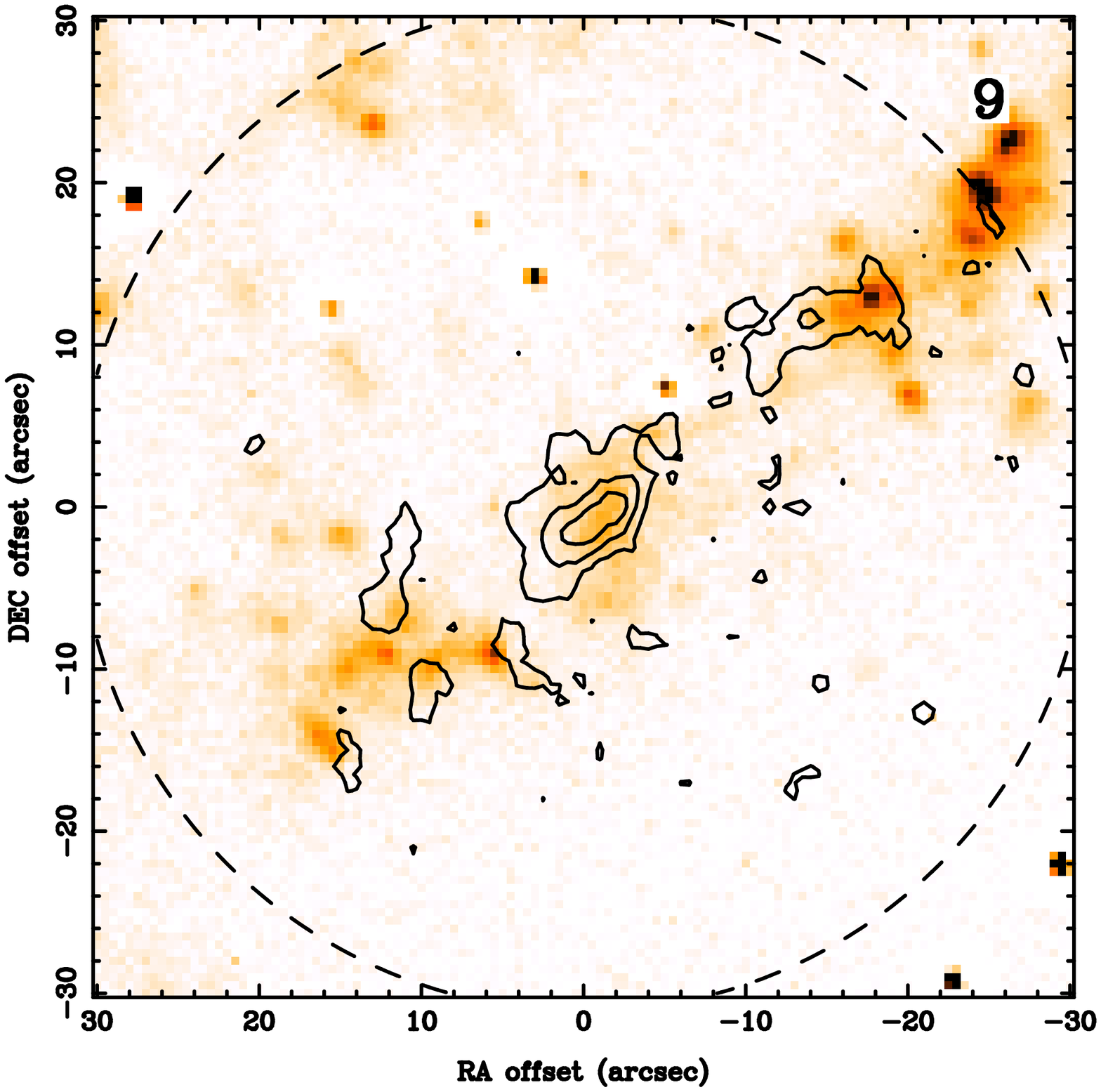,width=0.27\linewidth,angle=-90} \\
\epsfig{file=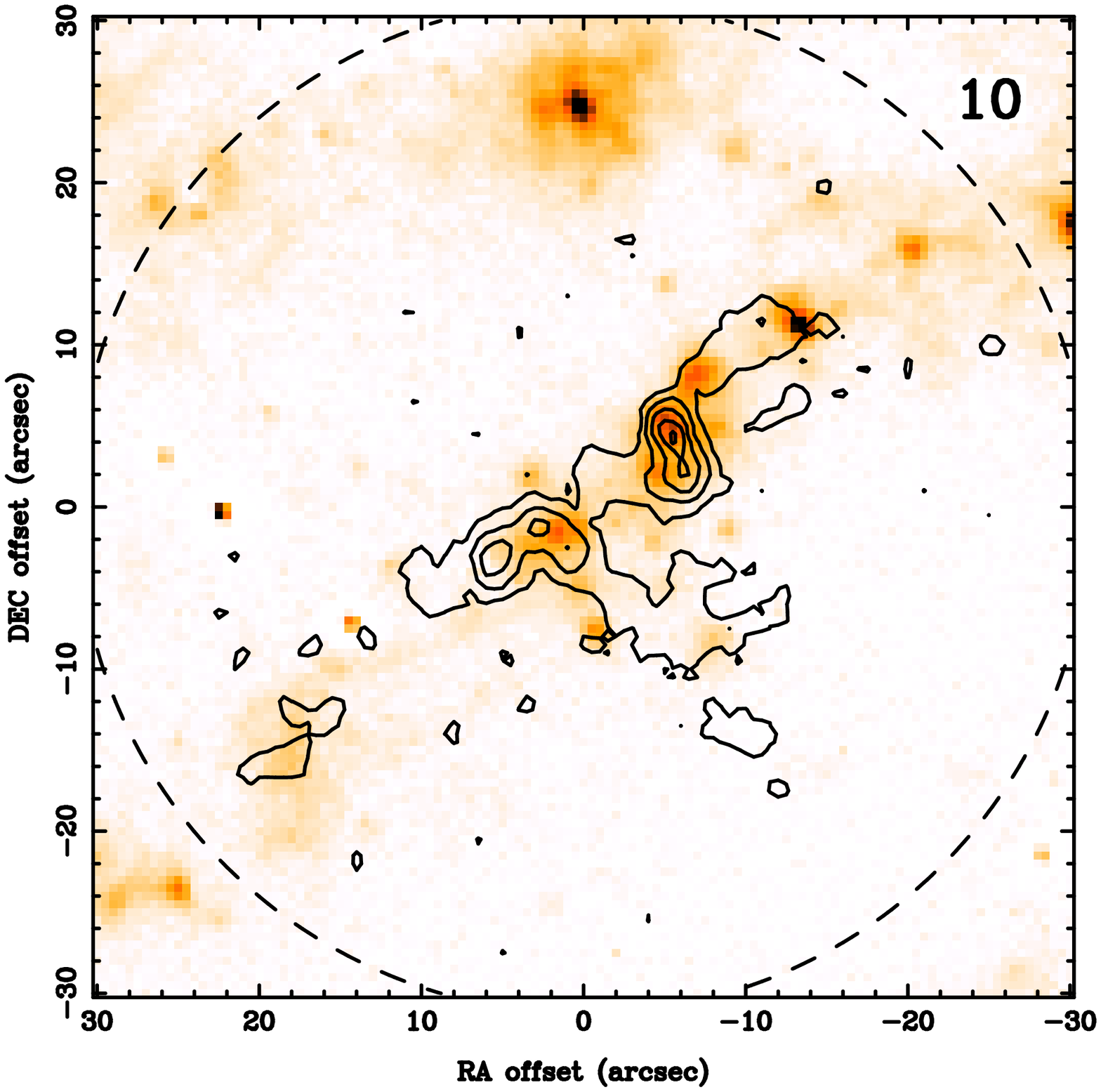,width=0.27\linewidth,angle=-90}
\epsfig{file=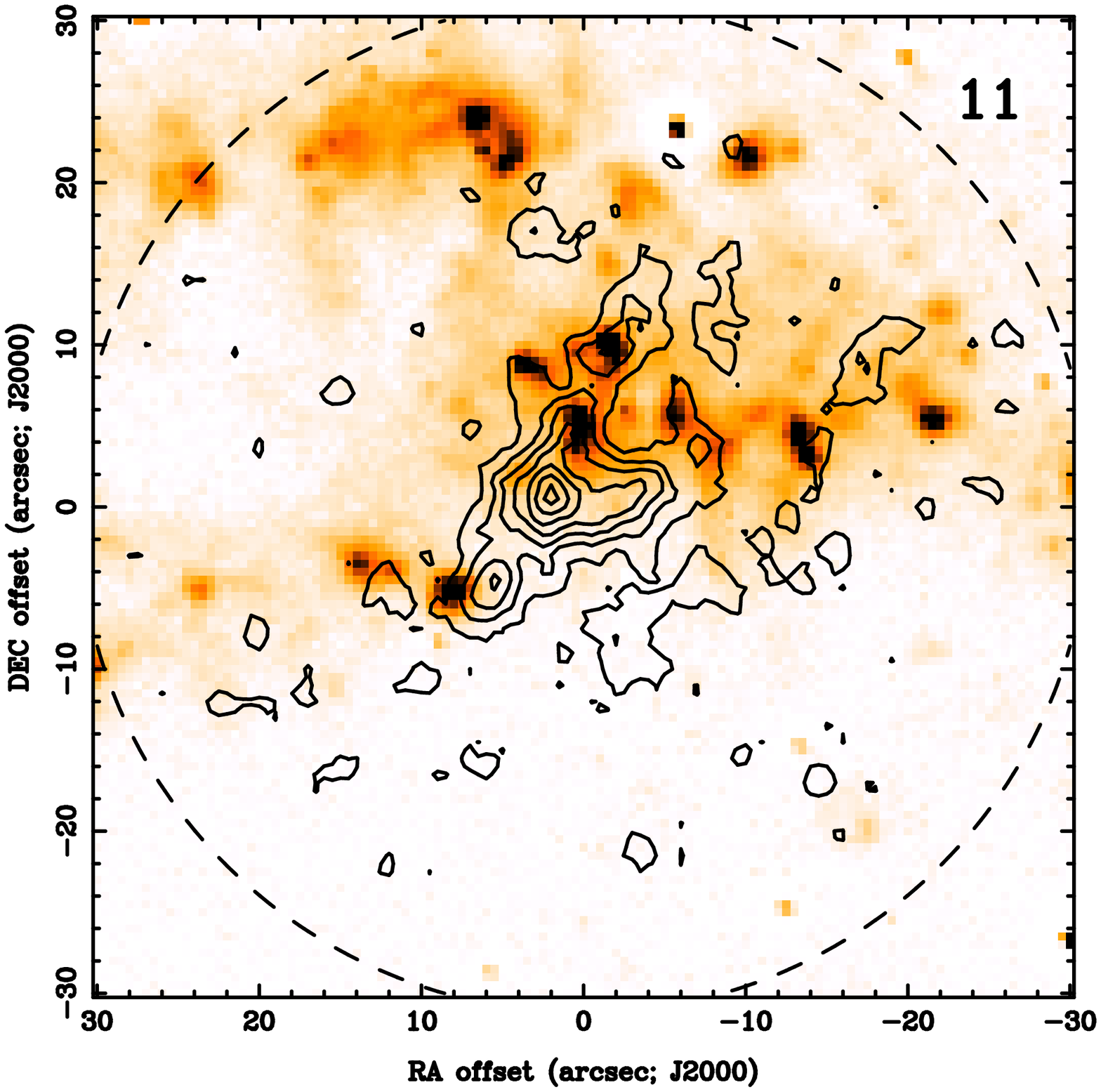,width=0.27\linewidth,angle=-90}
\end{tabular}
\caption{$\cotwo$ integrated intensity contours overlaid on the H$\alpha$ image for all the regions targeted in the eastern part of NGC 6946.  The number given in the top-right corner corresponds to the region number in the full mosaic as illustrated in the left panel of Figure \ref{figure_ngc6946e_13co}.  Contours begin at 5$\sigma$ and are spaced by 5$\sigma$.  The noise at the center of the map is 2.3 K km s$^{-1}$.}
\label{fig_co21maps}
\end{figure*}

\subsection{SINGS data}

\subsubsection{3.6 $\mu$m map}
We use 3.6 $\mu$m maps from the Infrared Array Camera (IRAC, \citealt{2004ApJS..154...10F}) onboard the {\it Spitzer Space Telescope} to estimate the stellar surface density (SINGS, \citealt{2003PASP..115..928K}).  We aim to identify the spiral arms through tracing the old stellar population.  Following a similar approach as taken by \citet{2010ApJ...725..534F}, we have applied a spatial filtering to the map to remove variations of 3.6 $\mu$m surface density which are not related to stellar mass density enhancements.  We have implemented this procedure by using the GIPSY task {\it mfilter}, which applies a median filtering to the pixels inside a given spatial box.  In this study we used a 6.75$\times$6.75 arcsec$^2$ box, which successfully removes the bright spots due to recent star formation in the eastern region of NGC 6946. 

\subsubsection{24 $\mu$m map}
To estimate the amount of obscured star formation by dust, we use the 24 $\mu$m data obtained by SINGS with the MIPS instrument of the {\it Spitzer Space Telescope} (\citealt{2004ApJS..154...25R}).  The angular resolution of MIPS is $5\farcs7$ (FWHM).  We perform a background estimation as follows.  We select a region away from the galaxy, and we did a median spatial filtering over this background region in order to remove bright spots and other features, as we did for the 3.6 $\mu$m map.  Then, the background value is represented by the mean flux over the filtered region.  The value obtained ($\sim$ 0.02 MJy/str) is negligible compared to the flux in the eastern region of NGC 6946, thus we did not apply a background subtraction to our map.

\subsubsection{H$\alpha$ map}
In order to estimate the star formation activity of the molecular structures at the resolution provided by $\cotwo$ ($\sim 2\arcsec$), we need a star formation tracer with resolution finer than the limiting resolution provided by 24 $\mu$m ($\sim 6\arcsec$).  With an angular resolution of $\sim 2 \arcsec$, H$\alpha$ maps provide a useful way to estimate star formation for the smallest identified structures.  An H$\alpha$ image of NGC 6946 is available in the SINGS fourth data release as part of the SINGS ancillary data program.  We used the stellar continuum-subtracted H$\alpha$ image, and we convert to flux units using the SINGS data release documentation.  In order to remove the [N II] contribution, we have assumed ratios of [N II]$\lambda$6584/ H$\alpha$=0.5 and [N II]$\lambda$6548/[N II]$\lambda$6584=0.335 (\citealt{2005ApJ...633..871C}).

\section{Cloud Properties}\label{cprops}

\subsection{Identification}\label{ident}

We identify molecular cloud structures and measure their properties using the cloud properties algorithm (CPROPS) described in \citet{2006PASP..118..590R}.  CPROPS identifies clouds by masking the spectral line cube to isolate regions of significant emission that are both spatially and kinematically connected, and assigns properties like cloud sizes, luminosities and line widths using moments of the emission inside the identified region. Those measurements can be corrected for the bias due to the finite resolution and sensitivity of the maps.  This methodology has already been applied in several extragalactic studies (\citealt{2007ApJ...654..240R}; B08; \citealt{2010MNRAS.406.2065H}, \citealt{2011ApJS..197...16W}).  

We identify regions of significant emission by selecting pixels with emission greater than a threshold of  $n_\mathrm{th}\times \sigma_\mathrm{rms}$ across two consecutive velocity channels.  Then, we extend the region of significant emission to the adjacent regions with emission greater than  $n_\mathrm{edge} \times \sigma_\mathrm{rms}$ in two adjacent velocity channels.  We probed a range of values of $n_\mathrm{th}$ and $n_\mathrm{edge}$ in each data set to suppress the presence of noise in our masked region.  Because the sensitivity in $\co$ map is better than $\cotwo$ cubes, we compare the regions identified in the latter with the emission found in $\co$ data cube.  Regions of marginal significance that do not present a counterpart in $\co$ are considered as spurious detections.  In this study we have used $n_\mathrm{th}=4$ and $n_\mathrm{edge}=3$ for both $\co$ and $\cotwo$ maps.

After regions of significant emission are identified, CPROPS will find individual clouds based on conditions that depend on the decomposition parameters used.  A cloud is considered as a separate structure if it is large enough to allow the algorithm to calculate properties, the contrast between the peak and the edge is larger than the minimum allowed, and the properties of the cloud change significantly when it is combined with a nearby identified cloud (see Appendix of \citealt{2006PASP..118..590R} for a complete description of the decomposition parameters).  Due to the observed substructure in the ISM, the parameters for decomposition must be based on the known properties of the objects we want to identify.  Previous studies have used the "physical priors" set of parameters presented by \citet{2006PASP..118..590R} which are based on the observed properties of GMCs in the Milky Way.  In that case, the decomposition is performed at a spatial resolution of 15 pc and velocity resolution of 2 $\kms$.  Because we are limited by the resolution of our maps, in our study we use parameters based on the spatial and spectral resolutions of the data cube.  Then, in finding local peaks in regions of significant emission we search in boxes of $\approx$ 110 pc (5.1$\arcsec$) and 2.5 $\kms$ in velocity for the $\co$ maps and $\approx$ 50 pc (2$\arcsec$) and 2.5 $\kms$ in the case of $\cotwo$.  Because the limiting resolution of the $\co$ maps likely does not allow us to resolve individual clouds, particularly the smallest ones, henceforth the structures identified in $\co$ maps will be referenced as ``molecular complexes''.  On the other hand, structures identified in $\cotwo$ will be called ``clouds''.  

\subsection{Size, line width and CO flux}\label{gmc-props}

The sizes, line widths and luminosities of the identified structures are derived using moments of the brightness distribution.  The rms size $\sigma_{r}$ is determined by the geometric mean of the second moments of emission along the major and minor axes.  The velocity dispersion $\sigma_{v}$ is determined by the second moment of the emission along the velocity axis.  The flux of the cloud is calculated using the zeroth moment of the emission in position and velocity.  As we noted at the beginning of this section, the sensitivity and resolution of the data can affect the derived properties of the clouds.  For instance, the size of the clouds will be underestimated by the second moment of the flux in a map with finite signal to noise.  CPROPS reduces this bias by extrapolating the size, the velocity width and the flux to the case of infinite sensitivity (i.e. brightness temperature of the edge of the cloud equal to 0 K).  The extrapolation for the size and the velocity is linear, while the extrapolation for the flux is quadratic.  \citet{2006PASP..118..590R} recommend correcting for resolution bias as well, which can be significant for extragalactic observations.  We correct for the effect of finite spatial and spectral resolution by subtracting in quadrature the beam size and the spectral channel profile from the extrapolated measurements of size and velocity width respectively.

The physical quantities are derived from the moment measurements following the standard procedure described in \citet{2006PASP..118..590R} and B08.  Following the definition of S87, the cloud size is defined as $R=1.91\sigma_r$.  Although this value is $\sim$ 10\% lower than the value we would obtain from the equivalent area of the cloud (assuming $\beta = 1$ in the density profile $\rho \propto r^{-\beta}$), we used this definition in order to compare our values to the sizes presented by S87 and B08.  The CO luminosity $L_\mathrm{CO}$ is given by

\begin{equation}\label{Lco-equ}
\frac{L_\mathrm{CO}}{\mathrm{K\ \kms pc^{2}}}=\frac{F_\mathrm{CO}}{\mathrm{K\ \kms arcsec^2}}\left(\frac{D}{\mathrm{pc}}\right)^{2}\left(\frac{\pi}{180 \times 3600}\right)^{2}
\end{equation}

\noindent where $D$ is the distance to NGC 6946 in parsecs.  

\subsection{Mass of the clouds}

Once we have determined the basic properties of the clouds (size, line width and luminosity), other cloud properties can be estimated by taking combinations of these basic properties.  For instance, the luminosity-based mass is obtained from $L_{\mathrm{CO}}$ by using  

\begin{equation}\label{Mco-equ}
\frac{M_\mathrm{lum}}{\mathrm{M}_\odot}=4.4 \frac{L_\mathrm{CO}}{\mathrm{K\ \kms pc^{2}}} \frac{X_\mathrm{CO}}{2\times10^{20}\mathrm{cm}^{-2}(\mathrm{K}\ \kms)^{-1}}.
\end{equation}

\noindent where $X_\mathrm{CO}$ is the assumed CO-to-H$_2$ conversion factor.  This conversion factor is thought to depend on local ISM properties such as metallicity or radiation field (\citealt{2006MNRAS.371.1865B}; \citealt{2011ApJ...737...12L}).  In this study we assume $X_\mathrm{CO}=2\times10^{20}\ \mathrm{cm}^{-2}(\mathrm{K}\ \kms)^{-1}$, which is broadly used in previous extragalactic studies (e.\ g., B08; L08).  In their study of GMCs within the central 5 kpc of NGC 6946, DM12 found $X_\mathrm{CO}=1.2\times10^{20}\ \mathrm{cm}^{-2}(\mathrm{K}\ \kms)^{-1}$ within a factor of 2 uncertainty, which is consistent with the value assumed here.  Equation \ref{Mco-equ} includes a factor of 1.36 to account for the presence of helium.  In this study, we have assumed $I(2 \rightarrow 1) = I(1 \rightarrow 0)$ in order to be consistent with the analysis presented in B08.

Virial masses are normally calculated under the assumption that molecular clouds are spherically symmetric.  However, the observed cloud shapes present a more complex structure.  Generalizing the standard virial analysis for spherical clouds, \citet{1992ApJ...395..140B} calculated the virial mass for spheroidal clouds.  Following Appendix A of  \citet{1992ApJ...395..140B} the gravitational energy of spheroidal clouds is given by 

\begin{equation}\label{grav-energy}
W=-\frac{3}{5}a_{1}a_{2}\frac{GM^{2}}{R}, 
\end{equation}

\noindent where $R$ is the radius perpendicular to the axis of symmetry of the cloud, while $a_{1}$ and $a_{2}$ are the density distribution and shape-dependent factors respectively.  For a $\rho(r) \propto r^{-\beta}$ power law density distribution, $a_1$ is given by $a_1=(1-\beta/3)/(1-2\beta/5)$.  On the other hand, $a_{2}$ will change if the cloud presents an oblate or prolate form.  If the cloud is oblate, $a_2$ is given by $\arcsin{\epsilon}/\epsilon$, where $\epsilon=(\sqrt{1-y^{2}})$ is the eccentricity of the cloud, and $y=Z/R$ is the ratio between the size along and perpendicular to the axis of symmetry.  If the cloud presents a prolate shape instead, $a_2=\mathrm{arcsinh}\ \epsilon/\epsilon$.  Ignoring the magnetic energy and external pressure terms in the virial equilibrium equation, and assuming $\beta=1$, the virial mass is given by

\begin{equation}\label{Mvir-equ}
\frac{M_\mathrm{vir}}{\mathrm{M}_\odot}=1040\left(\frac{\sigma_{v}}{\kms}\right)^{2}\left(\frac{R}{\mathrm{pc}}\right)\frac{1}{a_2},
\end{equation}

\noindent which corresponds to the expression for the virial mass assuming a spherically symmetric geometry, with a correction factor of 1/$a_2$.  The distribution of $a_2$ is observed to be narrow.  We found that the mean of $a_2$ for $\co$ complexes was 0.77, with a variance of 0.02 around the mean.  On the other hand, the mean of $a_2$ for $\cotwo$ clouds was 0.81, with a variance of 0.02.  Thus, the shape factor $a_2$ can account for a correction of the spherically symmetric virial mass by $\sim$ 30\% for complexes, and $\sim 20\%$ for $\cotwo$ clouds.

A complete description of the sizes and orientations of the clouds in three dimensions is preferable (e.g.\ \citealt{2010ApJ...712.1049S}), but that information is not available for observational data.  In particular, a correction for inclination can be only applied statistically, and not for individual clouds.  Thus, we assume here that the clouds can be fully described by the sky-projected major and minor axes.  We make the further assumption that the axis of symmetry is given by the major axis of the clouds, and the clouds are prolate.  The correction for the shape of the cloud can strongly affect the virial mass estimates for structures that present a large axis ratio.  Because we assume elongation in the plane of the sky, a structure elongated along the line of sight will appear roughly circular, and our correction will be inappropriate.  Nevertheless, given the sizes found in this regions ($\sim$ 50-150 pc), and the usual thickness of the molecular gas disk in galaxies (e.g., \citealt{2011AJ....141...48Y} find molecular gas thickness $\sim$200 pc for NGC 891), it is likely that the axes of symmetry of the clouds are directed along the disk, reducing the overcorrection we may be applying to our estimates.  Supporting this assumption, \citet{2006ApJ...638..191K} have provided observational evidence that molecular clouds are elongated along the Galactic plane. 

We followed the definition of the dimensionless virial parameter $\alpha_\mathrm{vir}$ given by \citet{1992ApJ...395..140B}, 

\begin{equation}\label{alpha}
\alpha_\mathrm{vir}=\frac{M_\mathrm{vir}}{M_\mathrm{lum}},
\end{equation}

\noindent which measures the relative importance of the cloud's kinetic energy compared to its gravitational energy.  This parameter is usually used to investigate whether the clouds are in virial equilibrium, $\alpha \simeq 1$; for $\alpha > 1$ the self-gravity is not important, and the clouds are confined due to an external pressure if the clouds are assumed to be in steady state.

The uncertainties assessed to the moment measurements (and thus to the physical quantities and their derivatives) are estimated using bootstrapping methods, and this is the only source of uncertainties that we include in our analysis.

\subsection{LTE masses}\label{lte-mass}

Additionally, we have used the $\cother$ line map to calculate LTE masses over the regions with significant emission.  If we assume that all energy levels are populated according to local thermodynamic equilibrium (LTE) at the temperature $T_\mathrm{ex}$,  the column density of the $^{13}\mathrm{CO}$ molecule in terms of the opacity of the $J=1\rightarrow0$ transition is given by (e. g., \citealt{2004tra..book.....R})

\begin{equation}\label{Nco-equ}
\frac{N(^{13}\mathrm{CO})}{\mathrm{cm}^2}=3.0\times 10^{14} \frac{T_\mathrm{ex}}{1-\exp(-5.3/T_\mathrm{ex})}\int{\tau_{13}}\mathrm{d}v,
\end{equation}

\noindent where $v$ is in $\kms$. In this study we have determined the excitation temperature by using the assumed optically thick line $\co$

\begin{equation}\label{tex-equ}
T_\mathrm{ex}=\frac{5.5}{\ln(1+5.5/T_\mathrm{B}^{12}+0.82)},
\end{equation}

\noindent and assuming a filling factor of unity.  The mean of $T_\mathrm{ex}$ is 4.2 K considering the regions of significant emission in the $\co$ map, while this mean increases to 5 K when considering the $\cother$ regions of detectable emission.  In section \ref{mvir-lum} we will investigate the effect of using higher values of $T_\mathrm{ex}$ in the LTE mass estimates.  The optical depth of the $\cother$ line is given by

\begin{equation}\label{tau13-equ}
\tau_{13}=-\ln\left[1-\frac{T_\mathrm{B}^{13}}{5.3}\left(\frac{1}{\exp(5.3/T_\mathrm{ex})-1}-0.16\right)^{-1}\right].
\end{equation}

By assuming that $\cother$ line is optically thin, the total molecular mass can be computed using an appropriate abundance ratio,

\begin{equation}\label{Mlte-equ}
\frac{M_\mathrm{LTE}}{\mathrm{M}_\odot}=\mu_{m}\left[\frac{\mathrm{H}_2}{\mathrm{^{13}CO}}\right] D^2 \int{N(^{13}\mathrm{CO})\mathrm{d}\Omega}
\end{equation}

\noindent where $\mu_{m}$ is the mean molecular mass per H$_2$ molecule, $D$ is the distance, and $[\mathrm{H}_2/\mathrm{^{13}CO}]$ is the abundance ratio.  In this study we have assumed the abundance ratio to be $7\times10^5$.  This value is close to the ratio given by the relation found by \citet{2005ApJ...634.1126M}, which yields $[\mathrm{^{12}CO/^{13}CO}]\sim 60$ for $R_\mathrm{gal} \sim 6$ kpc, and assuming a $\mathrm{H_2/^{12}CO}$ abundance ratio of 1.1 $\times\ 10^4$ (\citealt{1982ApJ...262..590F}).

\subsection{Mass surface density} 
Another important cloud property is the mass surface density.  Surface density has been estimated by dividing the luminosity-based mass by the corresponding area covered by the  $\cotwo$ cloud (or $\co$ complex), thus $\Sigma_\mathrm{H2}=M_\mathrm{lum}/(\pi R^2)$.  In general the areas for $\cotwo$ clouds are smaller than $\co$ complexes.  The fluxes and areas involved in these calculations are the non-extrapolated values.  We have made this choice, because we intend to compare molecular gas surface density with other tracers (HI, 24$\mu$m, H$\alpha$, FUV) over the boundary of the clouds given by the mask provided by CPROPS.  Because we do not have extrapolated or deconvolved measurements of the fluxes for these tracers, we have used the raw properties of the clouds.

\subsection{Star Formation surface density}\label{sfr-mol}
In this paper, we have estimated the star formation rate (SFR) using FUV, 24$\mu$m and H$\alpha$ images described in Section \ref{obs}.  The method to estimate the star formation activity is different for $\co$ complexes and $\cotwo$ clouds, mainly due to the limiting spatial resolution of the SFR tracer images.  The details of the calculations are given below.

\subsubsection{SFR in $\co$ complexes}
Following the approach given by L08, we estimate the star formation rate surface density in $\co$ complexes using {\it GALEX} FUV and {\it Spitzer} 24 $\mu$m maps.  By tracing photospheric emission of O and B stars, FUV emission measures unobscured star formation over time scales of 10-100 Myr (\citealt{2005ApJ...633..871C}).  On the other hand, 24 $\mu$m traces the small dust grains heated by UV photons radiated from young stars, thereby tracing obscured star formation over time scales of $\sim$10 Myr (\citealt{2005ApJ...633..871C}; \citealt{2007ApJ...666..870C}).  The use of 24 $\mu$m as tracer of obscured star formation was analyzed by \citet{2007ApJ...666..870C}.  They showed that the combination of H$\alpha$ (unobscured SFR) and 24$\mu$m (obscured) recovers the total SFR, which they estimated by using Paschen-$\alpha$ (Pa$\alpha$) emission as a tracer of ionizing photons unaffected by extinction.  L08 used a similar prescription to derive the total SFR, but using FUV maps instead of H$\alpha$ to trace unobscured SFR.  They found that the total star formation surface density $\Sigma_\mathrm{SFR}$ can be estimated by:

\begin{equation}\label{sfr}
\frac{\Sigma_\mathrm{SFR}}{ \Msun \mathrm{yr}^{-1} \mathrm{kpc}^{-2}}=(8.1\times 10^{-2} I_\mathrm{FUV} + 3.2\times 10^{-3} I_\mathrm{24\mu m}) \cos i 
\end{equation}

\noindent where $I_\mathrm{FUV}$ and $I_{24\mu m}$ are in units of MJy/sr.  

\subsubsection{SFR in \cotwo\ clouds}\label{sfr-co21}
The coarse resolution of the 24$\mu$m map ($5\farcs7$) does not permit us to use the SFR from Equation (\ref{sfr}) to perform a direct comparison between star-forming regions and the structures identified in $\cotwo$.  We therefore use the H$\alpha$ image to estimate the amount of star formation in $\cotwo$ clouds.  However, H$\alpha$ emission can be strongly affected by dust extinction, and a correction should be applied to H$\alpha$-based SFR estimates to account for embedded star formation.  Following \citet{2007ApJ...666..870C}, the star formation surface density can be estimated by

\begin{equation}\label{sfr_ha}
\frac{\Sigma_\mathrm{SFR}}{ \Msun \mathrm{yr}^{-1} \mathrm{kpc}^{-2}}=(5.3\times 10^{-42} L_\mathrm{H\alpha})\times 10^{A_{\mathrm{H}\alpha}/2.5} \cos i 
\end{equation}

\noindent where $L_\mathrm{H\alpha}$ is in units of ergs s$^{-1}$, and $A_{\mathrm{H}\alpha}$ is the H$\alpha$ extinction.  In the Section \ref{sfr-result}, we will show that applying a $A_{\mathrm{H}\alpha}=1.0$ mag roughly recovers the total star formation as traced by FUV+24$\mu$m.

\section{Results}\label{cl-prop}

\subsection{Uncorrected properties}

\begin{figure*}
\centering
\begin{tabular}{ccc}
\epsfig{file=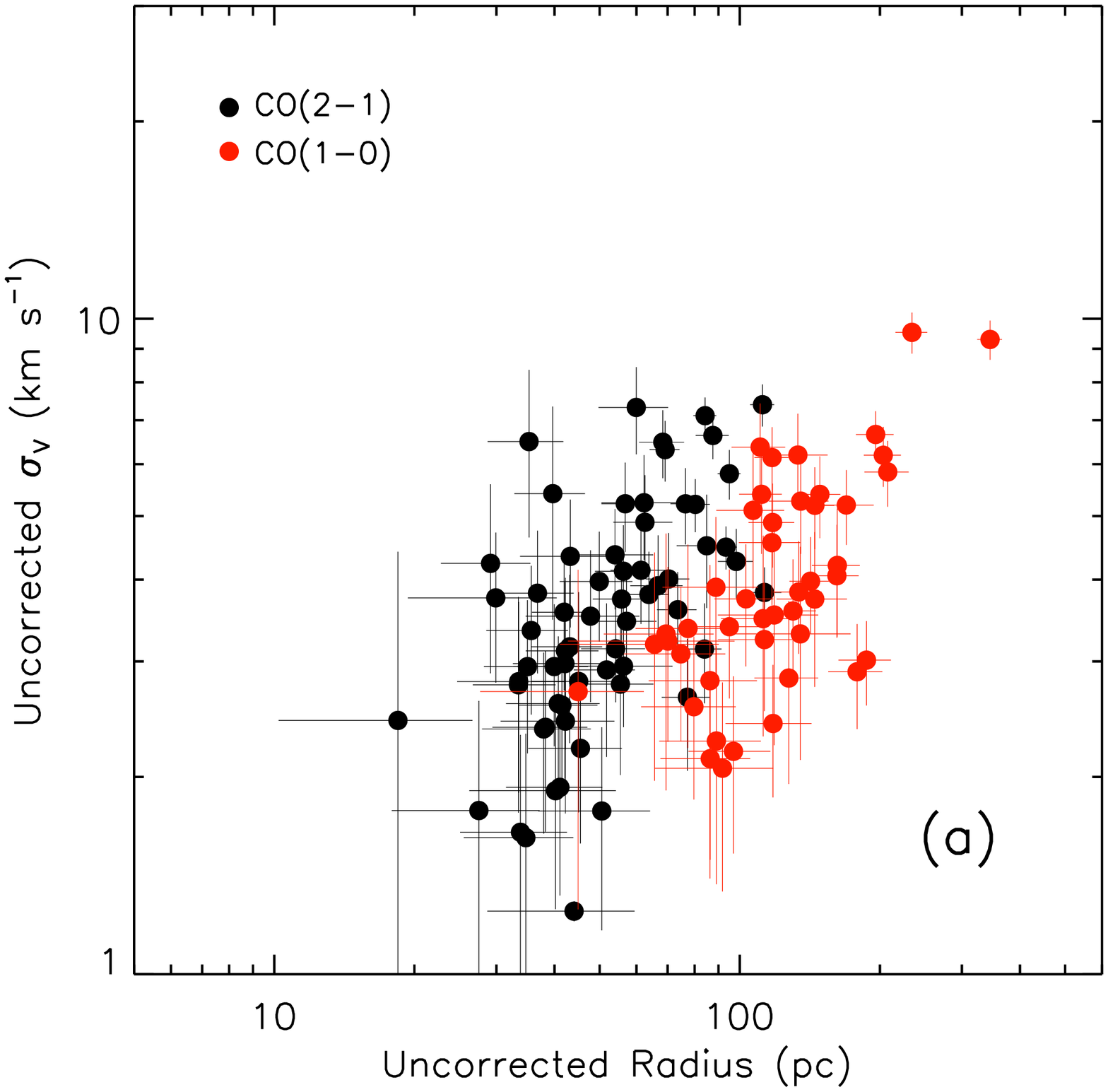,width=0.3\linewidth,angle=90}
\epsfig{file=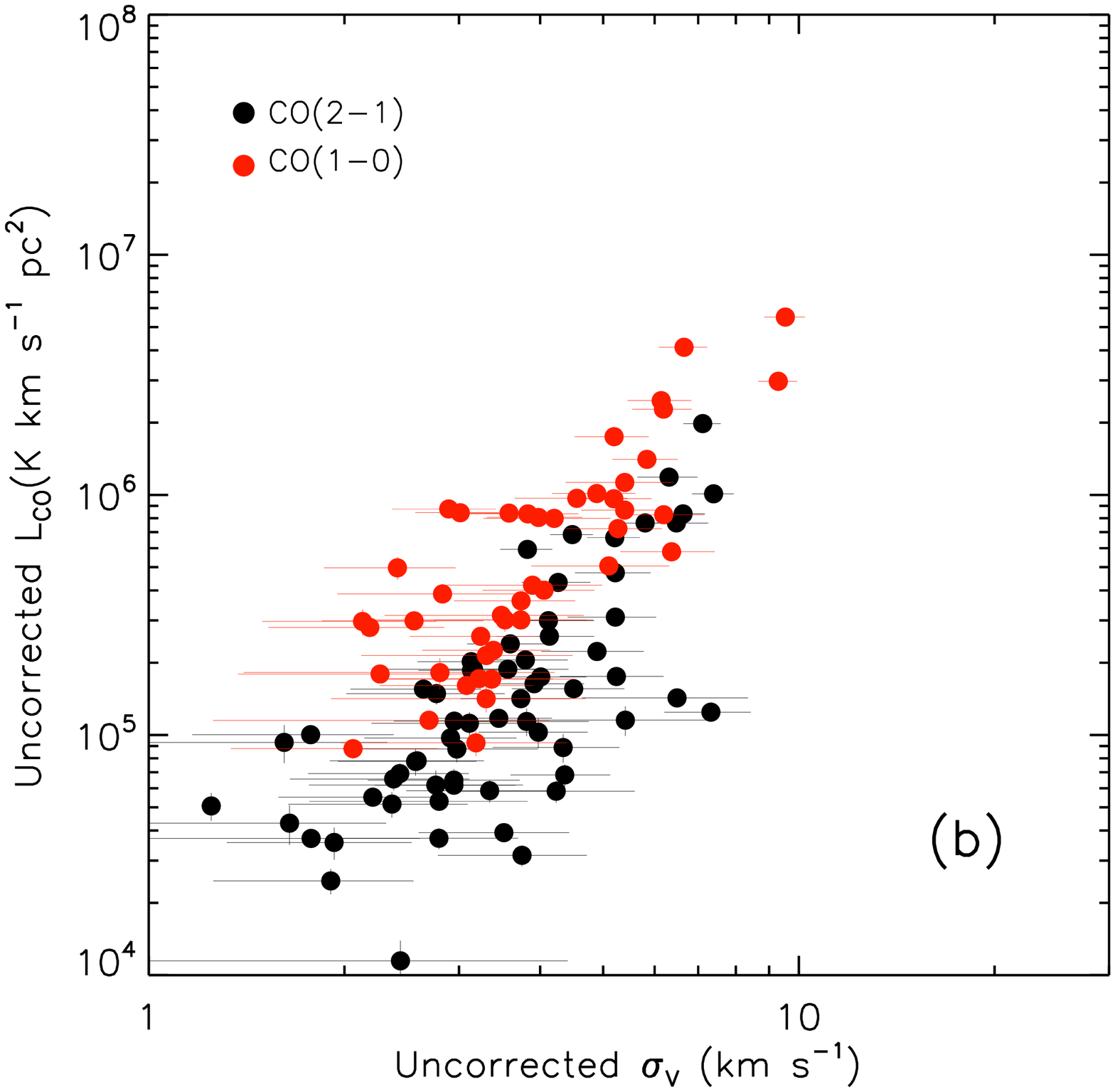,width=0.3\linewidth,angle=90}
\epsfig{file=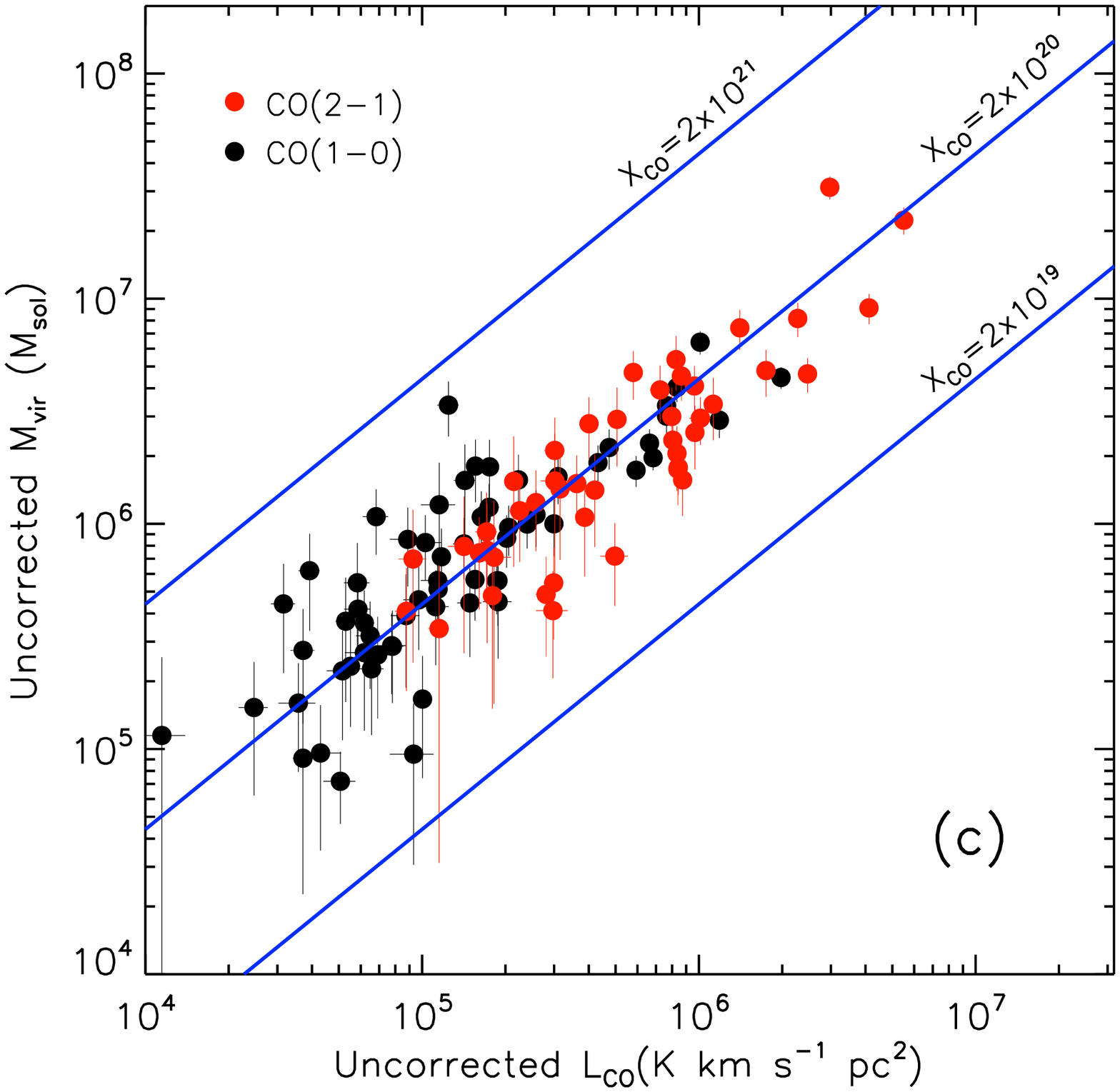,width=0.3\linewidth,angle=90}
\end{tabular}
\caption{Properties of the 45 molecular complexes and 64 clouds found in this study without sensitivity or resolution corrections.  {\bf (a)}: Size-line width relation.  Black dots represent the clouds identified in the $\cotwo$ maps, while red dots correspond to $\co$ complexes.  {\bf (b)}: Luminosity - line width relation.  {\bf (c)}: Virial mass - Luminosity relation.  Solid blue lines show different $X_\mathrm{CO}$ values.}
\label{fig_co10_siwi_raw}
\end{figure*}

Although the CPROPS method corrects for sensitivity and resolution bias, those corrections can introduce some additional scatter in the cloud property scaling relations.  Therefore, we will start this section by showing the scaling relations  without any correction to assess overall correlation strengths.  We have identified 45 complexes in the $\co$ channel map.  The mean uncorrected radius of the $\co$ clouds is 128.5 $\pm$ 3.1 pc, and the mean uncorrected line width is 4.2 $\pm$ 0.1 $\kms$.  On the other hand, the mean uncorrected luminosity of these complexes is (8.6 $\pm$ 0.1) $\times 10^5$ $\Kkmspc$, which corresponds to a mean luminosity-based mass of 3.8 $\pm$ 0.05 $\times 10^6 \Msun$.  Figure \ref{fig_co10_siwi_raw} shows the scaling relations for the raw properties.

We selected the 11 most massive $\co$ complexes to make follow-up observations with $\cotwo$, and they are shown in Figure \ref{figure_ngc6946e_13co}.  We identified 64 giant molecular clouds in the regions observed, and their scaling relations are shown in Figure \ref{fig_co10_siwi_raw} as well.  For this set of clouds, the mean radius is 54.9 $\pm$ 1.1 pc, the mean line width is 3.7 $\pm$ 0.1 $\kms$, the mean luminosity is (2.4 $\pm$ 0.03) $\times$ $10^5\ \Kkmspc$, and the mean luminosity-based mass is (1.04 $\pm$ 0.01)$\times$ $10^6\ \Msun$.

\begin{table}
\caption{Spearman rank correlation coefficient $r$ for both raw and bias corrected relations.\label{table-rank}}
\centering
\begin{tabular}{cccccc}
\hline\hline
&\multicolumn{2}{c}{$\co$ complexes}& &\multicolumn{2}{c}{$\cotwo$ clouds}\\
\cline{2-3} \cline{5-6} 
Relation & Raw & Bias Corrected & & Raw &  Bias Corrected \\
\hline
$\sigma_{v}-R $& 0.63 & 0.32 & & 0.60 & 0.17   \\
$L_\mathrm{CO}-\sigma_{v}$&  0.77 & 0.50  & &  0.74  & 0.64 \\
$M_\mathrm{vir}-L_\mathrm{CO}$ & 0.87 & 0.76 & & 0.86 & 0.74 \\
\hline
\end{tabular}
\end{table}

The significance of the correlations between the properties are estimated using the Spearman rank correlation coefficient $r$.  
We want to assess the change of this coefficient when both sensitivity and resolution bias are corrected for.  Table \ref{table-rank} shows the values found for the scaling relations before and after the corrections for resolution and sensitivity have been applied.  For the $\co$ complexes and $\cotwo$ clouds, the $\sigma_v-R$ relation has $r \sim 0.6$ for the uncorrected properties, while the correlation is weaker when corrections are applied ($r < 0.35$).  The extra scatter in the latter relation can be explained by the resolution bias correction.  This correction is sensitive to the axis ratio of the cloud being deconvolved, and can be extremely large if the minor axis is close to the beam size.  Thus, elongated clouds can deviate more from the original relation than more circular clouds of similar average size.  For the $L_\mathrm{CO}-\sigma_v$ relation, $\co$ complexes and $\cotwo$ clouds show a similar correlation $r \sim 0.76$, and the relations are relatively stable after corrections ($r \sim 0.65$ for $\cotwo$ clouds and $r \sim 0.5$ for $\co$ complexes).  Among all the scaling relations, the $M_\mathrm{vir}-L_\mathrm{CO}$ relation presents the highest $r$ values:  $r \sim 0.86$ for raw data, and $r \sim 0.75$ for the corrected values for both $\co$ and $\cotwo$.  This is expected, as both $M_\mathrm{vir}$ and $L_\mathrm{CO}$ involve the product of size and line width.

\subsection{Corrected properties and scaling relations}\label{cl-prop-corr}

Larson's laws reflect the dynamic state of the ISM, which is dictated by the interplay between self-gravity, turbulence and feedback processes from star formation.  Following Larson's pioneering work, and using an improved sample of Galactic clouds, S87 found that the line widths of the clouds increase as a power of their radius, i.e., the size-line width relation is given by $\sigma_v=0.72\ R^{0.5}\ \kms$, where $\sigma_v$ is the velocity dispersion of the clouds, and $R$ is the radius in parsecs.  Combining this relation with virial equilibrium, they showed that all GMCs have approximately the same mass surface density $\Sigma \approx 170\ \Msun$ pc$^{-2}$. Additionally, they found that the virial mass and the CO luminosity follow an approximately linear correlation, $M_\mathrm{vir}=39\ L_\mathrm{CO}^{0.81}\ \Msun$, where $L_\mathrm{CO}$ has units of $\Kkmspc$.  The third law relates the CO luminosity and the line width via $L_\mathrm{CO}=130\ \sigma_v^{5}\ \Kkmspc$.
 
As observational capabilities have improved in the past few years, several studies have established scaling relations for extragalactic clouds as well (\citealt{2007prpl.conf...81B}; B08; \citealt{2010MNRAS.406.2065H}).  In their study of molecular clouds in Local Group and dwarf galaxies, B08 found scaling relations similar to those found in Galactic cloud studies.  Nevertheless, they noticed that GMCs in low metallicity dwarf galaxies (such as the Small Magellanic Cloud) are larger than clouds found in the Milky Way, M31 and M33 for a given velocity dispersion or CO luminosity.  This observational trend was confirmed by \citet{2010MNRAS.406.2065H} in their study of the GMCs in the Large Magellanic Cloud.

To find the best fit relations between the properties of the clouds, in this paper we used the bisector linear regression method described in \citet{1990ApJ...364..104I},  which consists of the line that bisects the ordinary least-square regression (OLS) lines in both axes, i.e. OLS($Y$|$X$) and OLS($X$|$Y$) lines.  This method treats the variables symmetrically, by taking into account the uncertainties measured in the independent and dependent variables, along with the intrinsic scatter of the data.

\subsubsection{Size-line width relation}\label{size-width}
 
Figure \ref{figure_line-width}(a) shows the scatter plot of the line width and size of both complexes and clouds.  Along with the scatter plot, we have included the relations found by S87 and B08 for comparison.  The best-fitting relation for $\cotwo$ clouds is 

\begin{equation}\label{eq-size-width}
\frac{\sigma_v}{\kms}=(0.14 \pm 0.05)R^{0.88 \pm 0.08}
\end{equation}

\noindent where $R$ is in pc.  Similarly, for $\co$ complexes the best-fitting relation is

\begin{equation}\label{size-width10}
\frac{\sigma_v}{\kms}=(0.27 \pm 0.19)R^{0.65 \pm 0.14}.
\end{equation}

We observe that $\co$ complexes show smaller line widths than those found by S87 and B08 for a given radius.  On the other hand, $\cotwo$ clouds are roughly centered in the S87 and B08 fitting lines, but present a bigger scatter around those relations.  Additionally, in Figure  \ref{figure_line-width} we have included the GMCs found by DM12 in their study of the central part of NGC 6946 at a spatial resolution similar to our resolution for $\cotwo$ maps ($\sim$ 2\arcsec).  Although we have identified clouds with similar sizes and line widths, they found a set of clouds with larger velocity dispersion at sizes $\sim$100 pc.  These clouds happen to be located within $\sim$400 pc of the center of the galaxy.  Similar behavior is observed for clouds near the Galactic center (\citealt{2001ApJ...562..348O}).

\subsubsection{Luminosity-line width relation}\label{lum-width}

Figure \ref{figure_line-width}(b) shows the $L_\mathrm{CO}$-$\sigma_v$ relation for the clouds in our sample.  This figure reveals that in NGC 6946, clouds are generally located above the relation found by B08, i.e.\ they are overluminous for their velocity dispersion.  Nevertheless, this trend breaks down for clouds with luminosity smaller than $\sim 3\times 10^{5}\ \Kkmspc$, where the velocity dispersion spreads over a wide range.  The best-fit relation for $\cotwo$ clouds yields  

\begin{equation}\label{lco-width}
\frac{L_\mathrm{CO}}{\Kkmspc}=((4.99 \pm 2.99) \times 10^3)\sigma_v^{2.66 \pm 0.33}
\end{equation}

\noindent where $\sigma_v$ is in $\kms$.  The best-fit luminosity-line width relation for emitting complexes derived from $\co$ is given by

\begin{equation}\label{lco-width10}
\frac{L_\mathrm{CO}}{\Kkmspc}=((1.91\pm 1.15)\times 10^4)\sigma_v^{2.41 \pm 0.32}.
\end{equation}

Our results for $\co$ complexes are consistent with the values presented by DM12 for the central region of NGC 6946.  $\cotwo$ clouds, on the the other hand, show higher velocity dispersions for luminosities below $\sim 3 \times 10^{5}\ \Kkmspc$.

\subsubsection{Mass estimates of the observed molecular gas}\label{mvir-lum}

According to the conceptual model provided by S87, a mass-luminosity relation is a natural consequence of the internal structure and gravitational equilibrium of the observed clouds.   The empirical relation found by them ($M_\mathrm{vir}=39\ L_\mathrm{CO}^{0.81}$) can be explained by combining the size-line width relation with the assumption of virial equilibrium, plus assuming similar kinetic temperatures and area filling factors for the clouds.  This provides a framework for using the optically thick CO emission as a tracer of the molecular mass, in which the virial masses and CO luminosities of clouds are used to solve for the CO-to-H$_2$ conversion factor $X_\mathrm{CO}$.  Nevertheless, they emphasized that this relation depends on the temperature of cloud, and may be different in extragalactic sources, particularly at the centers of galaxies where temperatures and densities may deviate from values for molecular clouds in the disk of the Milky Way.  Surprisingly, recent extragalactic studies have found that values of the conversion factor are relatively close to those derived from Galactic studies.  For instance, for a sample of disk and dwarf galaxies in the Local Group, B08 found an $X_\mathrm{CO}$ consistent within a factor of two with the value derived for Galactic clouds.  Moreover, using high resolution observations of GMCs within the central 5 kpc of NGC 6946, DM12 found an $X_\mathrm{CO}$ average value of 1.2 $\times$ 10$^{20}\ \mathrm{cm}^{-2}(\mathrm{K}\ \kms)^{-1} $, just a factor of 2 below the value derived from S87.

\begin{figure*}[ht!]
\centering
\begin{tabular}{ccc}
\epsfig{file=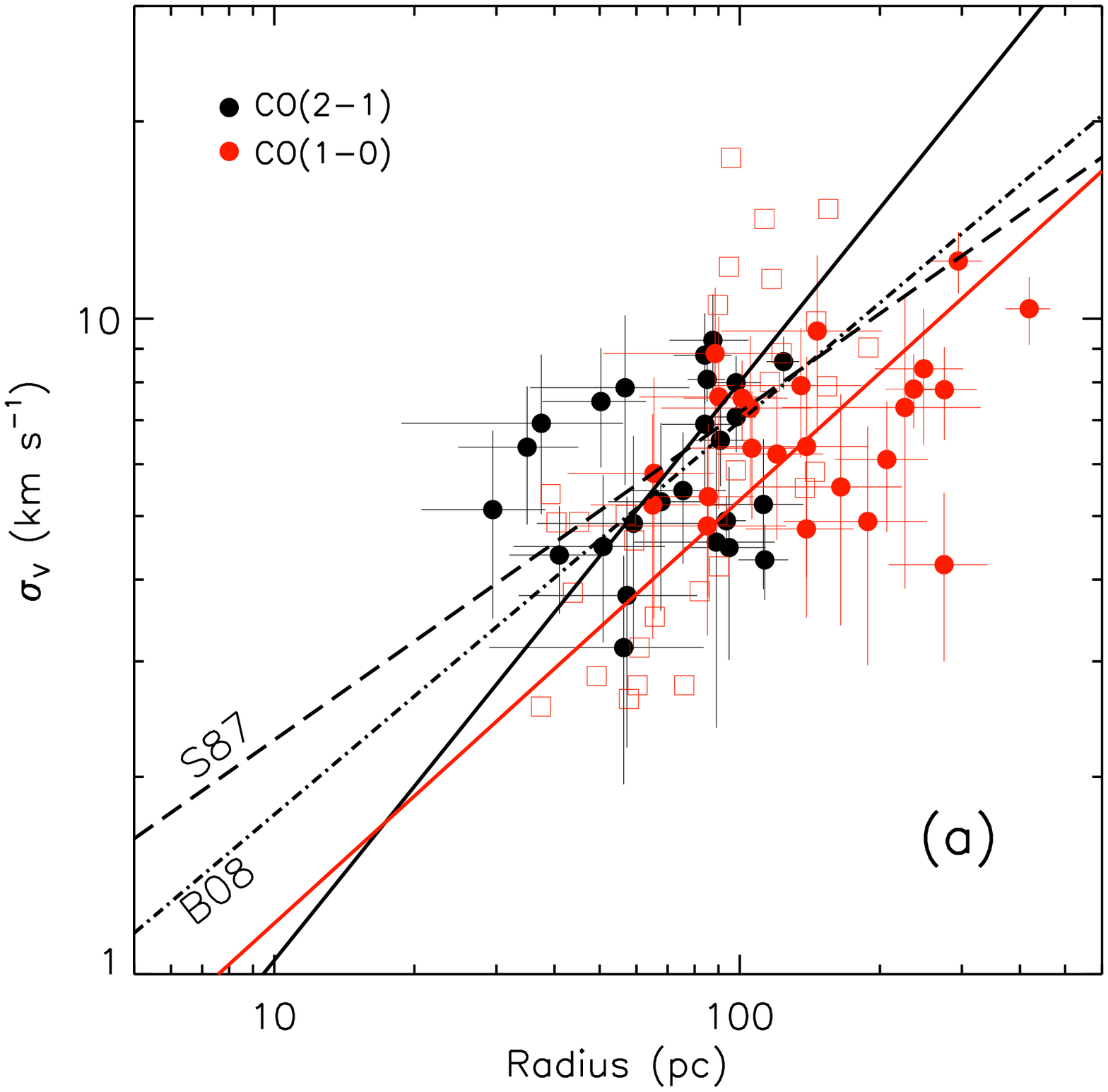,width=0.3\linewidth,angle=90}
\epsfig{file=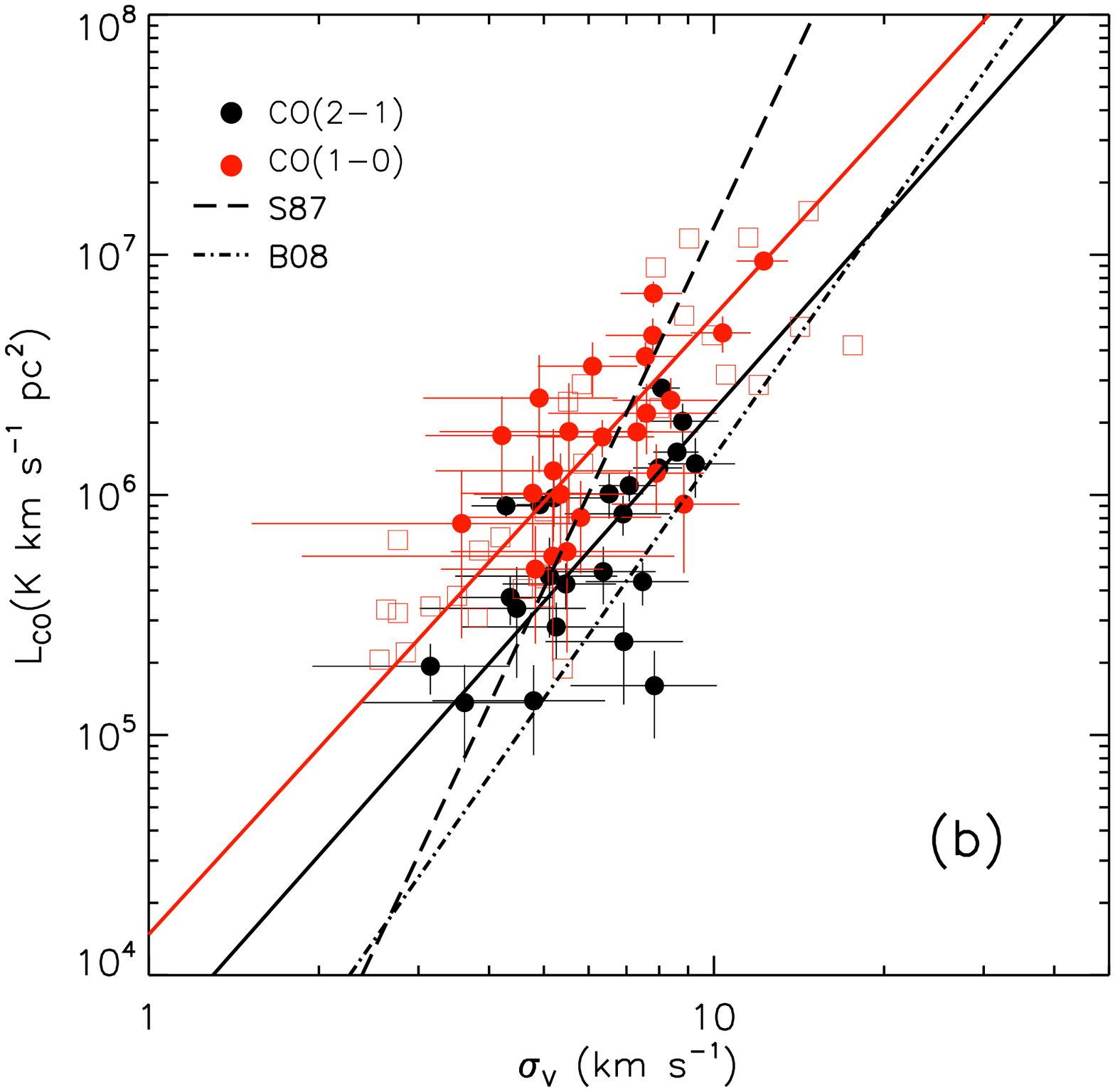,width=0.3\linewidth,angle=90}
\epsfig{file=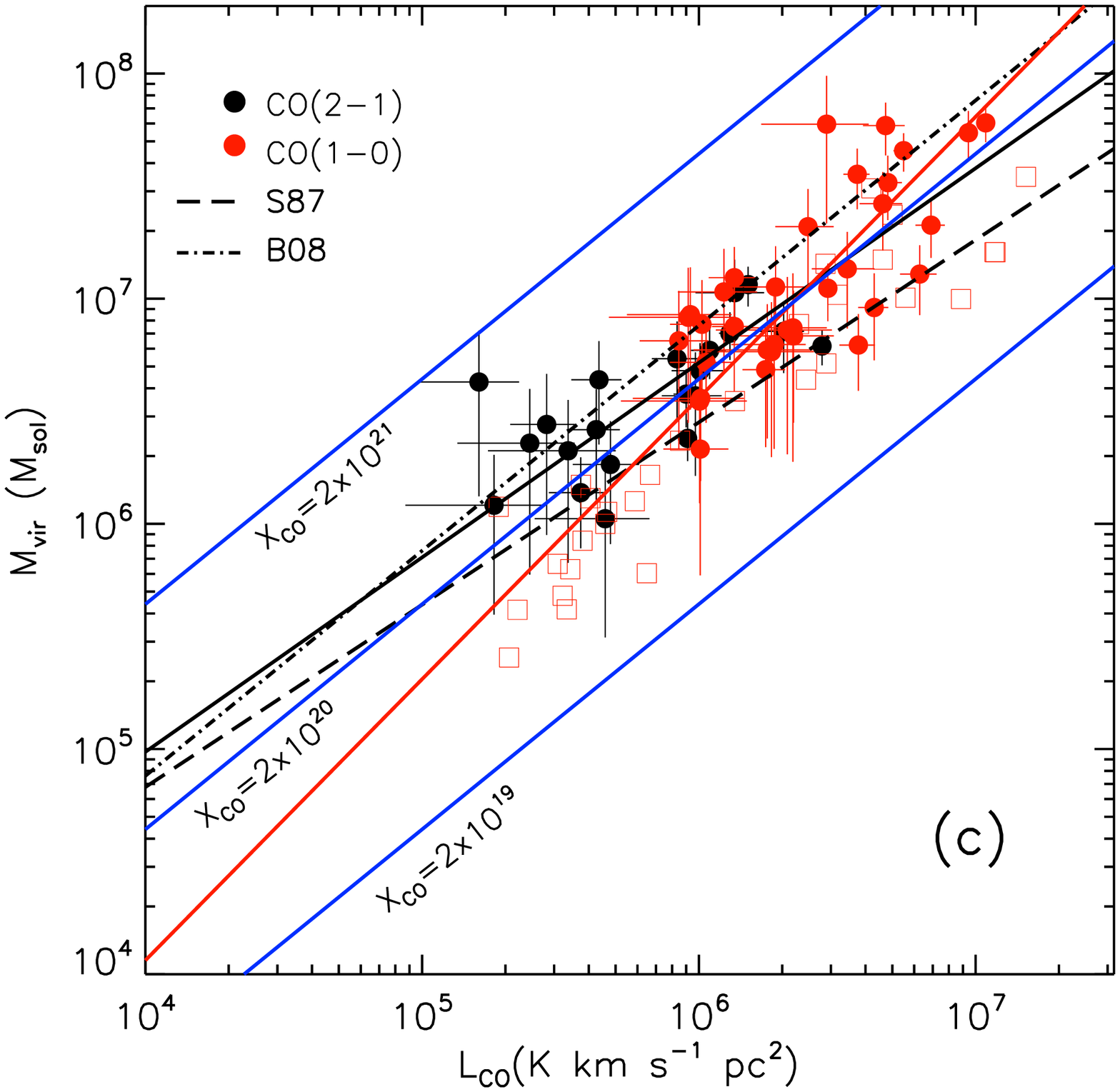,width=0.3\linewidth,angle=90}
\end{tabular}
\caption{Scaling relations for the corrected properties of the complexes and clouds in our sample.  After correction, we have 23 $\co$ complexes and 25 $\cotwo$ clouds in our sample.  As in Figure \ref{fig_co10_siwi_raw}, $\co$ complexes and $\cotwo$ clouds are illustrated with black and red circles respectively.  The black solid line represents the bisector fit for the $\cotwo$ clouds, while the red solid line represents the bisector fit for the $\co$ complexes.  The dashed line illustrates the relation found by S87, and the dashed dotted line represents the fit found by B08.  Red open squares show the values reported by DM12 in the central region of NGC 6946.  {\bf (a)}: Size-line width relation.  {\bf (b)}: Luminosity-line width relation.  {\bf (c)}:  Virial mass-Luminosity relation.  Solid blue lines show different $X_\mathrm{CO}$ values.}
\label{figure_line-width}
\end{figure*}

The virial mass-luminosity relation for our observations is shown in Figure \ref{figure_line-width}(c).  This figure shows that $M_\mathrm{vir}$ values are well predicted by the CO luminosity, supporting the use of a relatively constant CO-to-H$_\mathrm{2}$ factor $X_\mathrm{CO}=2\times10^{20}\mathrm{cm}^{-2}(\mathrm{K}\ \kms)^{-1}$ adopted in equation (\ref{Mco-equ}).  The best-fit relation for $\cotwo$ clouds yields

\begin{equation}\label{mvir-lco}
\frac{M_\mathrm{vir}}{\Msun}=(32.2 \pm 64.7) L_\mathrm{CO}^{0.87 \pm 0.15},
\end{equation}

\noindent where $L_\mathrm{CO}$ is in $\Kkmspc$.  On the other hand, the best fit relation for $\co$ emitting complexes is given by

\begin{equation}\label{mvir-lco10}
\frac{M_\mathrm{vir}}{\Msun}=(0.12 \pm 0.26) L_\mathrm{CO}^{1.24 \pm 0.14}.
\end{equation}

Once again, we have compared the values found in this paper to the values found by previous works.  The structures identified in this paper are located in the $M_\mathrm{vir}-L_\mathrm{CO}$ relation found by B08, but are systematically above the relation found by S87 for Galactic clouds, and the values found in the center of NGC 6946 by DM12.

\begin{figure*}
\centering
\begin{tabular}{cc}
\epsfig{file=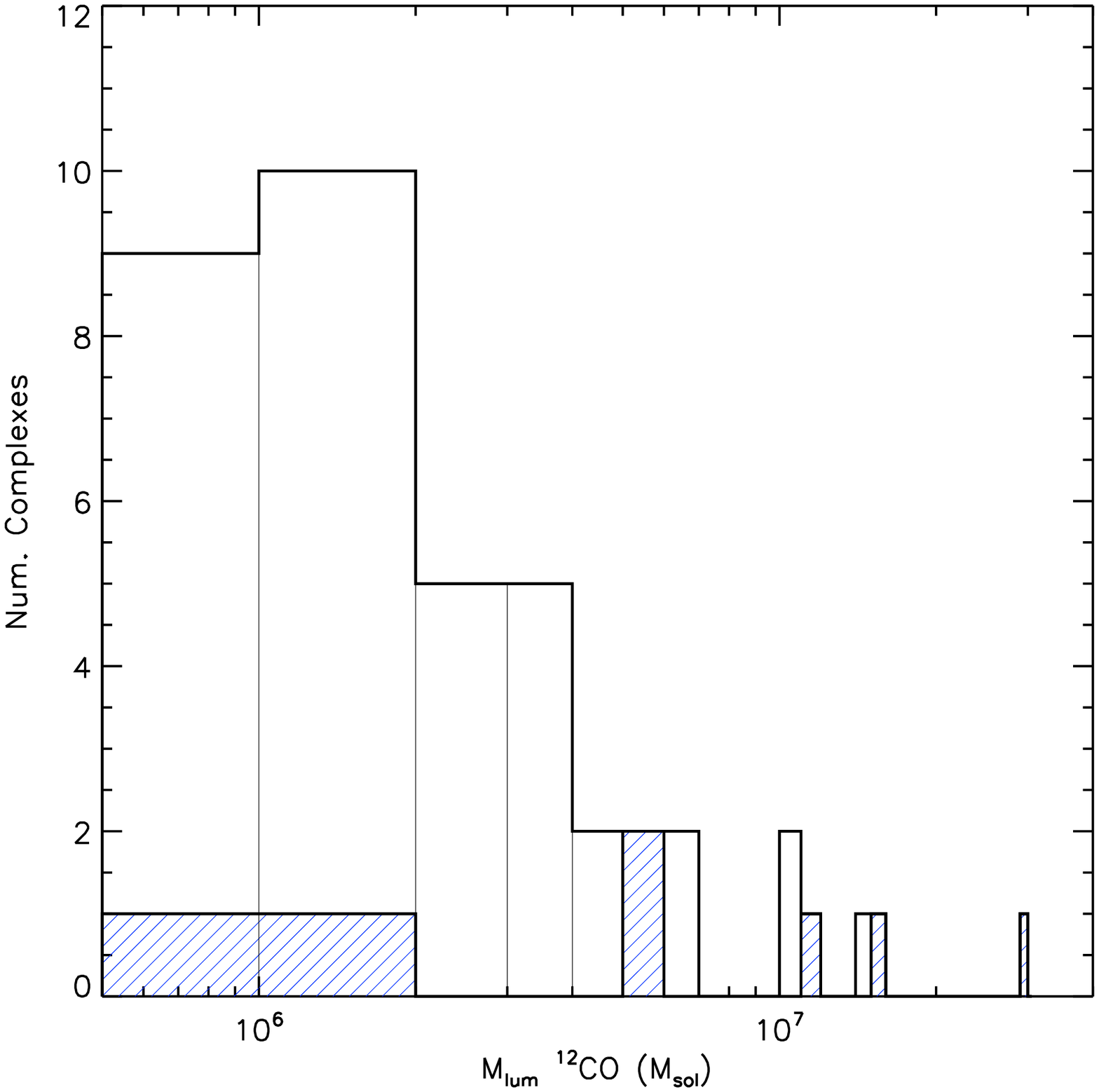,width=0.38\linewidth,angle=90}
\epsfig{file=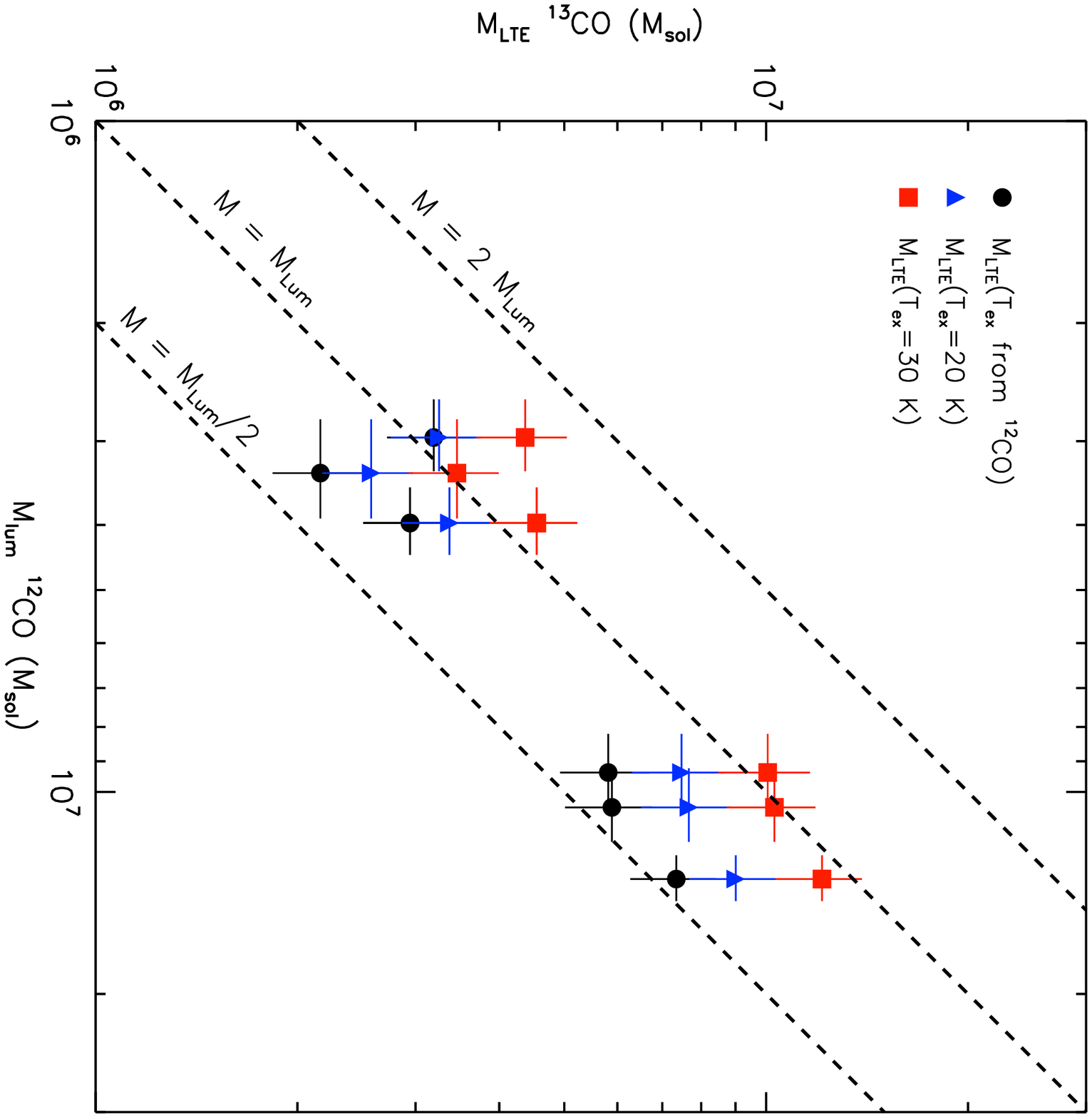,width=0.38\linewidth,angle=90}
\end{tabular}
\caption{{\bf Left}:  Histogram of luminosity-based mass for $\co$ complexes.  Black outlines represent the complete sample, while blue shaded regions represent the complexes which have $\cother$ detections.  Vertical lines illustrate the bin size used in the histogram ($10^6\ M_\odot$).  $\cother$ is detected in several of the most massive $\co$ complexes.  {\bf Right}:  Comparison between LTE masses of $\cother$ clouds and luminosity-based mass of the corresponding $\co$ complexes (black dots).  Blue triangles show LTE masses derived using $T_\mathrm{ex}$=20 K and red squares illustrate LTE masses derived assuming $T_\mathrm{ex}$=30 K.  We observe that $M_\mathrm{lum}$ is usually larger (by a factor of $\sim$ 2 or less) than $M_\mathrm{LTE}$ over the regions we detect in both $\co$ and $\cother$ lines.  The departure of luminosity-based masses from LTE masses is more pronounced for more massive clouds.}
\label{figure_13co}
\end{figure*}

In order to further test the feasibility of using CO luminosity to trace the molecular mass, we have taken advantage of our observations of $\cother$.  We have detected 6 clouds in the $\cother$ intensity map, and their properties are calculated as we did for $\co$ and $\cotwo$ clouds.  We have calculated their LTE masses ($M_\mathrm{LTE}$) using the approach explained in Section \ref{lte-mass}.  The left panel of Figure \ref{figure_13co} shows the histogram of the $\co$ complex luminosity-based masses, indicating the clouds with $\cother$ counterparts detections.  We noticed that significant $\cother$ emission is detected over several of the most massive $\co$ complexes.  The right panel of Figure \ref{figure_13co} shows the comparison between $\cother$ $M_\mathrm{LTE}$ and luminosity-based mass derived from $\co$, both calculated over the $\cother$ emission region.  We observed that $M_\mathrm{lum}$ is usually larger than $M_\mathrm{LTE}$ over the regions we have detections in both $\co$ and $\cother$ lines.  The departure of luminosity-based mass from LTE mass is more pronounced for more massive clouds.  Similar discrepancies between LTE masses and other masses tracers have been reported in the literature.  For instance, in their revision of the Larson's scaling relations, \citet{2009ApJ...699.1092H} found that LTE-derived masses estimated from $\cother$ observations of Galactic clouds are significantly smaller than the virial masses derived by S87.  This discrepancy can be explained by the systematic errors yielded by the assumptions in both methods.  \citet{2009ApJ...699.1092H} generated models of $^{12}$CO and $^{13}$CO line intensities for several physical conditions of the clouds.  They found that LTE derived column densities can both underestimate or overestimate the true column density, depending on the volume density regime, kinetic temperature, among other factors.  In addition, the assumption of a constant CO abundance inside the clouds is another source of uncertainty in calculating the LTE and luminosity-based masses.  The abundance of CO in the deeply embedded material of the cloud likely will be different from the the material being exposed to external UV radiation.  According to \citet{2008ApJ...680..428G} in their study of Taurus molecular cloud, the total cloud mass estimated considering the CO deficient envelopes can be a factor of two larger than the estimated mass assuming a constant CO abundance across the cloud structure.  Finally, we mention that some difference between $M_\mathrm{LTE}$ and $M_\mathrm{lum}$ is expected, since these two different lines, $\co$ and $\cother$, trace different density regimes in the molecular gas.

Along with the LTE masses from $\cother$ map, Figure \ref{figure_13co} shows the effect of varying the excitation temperature on the derived masses.  It shows that it would be necessary to increase the excitation temperature to 20 K in order to make the LTE mass agree with the luminosity-based mass derived from $\co$ for complexes with masses below $\sim 3 \times 10^6 \Msun$.  On the other hand, for clouds more massive than $\sim 3 \times 10^6 \Msun$, a $T_\mathrm{ex} = 30$ K is required to account for the difference between both mass tracers.  Similar $T_{ex}$ has been observed in nearby molecular clouds (\citealt{2009PASJ...61...39M}; \citealt{2009A&A...507..369W}).  Considering the results presented above, we conclude that while the LTE masses are smaller by a factor of 2 than luminosity-based masses in our sample of clouds, this discrepancy does not mean necessarily that optically thick $\co$ observations are not suitable to estimate true molecular cloud masses, but reflects the systematic error related to the assumptions we have made in calculating $M_\mathrm{LTE}$.  

\begin{figure}
\centering
\begin{tabular}{cc}
\epsfig{file=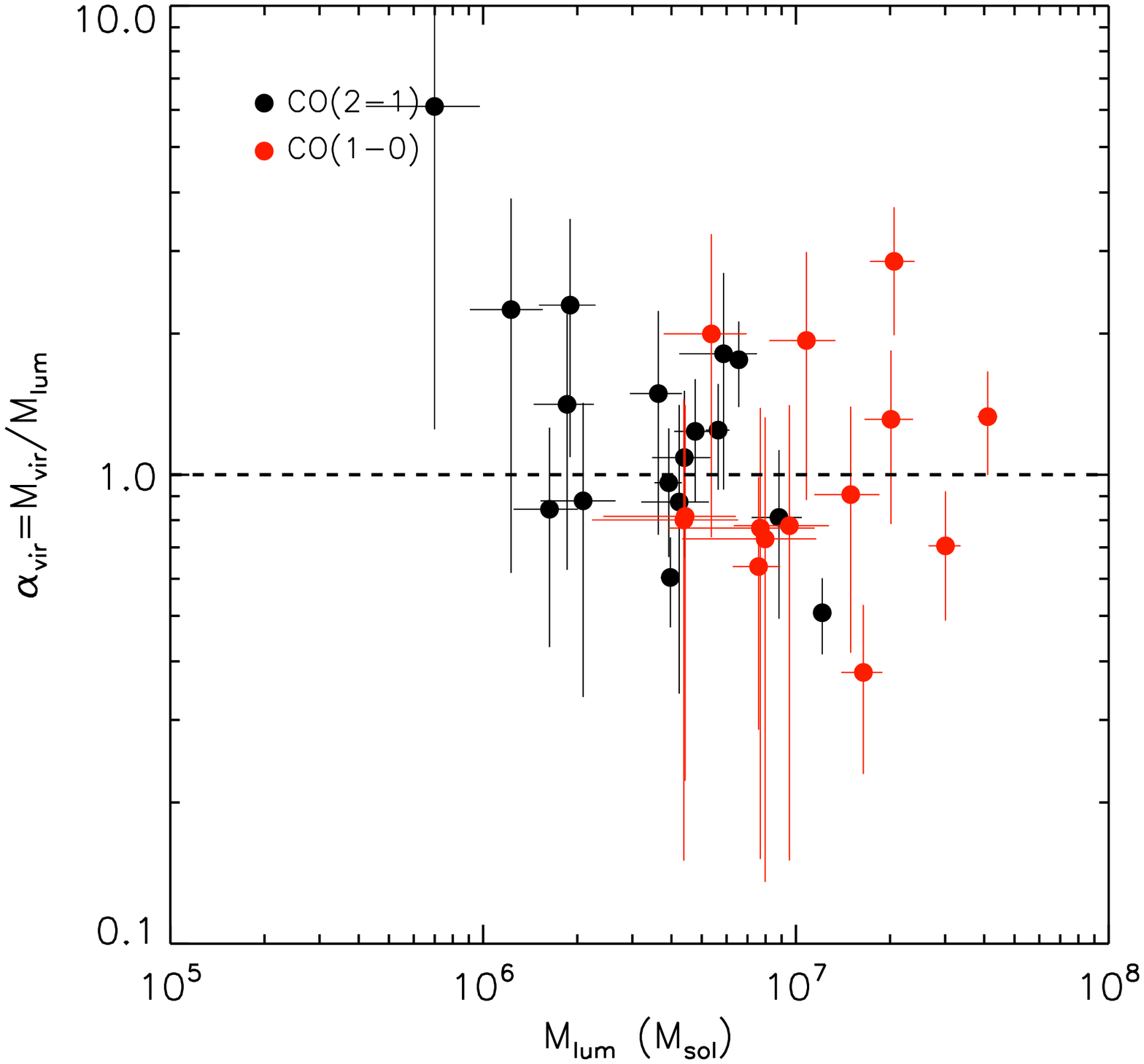,width=0.45\linewidth,angle=90}
\epsfig{file=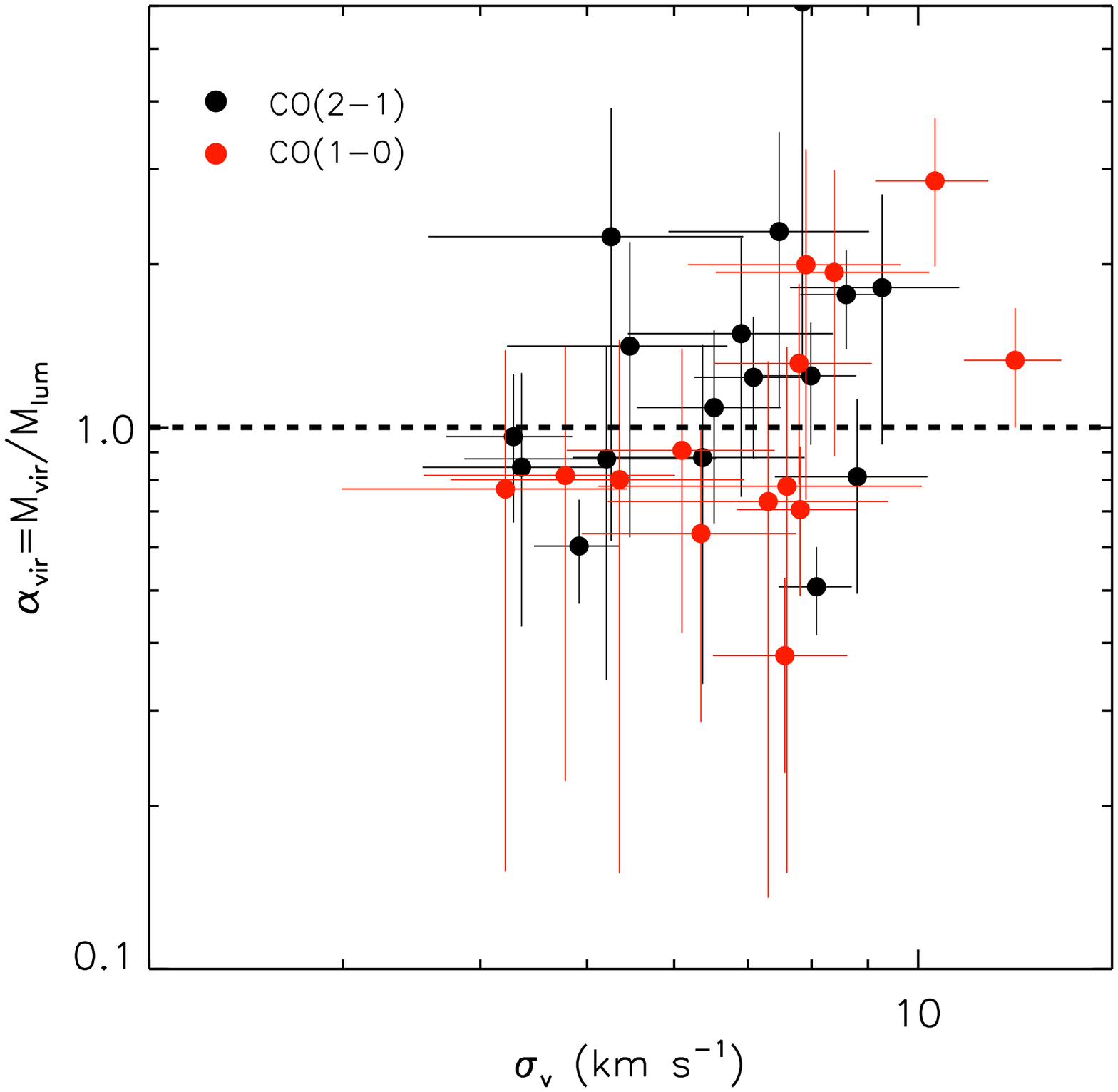,width=0.45\linewidth,angle=90}
\end{tabular}
\caption{{\bf Left}: Virial parameter vs.\ luminosity-based mass for the identified structures in $\co$ and $\cotwo$.  {\bf Right}:  Virial parameter vs. velocity line width.  While a clear correlation between $\alpha_\mathrm{vir}$ and $M_\mathrm{lum}$ is not observed, $\alpha_\mathrm{vir}$ shows a correlation with the velocity dispersion.  A similar correlation has been found by \citet{2010PASJ...62.1261H} in their observations of northeastern spiral arm segment of IC 342.  Nevertheless, a significant correlation is expected due to the presence of the velocity dispersion in both axes.}
\label{figure_vir}
\end{figure}

\subsection{Virial Parameter} 
Figure \ref{figure_vir} shows the virial parameter as a function of $M_\mathrm{lum}$ and $\sigma_v$ for the structures identified in both $\co$ and $\cotwo$ maps.  The values of $\alpha_\mathrm{vir}$ span a range of one order of magnitude (0.3 to 2), similar to values that have been observed in other extragalactic studies (e.g., \citealt{2010PASJ...62.1261H}; B08).  We observe no clear correlation between $\alpha_\mathrm{vir}$ and $M_\mathrm{lum}$, suggesting that the structures are bound, regardless of how massive they are.  Although \citet{2010PASJ...62.1261H} found a similar relation for the virial parameter and the luminosity-based mass in their study of the nearby galaxy IC 342, they reported a good correlation between $\alpha_\mathrm{vir}$ and the velocity dispersion.  They suggested that the degree of gravitational binding is set mainly by the turbulence within the clouds.  Additionally, they found that clouds associated with \hii\ regions possess smaller velocity dispersions and $\alpha_\mathrm{vir}$ than clouds lacking \hii\ regions, consistent with a dissipation of turbulence following passage of the spiral arm.  In the right panel of Figure \ref{figure_vir} we show the dependence of the virial parameter on the line width for our observations.  Although we observe a weak correlation between these two variables, it may be related to the fact that these variables are not completely independent ($\alpha_\mathrm{vir}= M_\mathrm{vir}/M_\mathrm{lum} \propto \sigma^2 R/\sigma T R^2 \propto \sigma $).

\subsection{Mass surface density} 
The mean surface density for $\co$ complexes is $50\ \Msun\ \mathrm{pc}^{-2}$, while for $\cotwo$ clouds, the mean is $70.1\ \Msun\ \mathrm{pc}^{-2}$.  The difference between the mean surface densities is not surprising, given that we have selected the brightest regions in the $\co$ map to observe $\cotwo$.  

We note that there is a sub-group of clouds, in both lines, that deviate from the main group, which is mostly in the regime of $\Sigma_{\rm H2} < 100\ \Msun\ \mathrm{pc}^{-2}$.  As we will discuss below, those clouds are located in two specific regions in the arms.  We will discuss the change in cloud properties across the disk in Section \ref{env-eff}.

\begin{figure}
\centering
\begin{tabular}{c}
\epsfig{file= 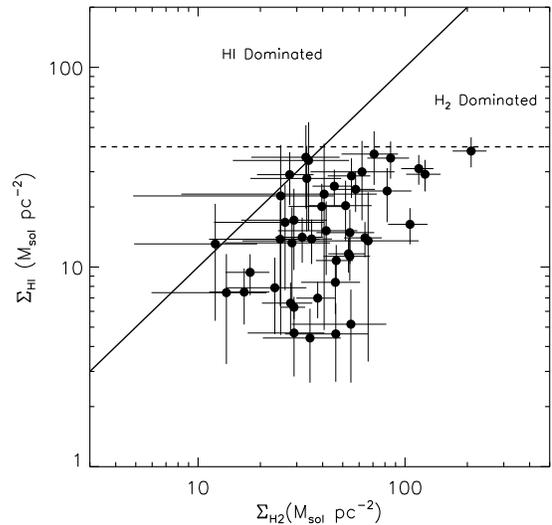,width=0.8\linewidth,angle=90}
\end{tabular}
\caption{Atomic gas surface density vs.\ molecular gas surface density for $\co$ complexes.  We observe a maximum value of $\Sigma_\mathrm{HI} \sim 40$ \Msun\ pc$^{-2}$ (black dashed line) above a molecular surface density of $\Sigma_\mathrm{H_2} \sim 80$ \Msun\ pc$^{-2}$.  The black solid line divides the \HI\ and H$_2$ dominated regions.}
\label{figure_hi_h2}
\end{figure}

\begin{figure*}
\centering
\begin{tabular}{cc}
\epsfig{file=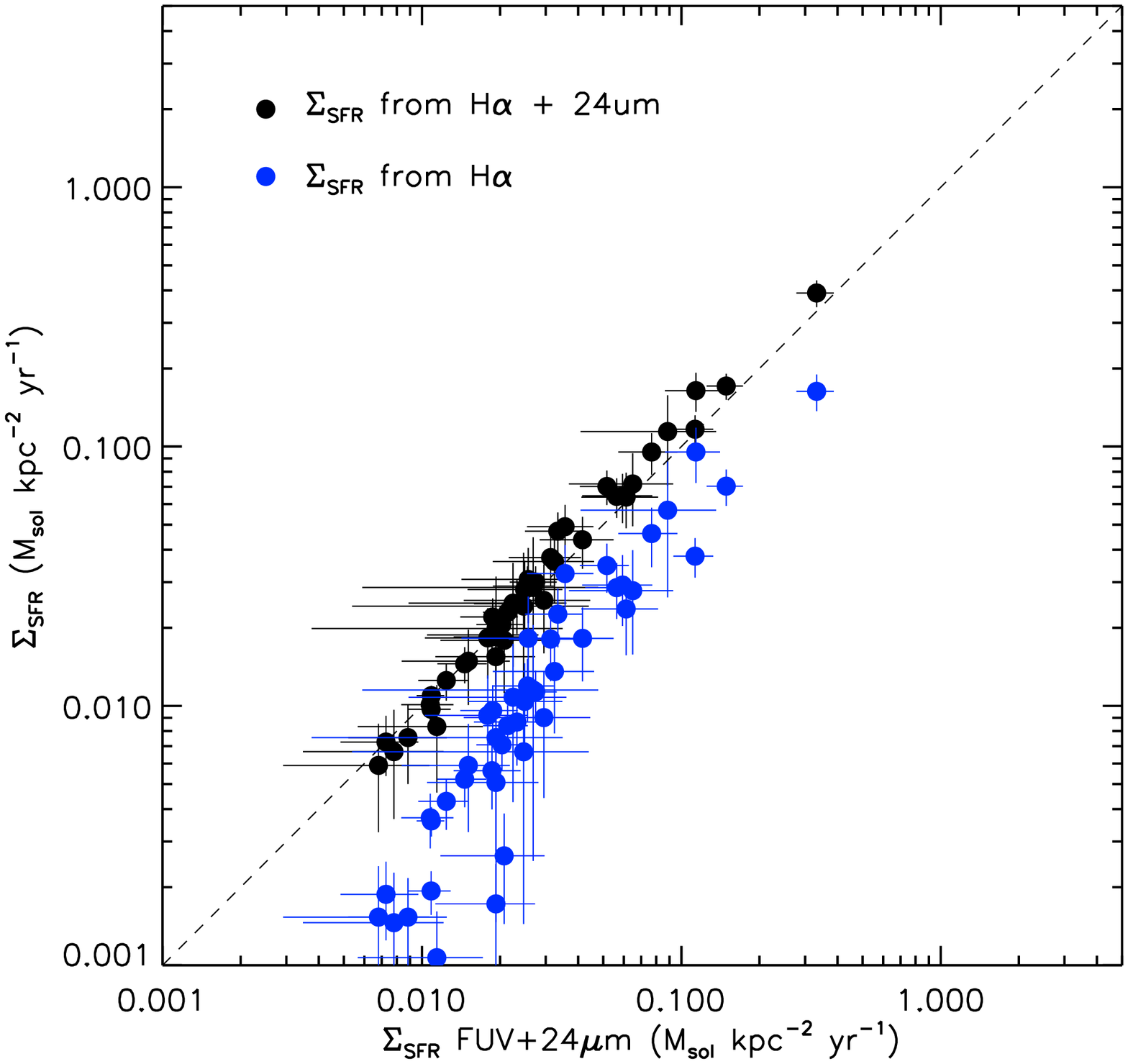,width=0.4\linewidth,angle=90}
\epsfig{file=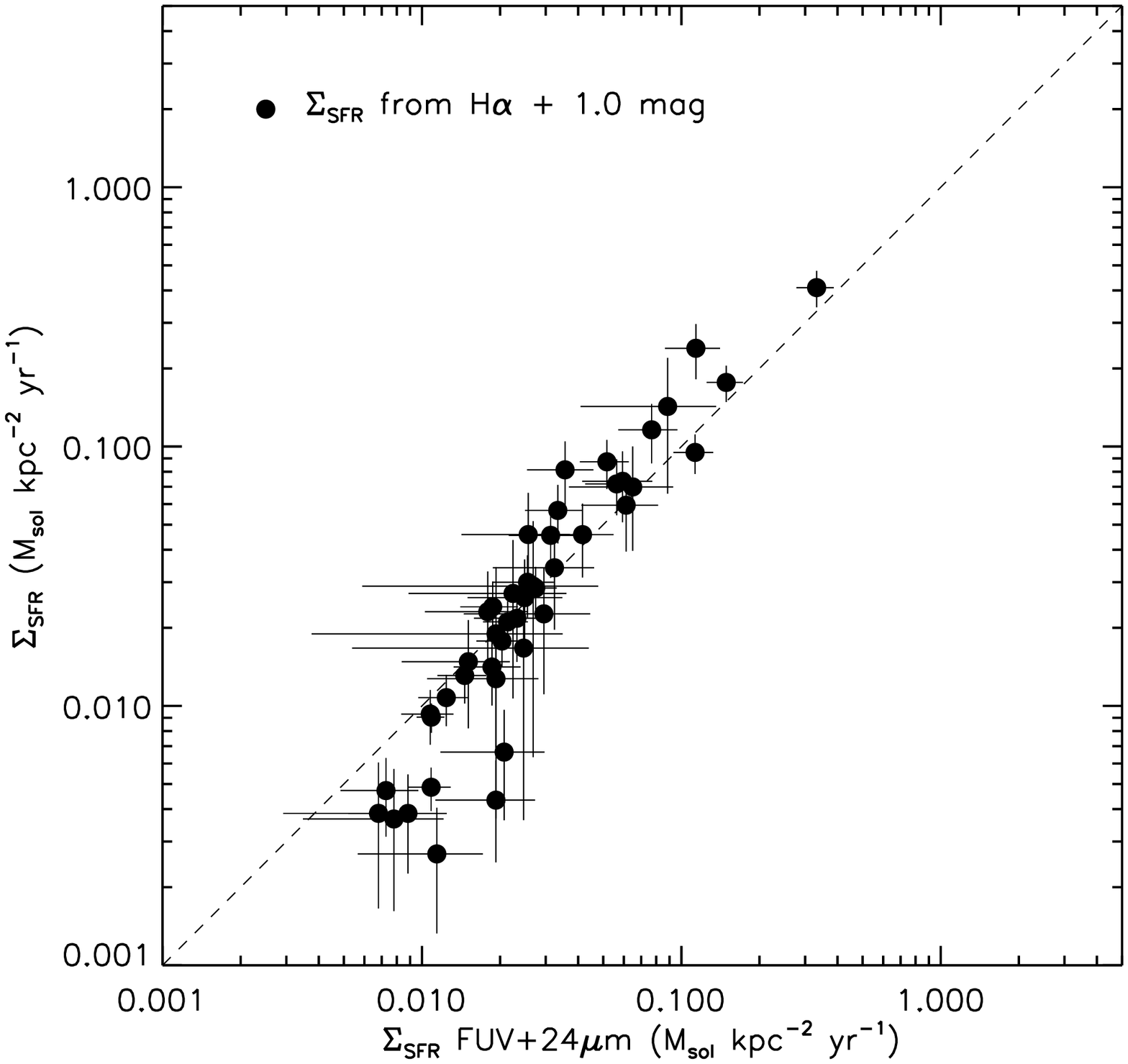,width=0.4\linewidth,angle=90}
\end{tabular}
\caption{{\bf Left}: $\Sigma_\mathrm{SFR}$ estimates  for $\co$ complexes as a function of $\Sigma_\mathrm{SFR}$ predicted using a combination of FUV + 24$\mu$m, following L08.  Black points represent the $\Sigma_\mathrm{SFR}$ derived from the combination of H$\alpha$ and 24$\mu$m as given by \citet{2007ApJ...666..870C}.   Blue points represent the $\Sigma_\mathrm{SFR}$ derived using H$\alpha$ only.  {\bf Right}:  $\Sigma_\mathrm{SFR}$ derived from H$\alpha$ corrected by A$_\mathrm{H\alpha}=1.0$ magnitude extinction.}
\label{figure_sfr10}
\end{figure*}

Additionally, we have estimated the atomic gas surface density for $\co$ complexes.  Unfortunately, the limited resolution of HI maps ($\sim 5\arcsec$) did not allow us to calculate the atomic gas component for the $\cotwo$ clouds, so we are reporting atomic gas surface densities just for $\co$ complexes.  The mean atomic gas surface density for $\co$ complexes is $18\ \Msun\ \mathrm{pc}^{-2}$.  Figure \ref{figure_hi_h2} shows the comparison between HI and H$_2$ surface density.  We observe a maximum value of $\Sigma_\mathrm{HI} \sim 40\  \Msun\ \mathrm{pc}^{-2}$.  Moreover, the molecular gas fraction, $f_\mathrm{H2}\equiv \Sigma_\mathrm{H2}/\Sigma_\mathrm{gas}$, inside the complexes is usually high ($\sim 0.7-0.9$), confirming the molecular character for most of the regions we are targeting.  Nevertheless, there are some complexes that show $f_\mathrm{H2}$ below 0.7.  Those regions are located beyond the radius $\sim 4$ kpc, where the atomic gas surface density starts becoming comparable to the molecular gas component.  Similar molecular gas fractions were reported by \citet{2009ApJ...700L.132K} in their study of molecular clouds in M51.

\subsection{Star Formation activity of the clouds}\label{sfr-result}
Having established the properties of the molecular gas structures, we now estimate the star formation activity inside these regions.  GMCs are the nurseries of massive stars in disk galaxies, so it is natural to think that the intrinsic physical properties of the clouds are tightly coupled to the observed rate of star formation.  The effect of star formation on the properties of GMCs has been analyzed through simulations (e.g. \citealt{2011ApJ...730...11T}; \citealt{2010ApJ...709..191M}), theoretical models (\citealt{2010ApJ...721..975O}; \citealt{2010ApJ...713.1120K}) and observations (e.g. \citealt{2010PASJ...62.1261H}; \citealt{2010MNRAS.406.2065H}; \citealt{2011ApJS..197...16W}) among others.  Observations have revealed that star formation is a slow process or that a small portion of available gas is involved in the formation of stars.  Several authors have interpreted this inefficiency in the formation the stars as evidence of stellar feedback processes, where the injection of energy and momentum into the ISM plays a key role in setting the rate at which gas turns into stars.  Unfortunately, observations of extragalactic molecular clouds usually lack the spatial resolution to relate the star forming regions and their progenitor molecular gas, thus making it difficult to interpret the effect of star formation in individual clouds. 

Figure \ref{figure_sfr10} shows the comparison between SFR surface density for $\co$ complexes derived from FUV + $24\mu$m and H$\alpha$ + $24\mu$m.  As was stated above, the amount of SF is calculated over the boundary of the identified cloud as yielded by CPROPS.  We observe that both SFR surface density estimations are well correlated.  Along with this comparison, we have included the unobscured SFR surface density derived from H$\alpha$ only.  The difference between SFR traced by FUV and $24\mu$m and the SFR traced by H$\alpha$ is more pronounced for regions of low SFR.  In the right panel of Figure \ref{figure_sfr10}, we show the SFR from H$\alpha$ corrected for A$_\mathrm{H\alpha}=1.0$ magnitude extinction.  We notice that this global correction for H$\alpha$ works reasonably well in recovering the total SFR (specially for higher SFRs), and we have corrected the Equation \ref{sfr_ha} accordingly.  This correction is consistent with values found by L08.

\subsubsection{$\Sigma_\mathrm{SFR}$ vs. $\Sigma_\mathrm{H2}$ relation for clouds}\label{k-s_law}

The star formation rate (SFR) is observed to correlate with the distribution of gas according to a power law at large scales ($\Sigma_\mathrm{SFR} \sim \Sigma_\mathrm{gas}^N $).  This relation was first proposed by \citet{1959ApJ...129..243S}, and tested against observations by many authors subsequently.  The most influential study of the Schmidt law was performed by \citet{1998ApJ...498..541K}, who studied the relation between the SFR and gas content averaged over the disk for a sample composed of normal spiral and starburst galaxies.  He found a power-law index $N=1.4$ for the entire sample, and a steeper index $N=2.47$ including just the normal galaxies.  As the capabilities of the telescopes have been improved over the last years, the study of the Kennicutt-Schmidt law has been performed within individual galaxies.  \citet{2002ApJ...569..157W} used azimuthal averages of the SFR and surface gas density for seven molecular gas rich galaxies, finding indices $N=1.2-2.1$.  Using 750 pc resolution data for a sample of 18 nearby galaxies, \citet{2008AJ....136.2846B} found an index $N=1.0$ between SFR and molecular surface gas density using $\cotwo$.  They interpreted this linear relation as evidence of a constant molecular gas depletion time ($\sim 2\times 10^9\ $ years) within GMCs.  
Nevertheless, the slope and coefficient obtained for the Kennicutt-Schmidt law is sensitive to methodology (\citealt{2007ApJ...671..333K}; \citealt{2011ApJ...735...63L}; \citealt{2011ApJ...730...72R}), including the fitting method, the spatial resolution and the tracers used to estimate the star formation rate and the gas density.  

\begin{figure*}
\centering
\begin{tabular}{cc}
\epsfig{file=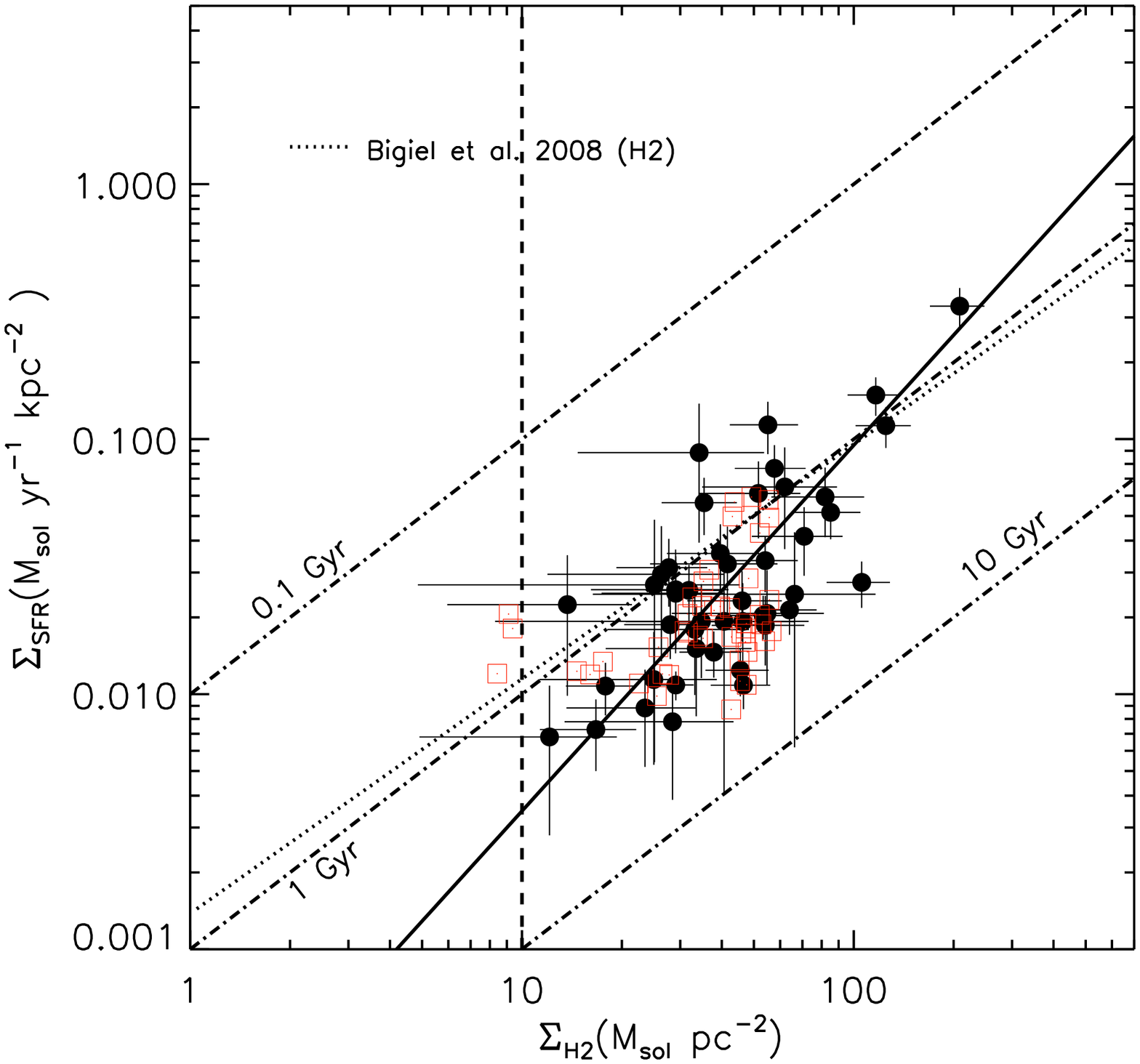,width=0.4\linewidth,angle=90}
\epsfig{file=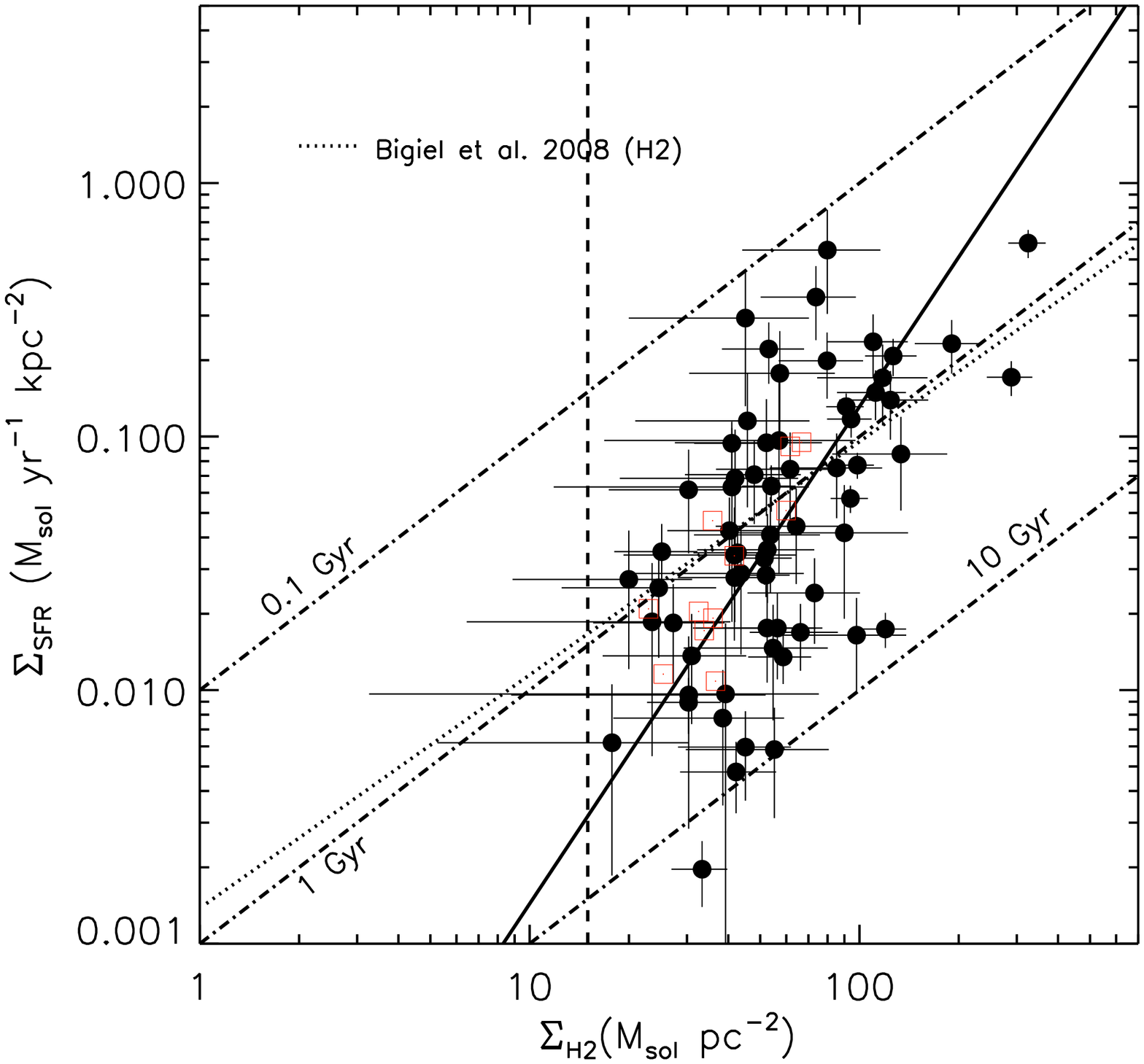,width=0.4\linewidth,angle=90}
\end{tabular}
\caption{{\bf Left}:  Star formation rate vs.\ molecular gas surface density for $\co$ complexes.  The SFR is calculated from the FUV + 24$\mu$m maps.  The black solid line illustrates the best-fit relation found for the complexes, while the black dotted line represents the relation found by \citet{2008AJ....136.2846B}.  The vertical dashed line illustrates the sensitivity limit of our map, $\sim$10 $\Msun$ pc$^{2}$.  Open red squares denote values calculated over 15$\arcsec$ (400 pc) apertures centered on the molecular complexes.  Black dot-dashed lines show constant molecular gas depletion times (SFE$^{-1}$) of 0.1, 1 and 10 Gyr.  {\bf Right}: Star formation rate vs. molecular gas surface density for $\cotwo$ clouds.  In this case, we have used H$\alpha$ to estimate the SFR, corrected by 1 mag as is suggested by the right panel of Figure \ref{figure_sfr10}.  Open red squares show the values calculated over apertures centered at the locations of $\cotwo$ pointings.  The other lines are the same as in the left panel, except the sensitivity limit is $\sim$15 $\Msun$ pc$^{2}$.  The fitted slope is steeper than that found for $\co$ complexes.}
\label{figure_sfr}
\end{figure*}

In the present paper, we have chosen to calculate the SFR and gas surface density inside the boundaries of the structures identified by CPROPS, similar to the approach followed in Galactic molecular cloud studies.  We aim to establish the $\Sigma_\mathrm{SFR}$-$\Sigma_\mathrm{H2}$ relation for scales close to individual molecular complex and cloud sizes.  We emphasize that this type of analysis, spatially biased to the most gas-rich regions, is different from the pixel-by-pixel or azimuthally averaged analyses which have been used to describe the Kennicutt-Schmidt law across galaxy disks (e.g., \citealt{2002ApJ...569..157W}; \citealt{2008AJ....136.2846B}; \citealt{2009ApJ...704..842B}).  
 
The relationship between SFR surface density and molecular gas surface density for $\co$ complexes is shown in the left panel of Figure \ref{figure_sfr}.  We see a clear correlation towards higher star formation activity for higher molecular gas surface density.  To find the best fit relation between $\Sigma_\mathrm{SFR}$ and $\Sigma_\mathrm{H2}$ we used the bisector linear regression method as we did for Larson's scaling laws.  The best fit relation is given by:

\begin{equation}\label{sfr10_fit}
\frac{\Sigma_\mathrm{SFR}}{ \Msun \mathrm{yr}^{-1} \mathrm{kpc}^{-2}}=((1.3 \pm 0.6) \times 10^{-4})\Sigma_\mathrm{H2}^{1.43 \pm 0.11},
\end{equation}

\noindent where $\Sigma_\mathrm{H2}$ is in $\Msun\ \mathrm{pc}^{-2}$.  This slope is similar to the value found by \citet{2008AJ....136.2846B} ($N=1.46$) for NGC 6946 at 750 pc resolution for the relation between SFR and the total gas component (\HI+H$_2$), but is steeper than the relation found for H$_2$ ($N=0.92$).  In Figure \ref{figure_sfr} we have included lines of constant molecular gas depletion time.  Molecular gas complexes are consistent with a depletion time of $\sim$ 1 Gyr, close to the value found by \citet{2008AJ....136.2846B}.

The right panel of Figure \ref{figure_sfr} shows the correlation between SFR and $\Sigma_\mathrm{H2}$ for $\cotwo$ clouds.  The best-fit relation is given by

\begin{equation}\label{sfr21_fit}
\frac{\Sigma_\mathrm{SFR}}{ \Msun \mathrm{yr}^{-1} \mathrm{kpc}^{-2}}=((1.6 \pm 1.1) \times 10^{-5})\Sigma_\mathrm{H2}^{1.96 \pm 0.18},
\end{equation}

\noindent which is steeper than the relation found for $\co$ complexes.  At these smaller scales, the $\Sigma_\mathrm{SFR}$-$\Sigma_\mathrm{H2}$ relation shows higher scatter due to local variations of the SFR and molecular gas densities.  The distribution of points is centered on a molecular gas depletion time $\sim$1 Gyr, and shows differences of more than 2 orders of magnitude in $\Sigma_\mathrm{SFR}$ between the lowest and highest gas surface density extremes.

\subsubsection{$\Sigma_\mathrm{SFR}$ vs. $\Sigma_\mathrm{H2}$ using apertures}\label{k-s_law-ap}
By calculating the SFR inside the complexes or clouds, we are estimating the star formation activity directly associated with these structures.  Thus, we may be missing some SFR activity {\it nearby} to the molecular gas considered in our calculations.  In order to estimate how the relation between $\Sigma_\mathrm{SFR}$ and $\Sigma_\mathrm{H2}$ changes when we use larger areas than individual clouds or complexes, we have taken an aperture like approach.  The idea is to obtain the enhancement in star formation activity associated to the region surrounding the molecular gas clouds or complexes, in contrast to the SFR estimated inside the clouds calculated in Section \ref{k-s_law}.

Along with the star formation rate and molecular gas surface densities calculated inside the structures identified by CPROPS, Figure \ref{figure_sfr} shows the $\Sigma_\mathrm{SFR}$ and $\Sigma_\mathrm{H2}$ values calculated over circular apertures.  These apertures have a radius of 15$\arcsec$ (400 pc), and are centered on the centroids of the $\co$ complexes, or at the pointing center of the fields observed for $\cotwo$ (red squares).  We observe that using 400 pc apertures, the $\Sigma_\mathrm{SFR}$-$\Sigma_\mathrm{H2}$ relation does not change significantly with respect to the relation found for individual clouds or complexes.  In the case of apertures in the $\co$ map, there are a few points with $\Sigma_\mathrm{H2} \sim 10\ \Msun$ pc$^{-2}$, but with SFR comparable to other regions with molecular gas several times larger.  Those regions correspond to complexes relatively isolated from the main spiral arm structures, with moderate star formation activity, and less extended molecular gas than SFR tracers.  By taking averages over large scales, while the molecular gas surface density decreases, the $\Sigma_\mathrm{SFR}$ remains roughly constant.  On the other hand, apertures calculated over $\cotwo$ fields (11 in total) roughly follow the relation found for individual clouds, although the points tend to lie to the left of the best-fitting relation.

\section{Environmental effect on molecular gas properties}\label{env-eff}

Having determined the properties of the molecular clouds in the eastern part of NGC 6946, we turn to the problem of investigating if those properties depend on local conditions of the gas.  More specifically, we search for differences in the properties according to the location of the molecular gas.  A simple way to search for such differences is to use the spiral structure of disk galaxies.  Observations have revealed a concentration of young stars in the arm regions of disk galaxies, and given that stars are formed inside GMCs, the properties of molecular gas may change once the material enters the arm region.  In their study of the northwestern spiral arm of IC 342, \citet{2010PASJ...62.1261H} found that GMCs with active star formation are more massive, have smaller velocity dispersions, and are more virialized than GMCs lacking star formation.  Considering that the spatial distribution of GMCs in IC 342 indicates that high-SFR molecular clouds are located downstream with respect to low-SFR clouds, they suggested that GMC properties change following passage through a spiral arm.  In this section, we will take advantage of our spatial resolution to compare the properties of the clouds according to their location on and off spiral arms.

\subsection{Inter-arm vs. on-arm regions}\label{in-on-arm}
Defining spiral arms can be done using multiple approaches and techniques.  In a recent work, \citet{2011ApJ...737...32E} utilized arm-inter arm contrast to trace arm structure and the spiral arm properties of 46 galaxies of different types.  They used {\it Spitzer} 3.6 $\mu m$ maps due to the small extinction at this frequency and because the old stellar population dominates the emission at this wavelength.  Alternatively, spiral arm amplitude can be traced by Fourier transform of the old stellar population brightness distribution (e.g., \citealt{2009MNRAS.397.1756D}; \citealt{1988A&AS...76..365C}).  \citet{2010ApJ...725..534F} used Fourier-decomposition of azimuthal scans of deprojected 3.6 $\mu$m maps to define the arm and inter-arm region.  In order to study how spiral arms affect the properties of GMCs, and subsequently the star formation, we have divided the sample of clouds into two sub-samples:  on-arm and inter-arm clouds.  We have defined on-arm and inter-arm regions as follows:  First, we transformed into polar coordinates the deprojected 3.6 $\mu$m image.  Table \ref{n6946-prop} shows the parameters used to deproject the images.  Then, we decomposed this image into annuli with a width of 5 pixels (3\farcs75).  Since the annuli are to be Fourier decomposed independently, we overlap the bins by 70\% of a bin width in order to maintain smoothness.
For each radial bin, the light distribution is expanded in a Fourier series by fitting the function: 

\begin{figure}
\centering
\begin{tabular}{c}
\epsfig{file=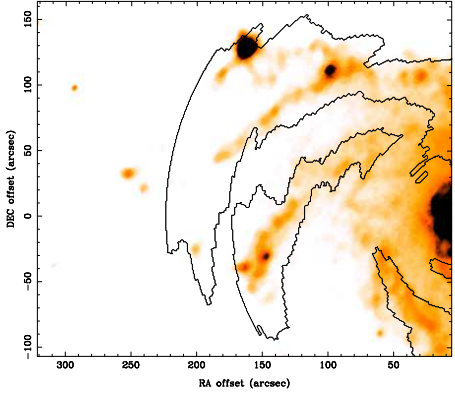,width=0.6\linewidth,angle=-90}
\end{tabular}
\caption{Identification of arm and inter-arm regions.  The image illustrates the 3.6 $\mu$m map (color map), and the Fourier-reconstructed mask (black contour) defined as enclosing the highest 30\% of pixels in each annulus.}
\label{figure_four}
\end{figure}

\begin{equation}\label{fourier}
I(r,\phi)=I_0(r)+\sum_{m=1}^{m_\mathrm{max}}  I_m(r) \cos [m(\phi-\phi_m)] 
\end{equation}

\noindent where $I_m$ and $\phi_m$ are the arm amplitude and phase for each Fourier component {\it m} respectively, and $m_\mathrm{max}$ is the maximum component considered in the expansion. The number of components considered depends on how complex the arm structure is.  For grand-design galaxies, spiral arms are well recovered by using up to $m=4$ (\citealt{2010ApJ...725..534F}).  On the other hand, flocculent galaxies may require expansion out to higher components.  In order to find the number of components that recovers successfully the spiral structure, in our study of NGC 6946 we have used a range of $m_\mathrm{max}=[5,9]$.  We observed that using $m_\mathrm{max}=7$ is enough to define the arm structure, and adding more components leads to basically the same model.  Having Fourier-reconstructed the 3.6 $\mu$m image, we defined the {\it arm region} as the area covered by the brightest 30\% of pixels in each annulus.  Figure \ref{figure_four} shows the 3.6 $\mu$m image (color map) overlaid with the spiral arm mask (black contour).  Once we have determined the spiral arm region, we can classify each GMC as on-arm or inter-arm, and search for differences in the properties of the two sets.  A cloud is defined as on-arm cloud if the center of the cloud is inside of the spiral arm region, and more than the 70\% of the cloud's pixels are inside the on-arm mask.  

\begin{figure}
\centering
\begin{tabular}{ll}
\epsfig{file= 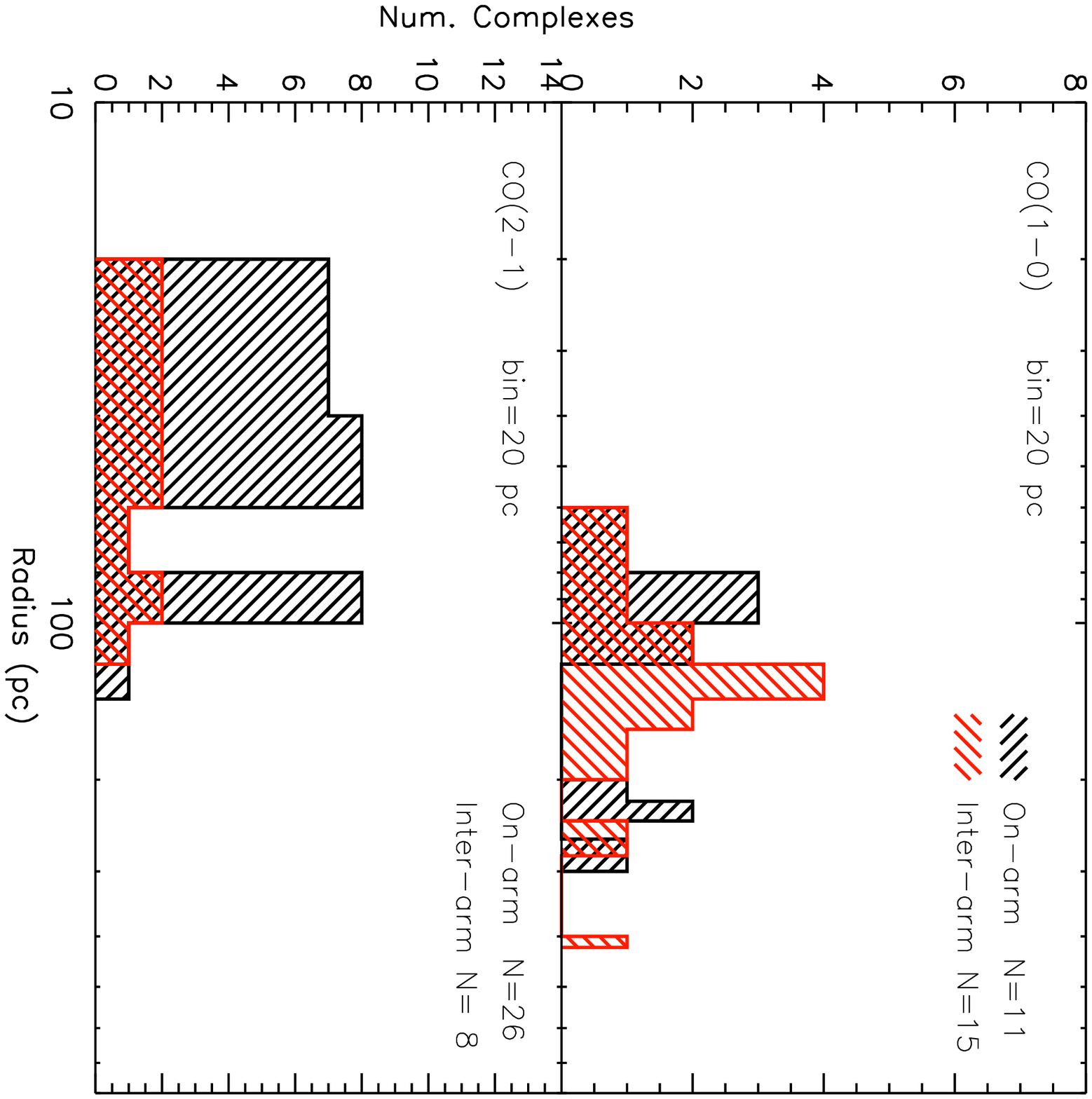,width=0.47\linewidth,angle=90} &
\epsfig{file= 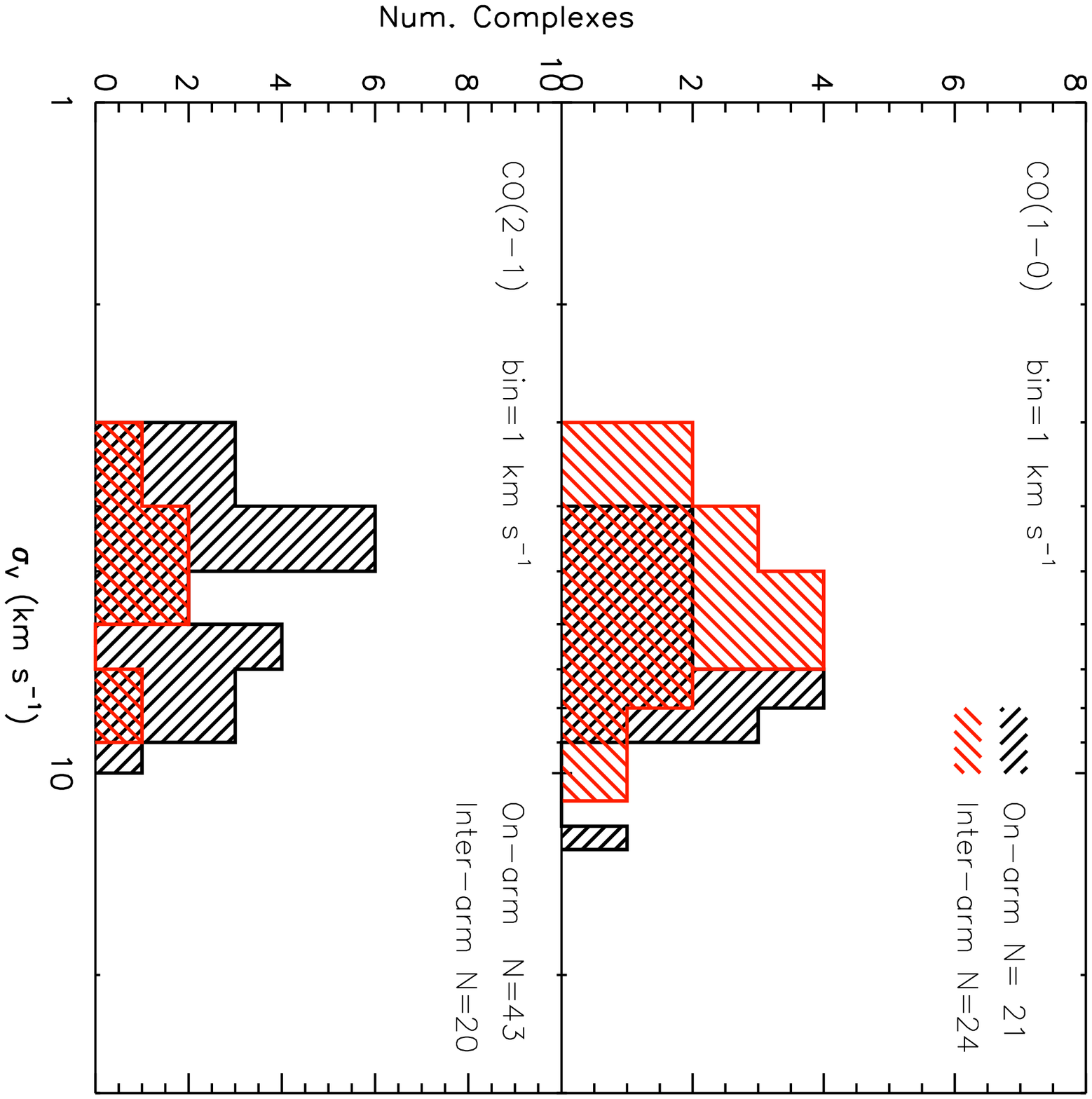,width=0.47\linewidth,angle=90}\\
\epsfig{file= 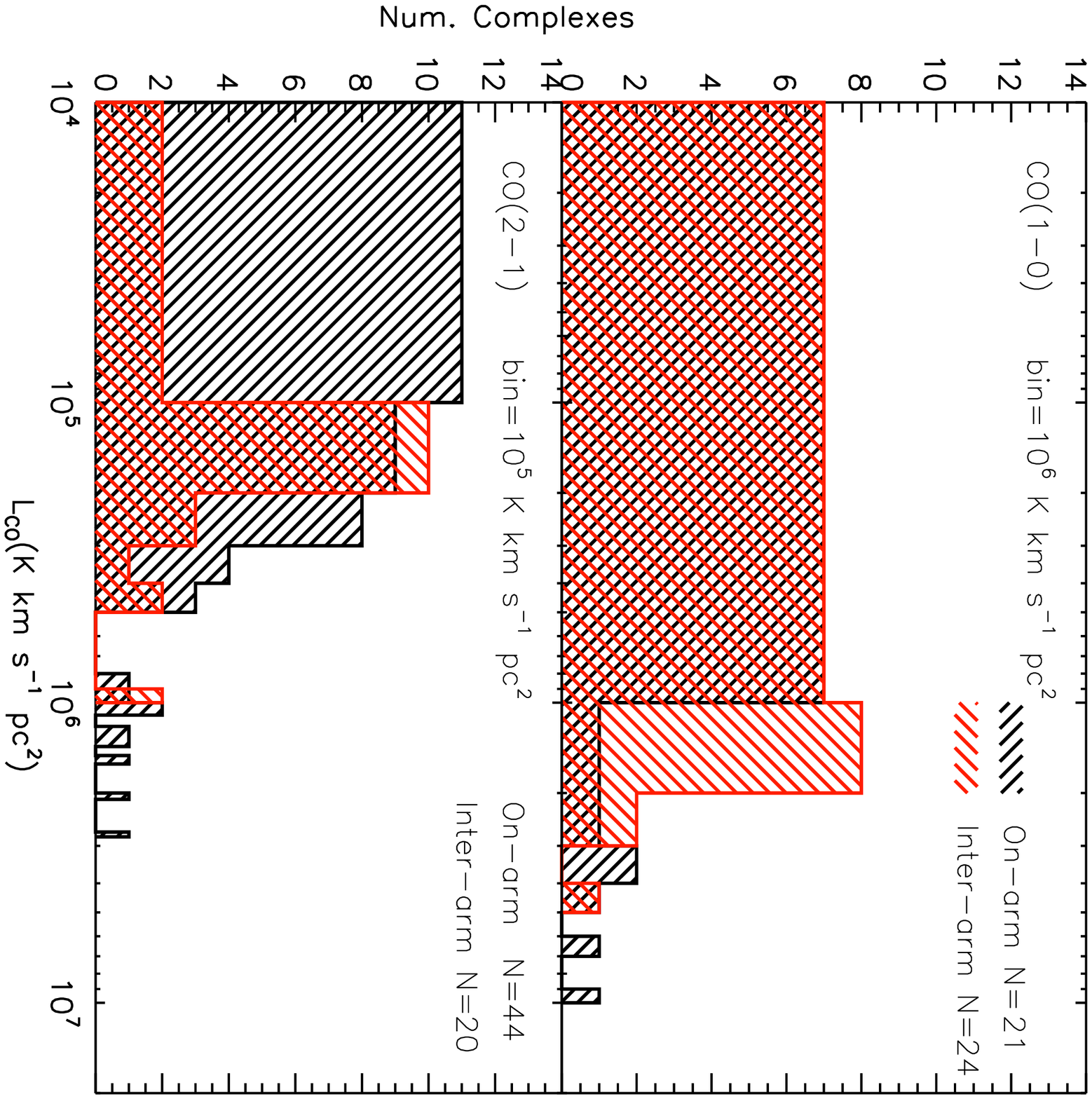,width=0.47\linewidth,angle=90} &
\epsfig{file= 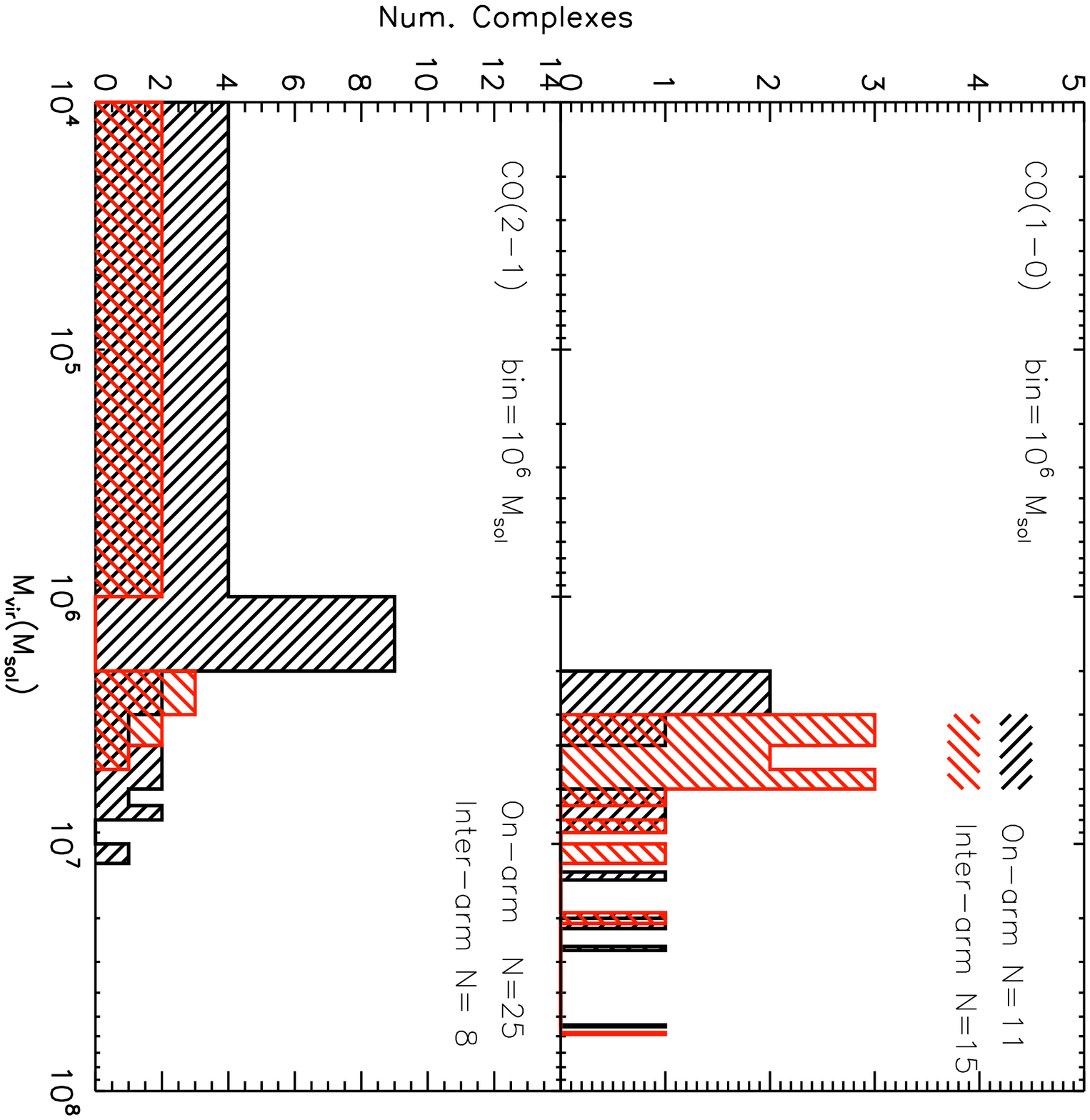,width=0.47\linewidth,angle=90}
\end{tabular}
\caption{Histograms of the size, line width, luminosity and virial mass for the $\co$ complexes and $\cotwo$ clouds classified as on-arm or inter-arm. Black shaded histograms represent structures identified as located on-arm, and red shaded histograms show the inter-arm structures.  Numbers at the top-right show the number of complexes or clouds associated to the on-arm or inter-arm sub-group.  We do not observe a clear distinction in the properties for on-arm and inter-arm $\co$ complexes and $\cotwo$ clouds.  Nevertheless, we observe that some $\cotwo$ clouds show higher luminosities and virial masses than other on-arm clouds (and inter-arm clouds).  Those clouds spatially match regions of higher star formation compared to other locations in the disk.}
\label{figure_arm10}
\end{figure}

\begin{figure*}
\centering
\begin{tabular}{cc}
\epsfig{file=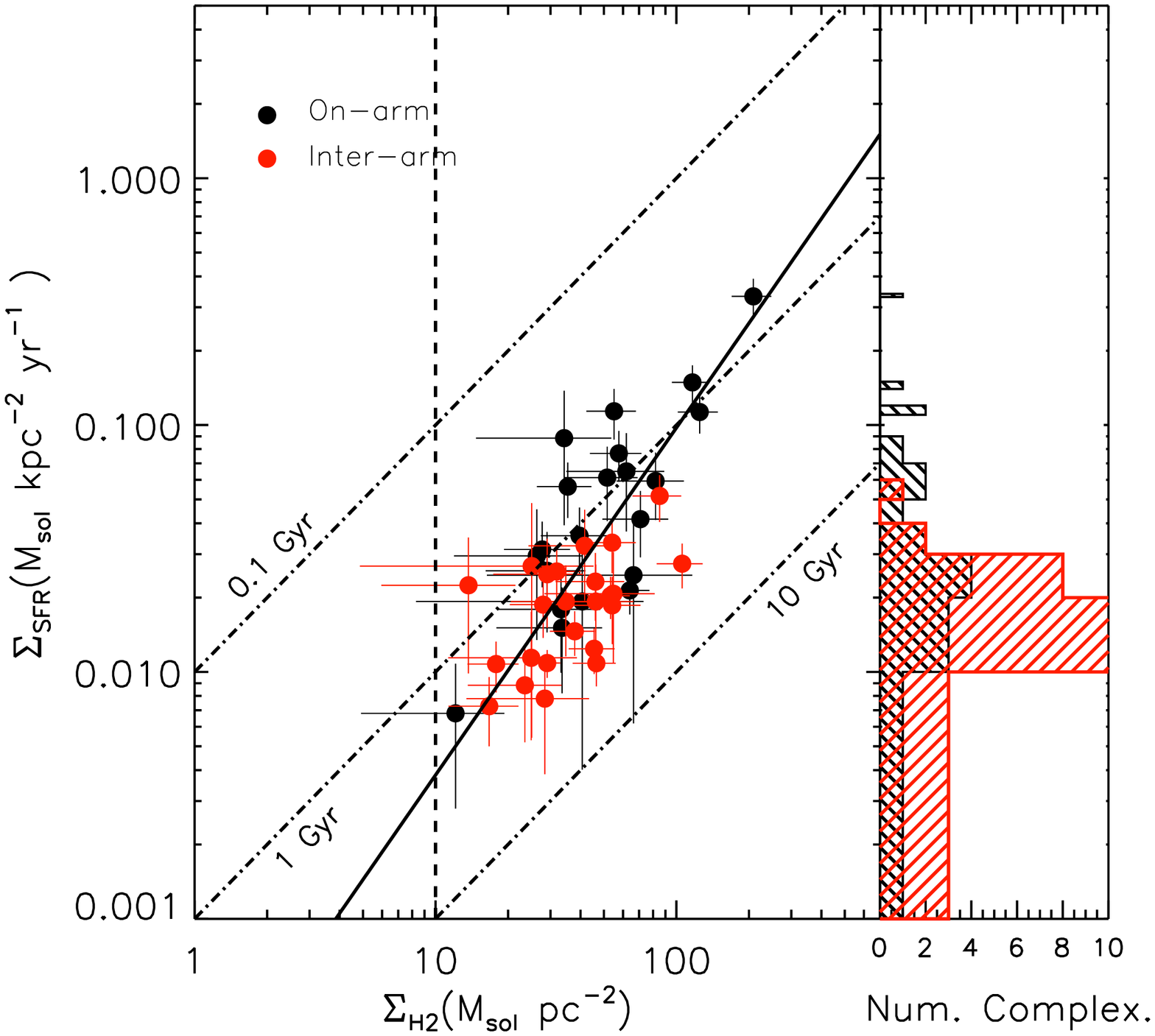,width=0.4\linewidth,angle=90} 
\epsfig{file=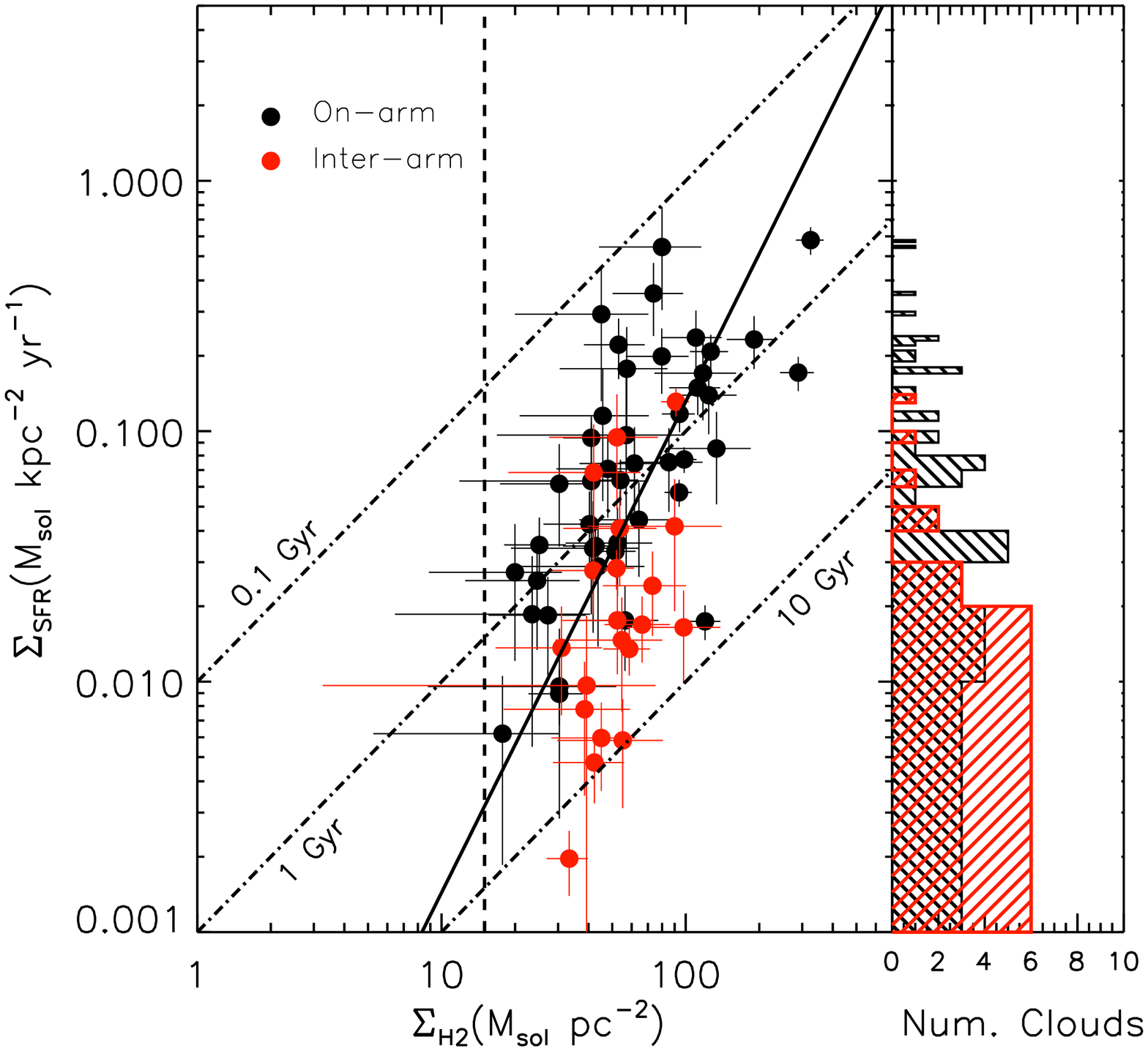,width=0.4\linewidth,angle=90}
\end{tabular}
\caption{$\Sigma_\mathrm{SFR}$ vs.\ $\Sigma_\mathrm{H2}$ relation for $\co$ complexes {\bf (left)} and $\cotwo$ clouds {\bf (right)} using the on-arm and inter-arm classification.  Lines are the same as in Figure \ref{figure_sfr}.  We observe that independent of the spatial resolution, the highest star formation rates and molecular mass surface densities are found in on-arm structures.  Nevertheless, the typical SFR in inter-arm regions is comparable to that in on-arm regions, especially for low surface density clouds.}
\label{figure_sf_arm}
\end{figure*}

\subsubsection{Scaling relations}\label{scal-rela-inon}
Figure \ref{figure_arm10} shows the histograms of the size, line width, luminosity and virial mass of the on-arm and inter-arm $\co$ complexes and $\cotwo$ clouds.  In the case of $\co$ complexes, we observed that the sizes and line widths show no significant differences between on-arm and inter-arm structures.  We emphasize, however, that due to our limiting resolution ($\sim$ 110 pc at 3 mm) broad line widths observed in $\co$ emitting complexes could be the result of high velocity gradient between unresolved clouds.  We do not observe a significant difference between on-arm and inter-arm complexes in the luminosity histogram either, perhaps with some on-arm complexes showing slightly larger luminosities than inter-arm complexes for a given line width.  Finally, the $\co$ virial masses are similar for both subsamples.  Similarly, we observe no significant differences between on-arm and inter-arm distributions of properties for $\cotwo$ clouds.  Nevertheless, we observe a group of clouds located in the high end of all the histograms shown for $\cotwo$ clouds, as they are more luminous, more massive and show broader line widths than the remaining clouds located in both on-arm and inter-arm regions.

\subsubsection{Star Formation for on-arm and inter-arm regions}\label{sf-inon}
Having studied the dependence of the properties of the clouds on the spiral structure, we will address the question whether the star formation process inside these clouds change if they are located at on-arm or inter-arm regions.  In their study of three disk galaxies (including NGC 6946), \citet{2010ApJ...725..534F} investigated the relationship between star formation and spiral arms using far-UV and 24$\mu$m images as star formation tracers.  Although they found that the SFR is more concentrated in spiral arms, the inter-arm star formation activity is still significant ($\sim 30\%$).  Moreover, they found that the fraction of gas in molecular phase is a factor of two larger in on arm regions than inter arm regions, although the regions of high molecular gas fraction are coincident with regions of high total gas surface density in the arms.  Thus, this behavior of star formation activity and molecular gas fraction provides no evidence for a shock-triggered molecular gas formation in spiral arms.  Nevertheless, the coarse resolution used by Foyle et.\ al ($\sim$13$\arcsec$) did not allow them to perform a cloud scale study of the properties.  

Figure \ref{figure_sf_arm} shows the same plot as Figure \ref{figure_sfr}, but using the on-arm/inter-arm classification.  We observe that on-arm sub group is shifted to higher SFR and higher molecular mass surface density with respect to the inter-arm population.  However, below $\Sigma_\mathrm{H2}  \simeq 70\ \Msun/\mathrm{pc}^2$, molecular clouds located in inter-arm regions are forming stars at a rate similar to the on-arm molecular gas.  In order to examine the significance of the difference between the SFR surface density of on-arm and inter-arm clouds, we have used a two-sample Kolmogorov-Smirnov (K-S) test.  By quantifying the difference between the cumulative distributions of two data samples, the K-S test allows us to test the null hypothesis that these arrays of data values are drawn from the same distribution.  A K-S test applied to $\Sigma_\mathrm{SFR}$ on-arm and inter-arm distribution of $\co$ complexes yields a $D=0.54$ (where $D$ is the maximum difference between the cumulative distributions), with a significance of $p=0.0015$, while the K-S test for $\cotwo$ clouds yields $D=0.54$ and a significance level of $p=0.0003$.  Thus, because both $p$-values are smaller than the default value of the level of significance, $\alpha=0.05$, we reject the null hypothesis, and we conclude that $\Sigma_\mathrm{SFR}$ for on-arm complexes and clouds are significantly different from the inter-arm molecular structures.

\subsection{Distribution of the properties in the disk}\label{rad-prop}
In this Section we examine whether the properties of the structures identified in this work vary across the disk.  Because we are not mapping the full extent of the disk of NGC 6946, we emphasize that the robustness of our analysis may be impacted by any large-scale asymmetry in the galaxy.

\begin{table*}
\caption{Mean properties of the complexes and clouds.\label{table-high-cld}}
\centering
\begin{tabular}{cccc}
\hline\hline
& $\langle\Sigma_\mathrm{H2}\rangle$ & $\langle\Sigma_\mathrm{SFR}\rangle$ & $\langle\Sigma_\mathrm{SFE}\rangle$ \\  
& $\Msun\ \mathrm{pc}^{2}$ & $\Msun\ \mathrm{yr}^{-1} \mathrm{kpc}^{-2}$ & $\times 10^{-3} \mathrm{Myr}^{-1}$ \\
\hline
$\co$ complexes in regions 6 ,7 and 11  & 121.1 $\pm$ 13.4 &  0.17 $\pm$ 0.02 & 1.3 $\pm$  0.44  \\
$\co$ complexes in the remaining regions & 42.9 $\pm$ 2.7 &  0.03 $\pm$ 0.002 &  0.59 $\pm$  0.06 \\
$\cotwo$ clouds in regions 6 ,7 and 11  & 100.3 $\pm$ 7.5 &  0.19 $\pm$ 0.01 & 2.3 $\pm$  0.43  \\
$\cotwo$ clouds in the remaining regions & 57.4 $\pm$ 3.1 &  0.048 $\pm$ 0.001 &  0.79 $\pm$  0.08 \\
\hline
\end{tabular}
\end{table*}

A full radial profile study of the gas and SFR over the optical disk of NGC 6946 can be found in \citet{2008AJ....136.2782L}.  We observe that, in general, the size, line width, virial mass and luminosity do not show any clear trends as we go from inner to outer part of the disk.  On the other hand, in Figure \ref{figure_rad_sfr} we present the variation of $\Sigma_\mathrm{H2}$, $\Sigma_\mathrm{SFR}$ and $\Sigma_\mathrm{SFE}$ across the disk.  In this plot we distinguish between on-arm and inter-arm classifications. While there is no overall trend with radius, we observe that both the molecular gas and the SFR surface density are enhanced for some structures located in on-arm regions.  This behavior is more pronounced for the SFR, which shows clearly a subgroup of clouds that depart from the remaining set of clouds, which shows SFR below $\sim$ 0.06 $\Msun$ yr$^{-1}$ pc$^{-2}$.  

\citet{2010ApJ...725..534F} found that SFR tracers are more concentrated to the spiral arm region than a uniform distribution for the grand design galaxies NGC 628 and NGC 5194, and for the galaxy being studied in this paper, NGC 6946.  They showed that this concentration depends on the pixel fraction attributed to arms, and can be 40\% using the brightest 30\% of pixels as the arm region.  A similar behavior for SFE is observed in the right panel of Figure \ref{figure_rad_sfr}.  Several on-arm structures are observed to have higher SFE than other structures located in on-arm or inter-arm regions.  Those regions correspond to the highest luminosity clouds identified in the histograms for $\cotwo$ showed in Figure \ref{figure_arm10}.  Spatially, those outliers correspond to $\co$ molecular emitting complexes (and the corresponding $\cotwo$ resolved inside them) located in regions 6, 7 and 11 illustrated in Figure \ref{figure_ngc6946e_13co}.  In order to have a clearer view of the location of these outlier structures, in Figure \ref{figure_map_sfe} we show the map of the SFE for both $\co$ complexes and $\cotwo$ clouds.  To create these maps, we have assigned the SFE value to the corresponding complex or cloud boundary given by CPROPS.  The black circles highlight the regions where we found the outliers in Figures $\ref{figure_arm10}$ and  \ref{figure_rad_sfr}, with black contours showing the $\co$ complexes with $\Sigma_\mathrm{H2} > 110\ \Msun$ pc$^{-2}$, and $\cotwo$ clouds with $\Sigma_\mathrm{H2} > 135\ \Msun$ pc$^{-2}$.

In order to quantify the difference between the structures located in regions 6, 7 and 11 and the molecular gas in the other parts of the regions observed, in Table \ref{table-high-cld} we show the average of $\Sigma_\mathrm{H2}$, $\Sigma_\mathrm{SFR}$ and $\Sigma_\mathrm{SFE}$ for this sub-group of clouds and the remaining set of structures.  We observe that the mean of $\Sigma_\mathrm{H2}$ calculated over the regions located in regions 6, 7 and 11 is a factor of 2 higher than the average value calculated in the remaining disk.  Also, the average $\Sigma_\mathrm{SFR}$ and $\Sigma_\mathrm{SFE}$ for this sub-group are a factor of 4 and 3 higher than the other regions respectively.

\section{DISCUSSION}\label{discuss}

\subsection{Molecular cloud properties}
Despite the long-standing acceptance of the Larson scaling relations, recent studies have raised questions about the form and universality of the scaling relations.  
For instance, \citet{2009ApJ...699.1092H}, reviewing the properties of the Galactic molecular clouds studied by S87, found LTE masses that are typically a factor of a few smaller than the virial masses derived by S87.  However, maybe a more remarkable finding is that the size-line width relation coefficient ($v_\mathrm{o} \equiv \sigma_v/R^{1/2}$) is not universal, but depends on the mass surface density of the clouds.  \citet{2009ApJ...699.1092H} attributed the variation in $v_\mathrm{o}$ to differences in magnetic field strength among the clouds.  Alternatively, \citet{2011MNRAS.411...65B} proposed that the dependence of $v_\mathrm{o}$ on the mass surface density is consistent with molecular clouds in a process of hierarchical and chaotic gravitational collapse.  They suggested that, although hydrodynamic turbulence is required to induce the first dense condensations in the ISM, once the structures become bound, the gravity is the main driver of the internal motions.

\begin{figure}
\centering
\begin{tabular}{c}
\epsfig{file=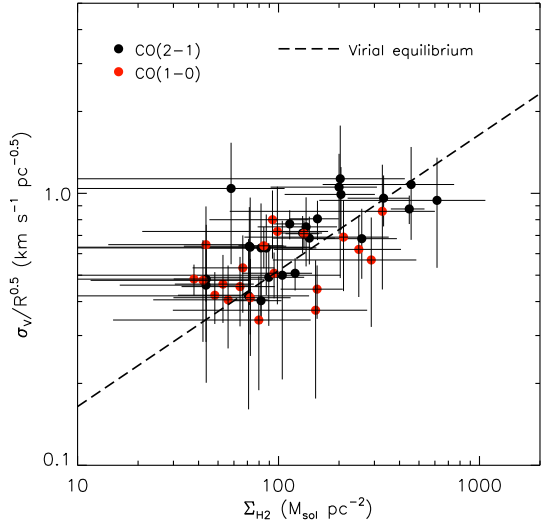,width=0.8\linewidth,angle=90}
\end{tabular}
\caption{Variation of the size-line width coefficient $v_\mathrm{o}=\sigma_v/R^{1/2}$ with the mass surface density of the structures derived in this study.  The dashed line represents the locus of clouds in virial equilibrium.}
\label{figure_co_fact}
\end{figure}

Figure \ref{figure_co_fact} shows the relation between $v_\mathrm{o}$ and the molecular gas surface density $\Sigma_\mathrm{H2}$ for the $\co$ complexes and $\cotwo$ clouds.  The structures we have identified appear to follow the relation shown in \citet{2009ApJ...699.1092H}, where $v_\mathrm{o} \propto \Sigma_\mathrm{H2}^{1/2}$ is expected for clouds in gravitational equilibrium, although these two variables are no independent ($\Sigma_\mathrm{H2} \propto \sigma_v$).

\begin{figure*}
\centering
\begin{tabular}{ccc}
\epsfig{file= 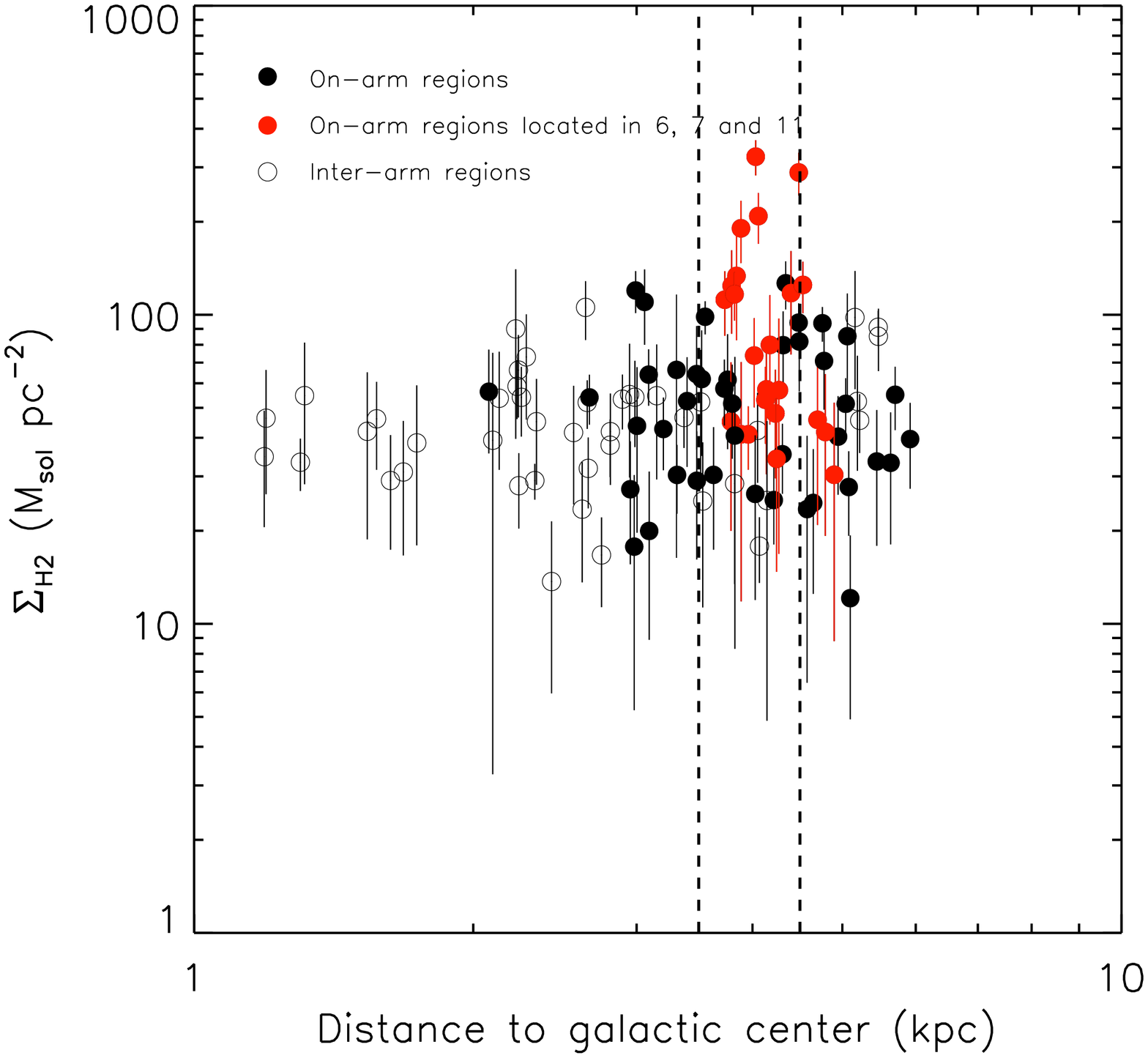,width=0.3\linewidth,angle=90} 
\epsfig{file= 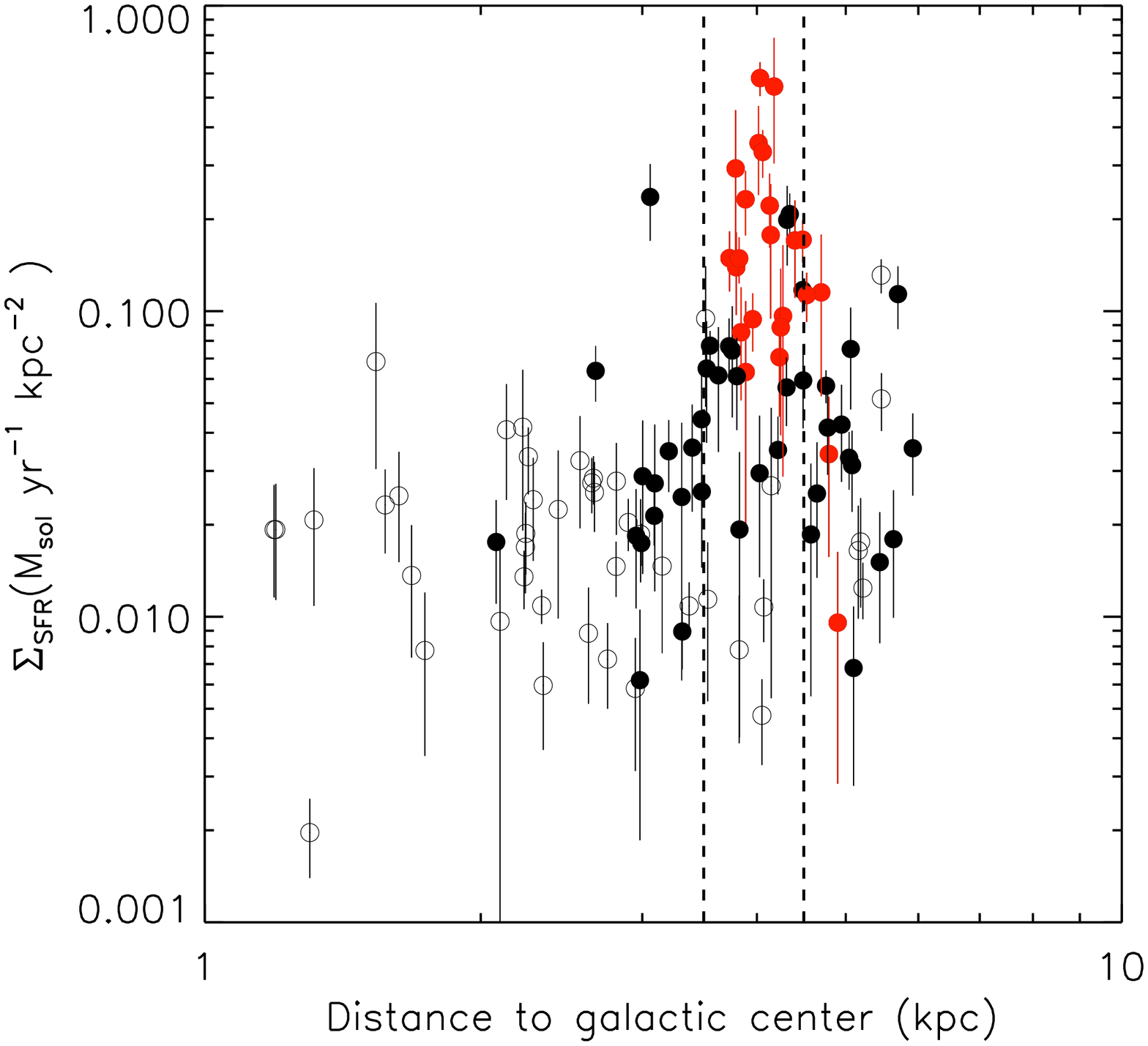,width=0.3\linewidth,angle=90} 
\epsfig{file= 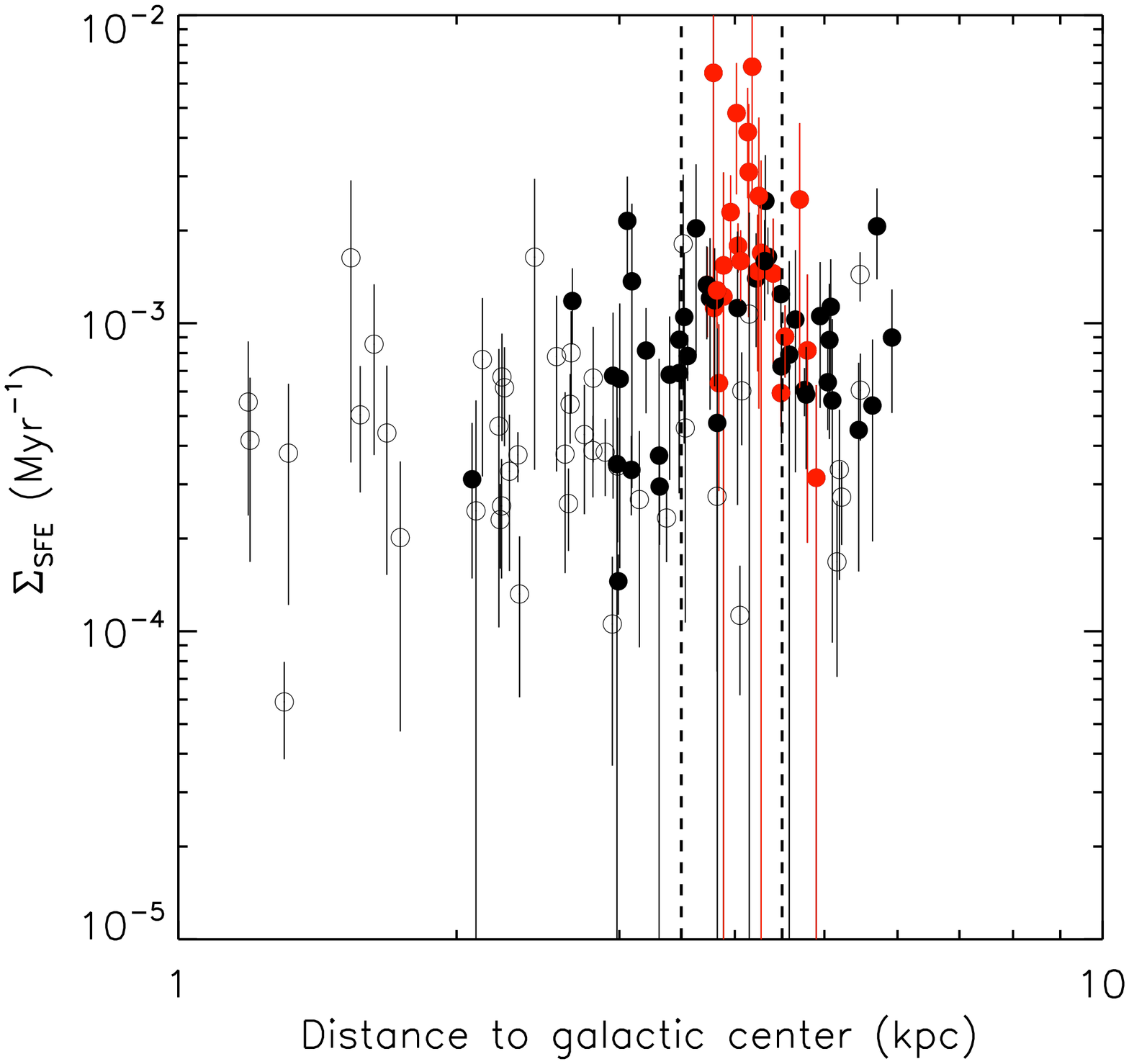,width=0.3\linewidth,angle=90} 
\end{tabular}
\caption{Radial distributions of $\Sigma_\mathrm{H2}$ {\bf (left)}, $\Sigma_\mathrm{SFR}$ {\bf (center)} and $\Sigma_\mathrm{SFE}$ {\bf (right)} of the regions observed over the disk, including both complexes and clouds.  Filled black dots represent on-arm complexes and clouds, while filled red dots illustrate the on-arm complexes and clouds located in regions 6, 7 and 11 (see Figures \ref{figure_ngc6946e_13co} and \ref{figure_map_sfe}).  Open circles illustrate inter-arm complexes and clouds.  Most of the on-arm structures that present higher $\Sigma_\mathrm{H2}$, $\Sigma_\mathrm{SFR}$ and $\Sigma_\mathrm{SFE}$ than the other complexes or clouds, particularly for the star formation rate, are located in regions 6, 7 and 11.  These structures are inside the region enclosed by $r=3.5$ kpc and $r=4.5$ kpc (illustrated by the black dashed lines, see Figure \ref{figure_map_sfe}).}
\label{figure_rad_sfr}
\end{figure*}

How are the GMCs formed in non-grand design galaxies like NGC 6946? In flocculent galaxies (or galaxies with active potentials) the formation of inter-arm molecular gas structures is proposed to be the result of local instabilities rather than being the fragmented residuals left behind by the kinematic shear of the GMAs in spiral arms of grand design galaxies.  In this scheme, the properties of the clouds are approximately similar across the disk, and no offsets between gas and star formation are expected.  High resolution observations of flocculent galaxies have provided evidence in support of this scenario.  For instance, \citet{2003PASJ...55..605T} mapped the $\co$ over the southern arm of the flocculent galaxy NGC 5055.  They found no obvious offset between H$\alpha$ and the molecular gas, and that on-arm and inter-arm clouds do not have significant differences in their properties.  Similarly, in the present study we have identified structures with similar properties across the observed portion of the disk.  Nevertheless, we have found that some of the on-arm clouds are more massive and have broader line widths than other clouds located in both on-arm and inter-arm regions.  Why are clouds departing from the mean properties observed elsewhere?  According to simulations of gas in galaxies with an active potential (\citealt{2006MNRAS.371..530C}; \citealt{2008MNRAS.385.1893D}; \citealt{2011ApJ...735....1W}) the most massive structures are observed to be produced in regions where collision or merging between spiral arms occur, yielding an enhanced star formation in those overdense regions.  We have found that the more massive clouds spatially match sites of higher SFR, which may provides evidence in favor of these models.  A detailed description of the velocity field is needed to examine the potential presence of colliding flows of gas in this type of galaxy (see Section \ref{velo-field} below).

\begin{figure*}
\centering
\begin{tabular}{cc}
\epsfig{file=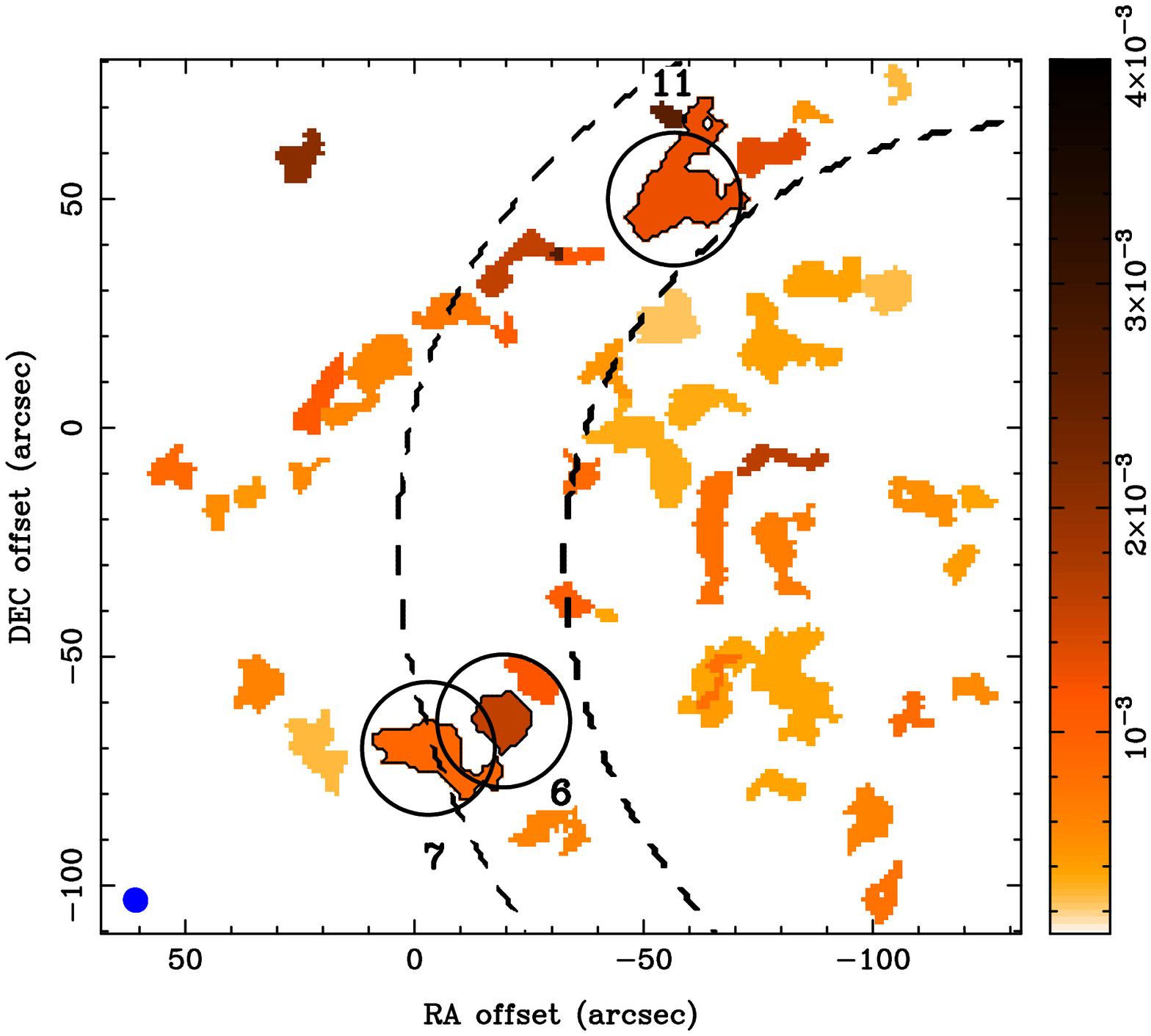,width=0.4\linewidth,angle=0} 
\epsfig{file=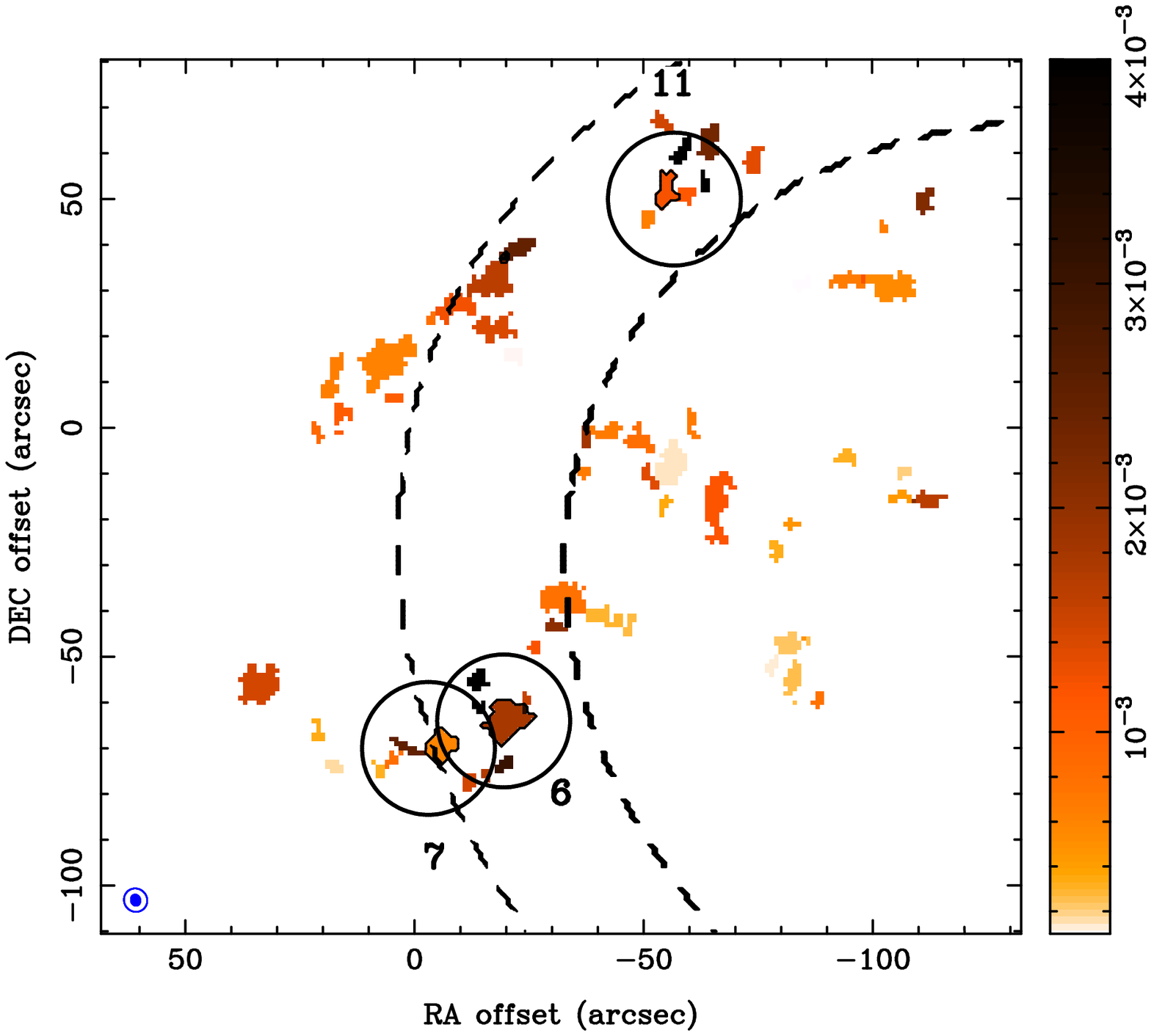,width=0.4\linewidth,angle=0} 
\end{tabular}
\caption{{\bf Left}: Map of the Star Formation Efficiency derived for individual $\co$ complexes.  The color bar is in units of Myr$^{-1}$.  The black contours highlight the complexes with $\Sigma_\mathrm{H2} > 110$ $\Msun$ pc$^{-2}$.  {\bf Right}: Map of the SFE for the $\cotwo$ clouds.  In this case, black contours highlight clouds with $\Sigma_\mathrm{H2} > 135$ $\Msun$ pc$^{-2}$.  As in the left panel, the color bar is in units of Myr$^{-1}$.  Circles illustrate regions where we found structures that deviate from the other identified structures in Figure \ref{figure_rad_sfr}.  Dashed lines denote radii of $r=3.5$ kpc and $r=4.5$ kpc (see Figure \ref{figure_rad_sfr}).}
\label{figure_map_sfe}
\end{figure*}

In Figure \ref{figure_arm10} we found that some of the inter-arm $\co$ complexes can have size and mass comparable to the most massive complexes found in on-arm structures.  However, taking advantage of the finer resolution provided by $\cotwo$ observations, we have found that while the most massive on-arm clouds remain equally massive as we resolve the smaller structures, the inter-arm complexes are decomposed into several less massive components.  Nevertheless, the sizes, line widths, luminosities and masses are similar for most the structures observed in both on-arm and inter-arm regions.  Thus, the properties of the giant molecular clouds in NGC 6946 do not change substantially when they enter the spiral arms.  However, we observe two regions that show significant higher surface density and more massive GMCs than in other regions across the disk.

A natural comparison to our study of resolved GMCs in the disk of NGC 6946 may be made with the $\co$ observations of the central 5 kpc of NGC 6946 presented in DM12. As shown in Figure \ref{figure_line-width}a, despite differences in resolution and cloud identification algorithms, the sizes and velocity dispersions of the DM12 GMCs and our complexes (as traced by $\co$) follow a similar trend (with the exception of the handful of DM12 clouds within 400 pc of the galaxy center), which is slightly steeper than that measured by S87 for Galactic disk GMCs. Figure \ref{figure_line-width}b shows that the trends defined by the luminosities of the $\co$ complexes and DM12 GMCs are also consistent. In Figure \ref{figure_line-width}c, our observations indicate that a constant (i.e., not radially varying) value of $X_\mathrm{CO}$ of $2 \times 10^{20}$ is appropriate, which is consistent with the value typically assumed for the Milky Way disk clouds. This value is slightly higher than (but within the quoted errors of) DM12, who derive an average, non-radially varying value of $1.2 \times 10^{20}$ in the central kiloparsecs of NGC 6946. We note that adopting an $I(2 \rightarrow 1) = I(1 \rightarrow 0)$ ratio of $< 1$ would reduce the best-fit value of $X_\mathrm{CO}$ for the $\cotwo$ clouds by the same factor.  Overall, the trend for the $\co$ detections in Figure \ref{figure_line-width}c is consistent between the two samples, but they begin to diverge at the highest cloud masses and luminosities, likely because the complexes presented here are blends of multiple GMCs.

\subsection{Comparison with Galactic studies}\label{discuss-gal}
Recent studies of star formation rate (commonly traced by young stellar objects or YSOs) and column densities of gas (usually traced by near-IR extinction) in nearby Galactic clouds have compared Galactic and extragalactic gas and SFR surface density relations (\citealt{2009ApJS..181..321E}; \citealt{2010ApJ...723.1019H}; \citealt{2011ApJ...739...84G}).  For instance,  \citet{2010ApJ...723.1019H} found that the prescription given by  the Kennicutt-Schmidt law underpredicts the values found in nearby molecular clouds by factors that range from 21 to 54.  They found that such differences emerge from the different scales involved in calculating the SFR and the gas surface densities.  While the gas tracers used in Galactic studies are usually probing the denser component of the gas, the extragalactic observations have averaged over scales that include both dense and a more diffuse gas component which may not be related to the star formation process.  

\begin{figure}
\centering
\begin{tabular}{c}
\epsfig{file=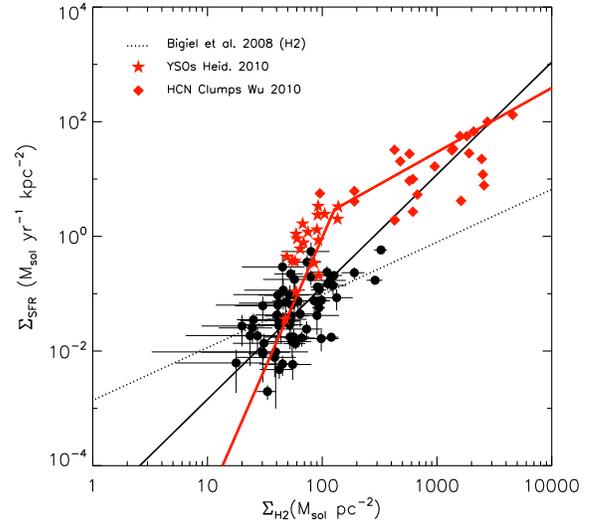,,width=0.8\linewidth,angle=90} 
\end{tabular}
\caption{Comparison between the $\Sigma_\mathrm{SFR}-\Sigma_\mathrm{H_2}$ values found in this work for NGC 6946 GMCs from $\cotwo$ (black dots) and the values presented by \citet{2010ApJ...723.1019H} for Galactic star forming regions (red symbols).  Stars show low-mass star formation regions, with $A_V$ maps and counts of YSOs used as tracers of gas and SFR respectively.  Diamonds illustrate the most massive star-forming clumps from \citet{2010ApJS..188..313W} traced by HCN gas maps.  Red solid lines show the broken power law found by \citet{2010ApJ...723.1019H}, while the black solid line illustrates the relation found in this present paper for $\cotwo$ clouds.  Galactic low-mass star-forming regions overlap with the NGC 6946 GMCs for $\Sigma_\mathrm{H2} \sim 80\ \Msun\ \mathrm{pc}^{2}$, although we emphasize that our estimates of SFR trace massive star formation.  Thus, HCN massive clumps represent a more direct comparison to our GMC values.  We observe that the massive dense clumps fall along a crude extrapolation of the $\Sigma_\mathrm{SFR}$-$\Sigma_\mathrm{H2}$ relation found for NGC 6946 GMCs (Equation \ref{sfr21_fit}) to higher gas surface densities.}
\label{figure_sfr_heid}
\end{figure}

Figure \ref{figure_sfr_heid} compares the values found in this work for NGC 6946 GMCs and the values presented by \citet{2010ApJ...723.1019H} for Galactic star forming regions.  Besides the $\Sigma_\mathrm{gas}$ and $\Sigma_\mathrm{SFR}$ values for low-mass star formation regions traced by $A_V$ maps and counts of YSOs respectively, they included the most massive star-forming clumps that \citet{2010ApJS..188..313W} studied with HCN gas maps.  We note that low-mass star-forming regions (red stars) overlap with the GMCs for $\Sigma_\mathrm{H2} \sim 80\ \Msun\ \mathrm{pc}^{2}$.  Nevertheless, this apparent coincidence has to be considered carefully, as we are comparing low-mass (galactic) and massive star formation (extra-galactic).  Additionally, the masses of the large molecular clouds in \citet{2010ApJ...723.1019H} study are two order of magnitudes smaller than the masses of the GMCs reported in this study.  Probably a more direct comparison to our GMCs would be provided by the massive HCN clumps, since extragalactic SFR tracers are exclusively sensitive to massive star formation.  In Figure \ref{figure_sfr_heid} we have included the massive HCN clumps as red diamonds.  Surprisingly, if we extrapolate the $\Sigma_\mathrm{SFR}$-$\Sigma_\mathrm{H2}$ relation we found for $\cotwo$ GMCs (Equation \ref{sfr21_fit}) to higher gas surface densities, we observe that the massive dense clumps fall roughly along the relation.  Thus, the $\Sigma_\mathrm{SFR}$-$\Sigma_\mathrm{H2}$ relation for massive star formation regions is consistent with a quadratic relation.  Nevertheless, a more complete dynamical range in molecular gas and star formation surface density is needed to fill the gap between Galactic and extragalactic observations, and assess the intrinsic $\Sigma_\mathrm{SFR}$-$\Sigma_\mathrm{H2}$ relation.

\subsection{Star formation and evolution of GMCs in NGC 6946}\label{discuss-sfr}
Although a complete unbiased survey of the GMCs population and their star formation activity in different environments is needed to properly investigate the $\Sigma_\mathrm{SFR}-\Sigma_\mathrm{H2}$ relation, our analysis allows us to shed light on the difference in star formation activity between on-arm and inter-arm clouds.  In Sections \ref{sf-inon} and \ref{rad-prop}, we observed a clear enhancement of SFR in on-arm clouds.  This enhancement is more pronounced for structures located in regions we suspect are the result of recent convergence of gas flows (regions 6, 7 and 11 in Figure \ref{figure_ngc6946e_13co}).  In fact, it is in those regions where we find significant $\cother$ emission, giving further observational evidence for the presence of denser gas in those structures. 

We have found steeper slopes than previous studies on NGC 6946 for the $\Sigma_\mathrm{SFR}-\Sigma_\mathrm{H2}$ relation for both $\co$ complexes and $\cotwo$ clouds.  As was stated above, the slope of the relation $\Sigma_\mathrm{SFR}-\Sigma_\mathrm{H2}$ can be affected by several factors, including the method used to perform the linear regression between these two quantities, the tracer used to estimate gas densities and star formation activity, selection effects and the resolution of the maps used.  For the finest resolution in this paper, which is given by the $\cotwo$ observations towards the brightest molecular gas regions, we have found a $\Sigma_\mathrm{SFR}-\Sigma_\mathrm{H2}$ relation that is almost quadratic.  A quadratic relation for $\Sigma_\mathrm{SFR}$-$\Sigma_\mathrm{H2}$ has been proposed by \citet{2011ApJ...739...84G} for Galactic clouds.  They show that their data are consistent with a star formation law in which the $\Sigma_\mathrm{SFR}$ is proportional to the $\Sigma_\mathrm{gas}^2$.  Deviations from the power law are attributed to evolutionary stage differences in the local gas: in some older regions the gas can be disrupted by stars that have been formed there, and other younger regions can contain star clusters deeply embedded in dense gas at the onset of star formation.  

\begin{figure}
\centering
\begin{tabular}{c}
\epsfig{file=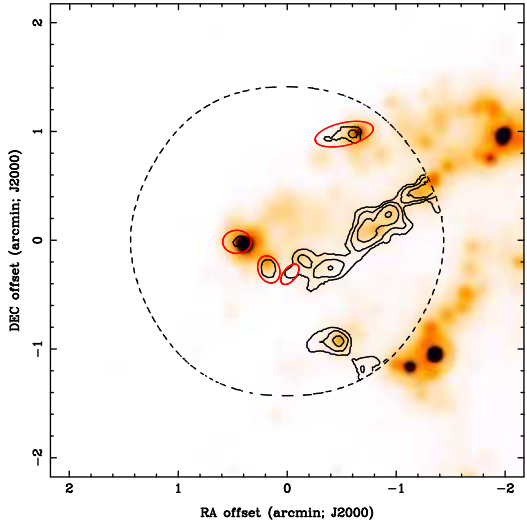,width=0.8\linewidth,angle=-90} 
\end{tabular}
\caption{This figure illustrates the region of the eastern part of NGC 6946 we have observed with higher sensitivity using CARMA in E array.  The dashed line represent the 7-pointing mosaic we used to cover the region.  The contour map illustrates the $\co$ intensity map, with contour levels of 2$^n$ $\times$ 1.5 K $\kms$ ($n \ge 1$).  The color map shows the FUV+24$\mu$m composite SFR map of the same region.  Red ellipses illustrate the molecular gas complexes with low $\co$ emission, but associated SFR.}
\label{figure_mom0_e}
\end{figure}

\begin{figure}
\centering
\begin{tabular}{ll}
\epsfig{file=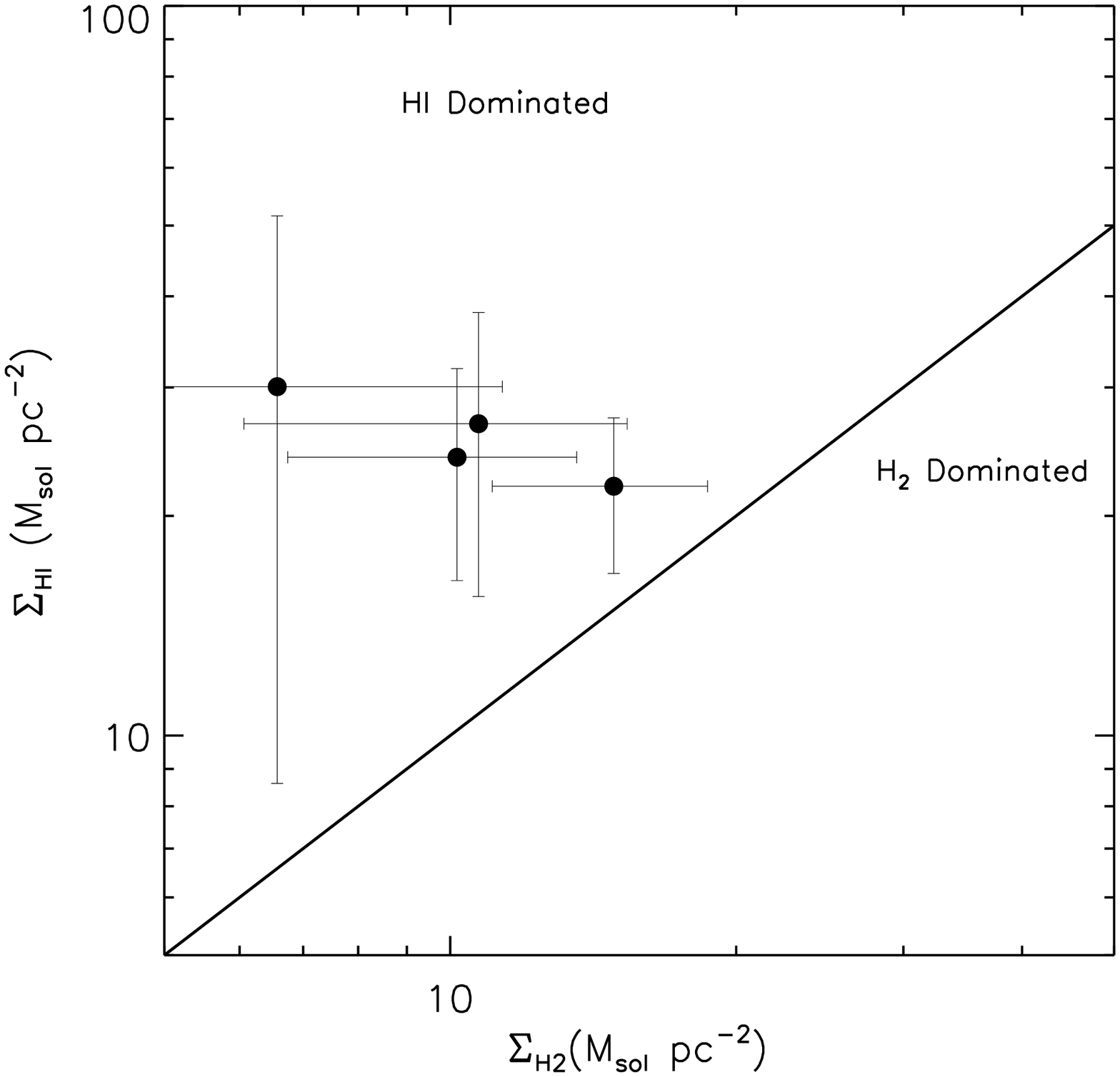,width=0.45\linewidth,angle=90} &
\epsfig{file=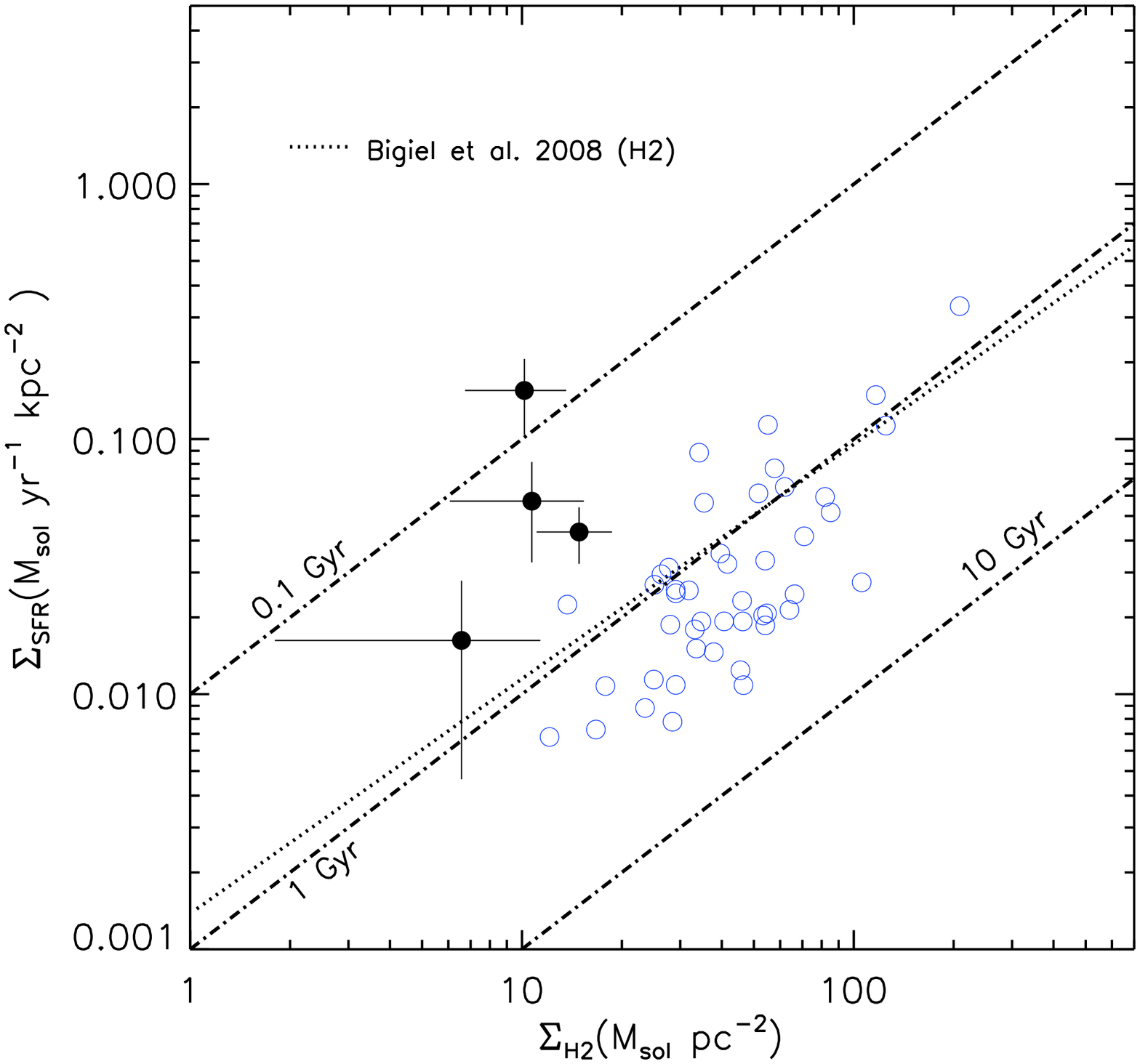,width=0.45\linewidth,angle=90} 
\end{tabular}
\caption{{\bf Left}: Atomic gas surface density vs.\ molecular gas surface density for $\co$ complexes detected with higher sensitivity (red ellipses in Figure \ref{figure_mom0_e}).  We observe that these complexes are located in regions where the gas is dominated by the atomic gas phase.  Symbols and labels are the same as Figure \ref{figure_hi_h2}.  {\bf Right}: Star formation rate vs.\ molecular gas surface density for the $\co$ complexes presented in the left panel (black circles).  As a comparison, blue open circles show the $\co$ complexes found in the complete mosaic (left panel of Figure \ref{figure_sfr}).  The complexes observed with higher sensitivity appear to deviate from the trend found by \citet{2008AJ....136.2846B} due to their high star formation activity but low molecular gas content, as is shown in Figure \ref{figure_mom0_e}.  These cases may represent complexes that have been consuming their molecular gas by forming stars, yielding HI from photodissociation of H$_2$ by the ambient radiation field.}
\label{figure_sfr_e}
\end{figure}

We noticed in Figure \ref{figure_ngc6946e} that a few regions located in the outskirts of the disk present weak or undetectable emission in $\co$ at our sensitivity level, but are bright in 24$\mu$m, FUV, H$\alpha$ and HI.  In order to detect the gas in some of those regions, we performed deeper low resolution observations of $\co$ using CARMA in E array configuration towards the small region illustrated in Figure \ref{figure_mom0_e}.  The sensitivity reached by these $\co$ observations was $\sigma \sim$ 3 $\Msun$ pc$^{-2}$ for a channel width of 2.5 $\kms$ and resolution $8\farcs45$ $\times$ $7\farcs29$ (a factor of $\sim$ 4 higher than the sensitivity yielded by our extended mosaic).  We have applied the same procedure explained in Section \ref{cprops} to identify discrete structures and estimate their properties.  Nevertheless, in this case we are only interested in the $\Sigma_\mathrm{H2}$, $\Sigma_\mathrm{HI}$ and the $\Sigma_\mathrm{SFR}$.  Figure \ref{figure_sfr_e} shows the relation between these quantities for the complexes found in this region.  The right panel shows that these complexes exhibit low molecular gas content, but SFR comparable to denser molecular clouds observed in the complete mosaic (shown by open blue circles).  Additionally, in the left panel we observe that the atomic content is comparable to the molecular gas in these low CO-luminosity complexes.

A possible way to explain structures with high SFR and low molecular gas surface density is to consider an evolutionary scenario similar to that proposed by \citet{2011ApJ...739...84G} for a sample of large Galactic molecular clouds.  Given that these molecular complexes in NGC 6946 appear to be isolated from the spiral arms, they could be GMCs that escaped from the gravitational potential minima, and consumed some of their molecular gas through forming stars.  That atomic gas peaks in these regions can be then naturally explained as a result of the photo-dissociation of the molecular gas by the newborn star population.  We observed peak atomic gas surface densities of $\sim$ 35 $\Msun$ pc$^{-2}$ in these regions, similar to the peaks found in the arm regions, and a factor of 3 larger than the azimuthally averaged value at that distance from the galactic center.  We postulate that, these structures deviate from the main $\Sigma_\mathrm{SFR}$-$\Sigma_\mathrm{H2}$ relation found for the remaining complexes or GMCs due to differences in evolutionary stage.  Higher sensitivity and resolution observations of the gas in these regions, as well as age estimates are needed to corroborate this scenario.

\subsection{Velocity field and comparison with models}\label{velo-field}
Although a detailed comparison between our observations and simulated non-grand design galaxies is beyond the scope of this paper, in this section we perform a simple comparison between the kinematics of the ISM observed in NGC 6946 and what is predicted by simulations.  In their simulation of a galactic disk with a live spiral model, \citet{2011ApJ...735....1W} presented a dynamical picture of the behavior of the ISM.  They showed that both gas and the stellar arms roughly follow the local galactic rotational velocity, and they predicted that the relative velocities of the gas with respect to the mean rotational speed should be $\lesssim$ 15 $\kms$.  In this scenario, the cold gas does not follow the organized flow pattern predicted by the conventional density-wave theory and shows rather chaotic velocity structures around spiral arm potential minima, especially when the time dependence of stellar spiral potential is also calculated (\citealt{2011ApJ...735....1W}). This effect was also observed by \citet{2008MNRAS.385.1893D} in their simulation of spiral galaxies with an active potential.  

\begin{figure}
\centering
\begin{tabular}{c}
\epsfig{file=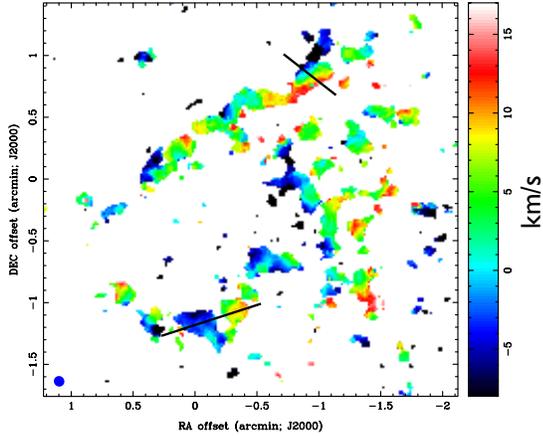,width=0.85\linewidth,angle=0} 
\end{tabular}
\caption{Residual velocity field of the region observed in NGC 6946.  The color bar is in units of $\kms$.  Black solid lines illustrate the slices used to generate the position-velocity diagrams shown in Figure \ref{figure_pv}.  We have subtracted a model of the rotation velocity from the observed $\co$ velocity field.  The model of the rotation velocity was created using the rotation curve derived by \citet{2008AJ....136.2648D}.   We observe that there is a significant velocity gradient in one of the northern regions where we detect high SFE (region 11 in Figure \ref{figure_map_sfe}).  In the southern region (region 6 and 7 in Figure \ref{figure_map_sfe}), on the other hand, we do not observe a significant velocity gradient inside the individual CO complexes, but there is a $\sim$ 15 $\kms$ discontinuity between the gas in regions 7 and 6.}
\label{figure_mom1_e}
\end{figure}

\begin{figure}
\centering
\begin{tabular}{cc}
\epsfig{file=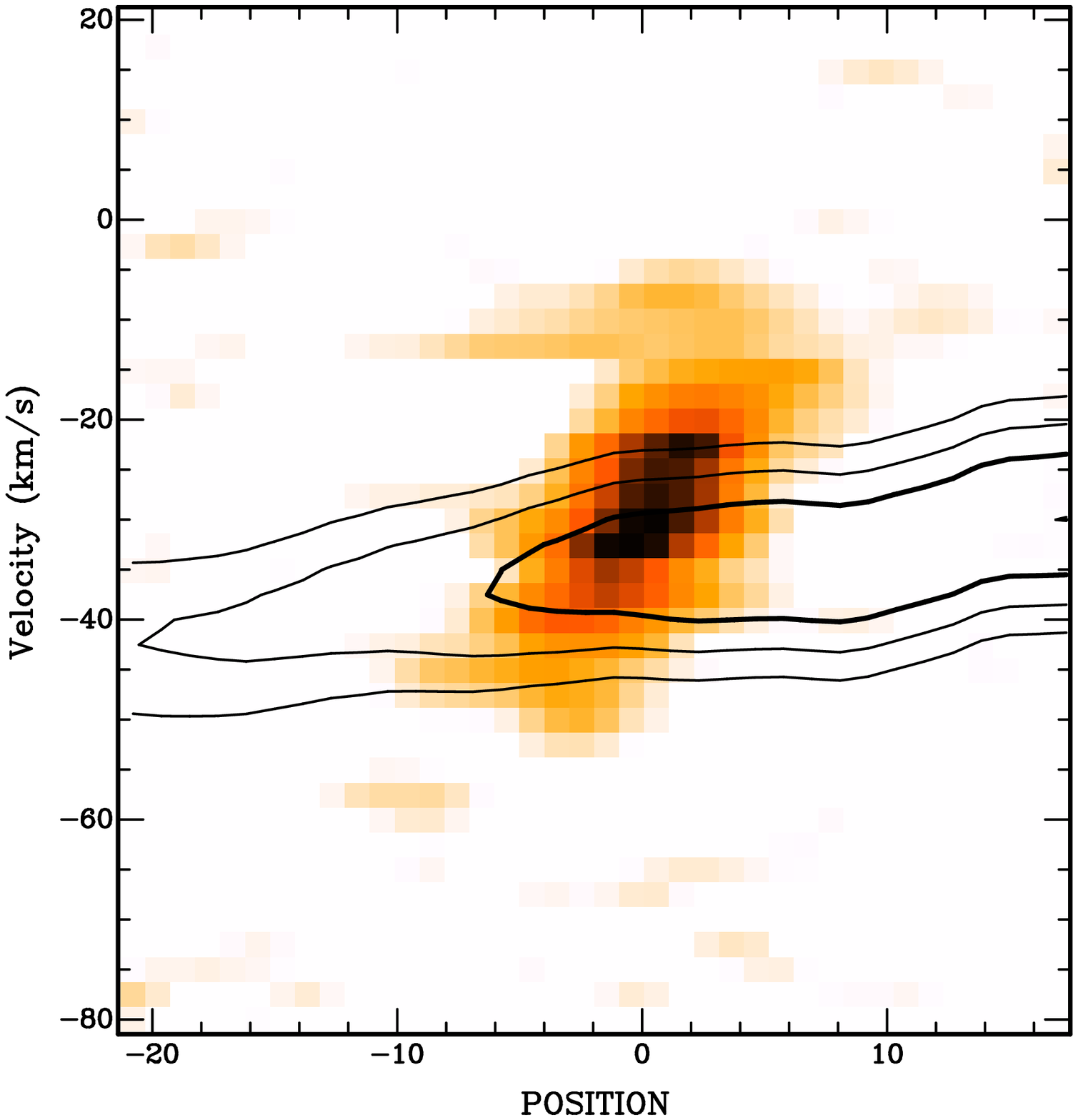,width=0.45\linewidth,angle=0} 
\epsfig{file=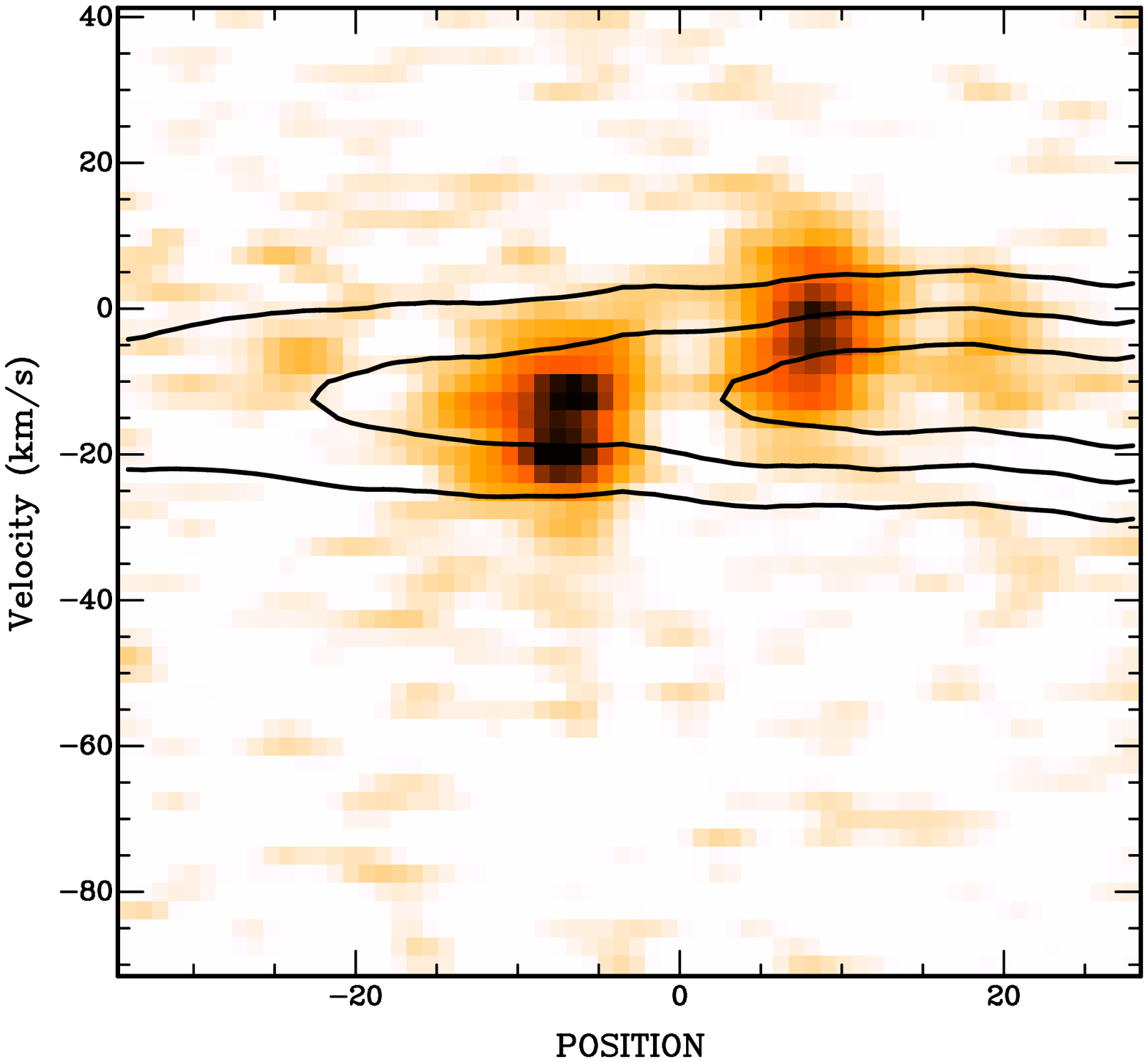,width=0.5\linewidth,angle=0} 
\end{tabular}
\caption{Position-velocity diagrams of the regions showing velocity features in the residual velocity field.  The extent and orientation of the slices are shown by the black solid lines in Figure \ref{figure_mom1_e}.  Color map shows the $\co$ data, and the black contours illustrate the model of the rotation velocity created using the rotation curve by \citet{2008AJ....136.2648D}.  {\bf Left}:  Position-velocity diagram of region 11.  We observe that the velocity gradient is sharper in the $\co$ data than the global rotation velocity model.  {\bf Right}:  Position-velocity diagrams of the regions 6 (peak on right) and 7.  We do not observe a significant velocity gradient inside the individual CO complexes, but the molecular gas in region 6 is moving about $\sim$ 15 $\kms$ away from the contiguous region 7.}
\label{figure_pv}
\end{figure}

In order to investigate the kinematics of gas flows in the ISM, we have created a velocity residual map in the region of NGC 6946 observed in this paper.  The velocity residual map was created by subtracting the local observed velocity field of the molecular gas traced by $\co$ by a model of the global circular motion of the gas.  We aim to identify the non-circular motions of the molecular gas, and assess whether the places with significant velocity gradient are related to regions of dense gas and high star formation.  The model of the velocity field was created using the rotation curve derived by \citet{2008AJ....136.2648D} for the HI map from THINGS.  We have used the GIPSY task {\it velfi} to create the velocity field model from the parameters provided by \citet{2008AJ....136.2648D} such as inclination, position angle, and the rotation velocity at a given radius using a tilted ring approach.  The final residual velocity map is shown in Figure \ref{figure_mom1_e}.  We observe that the molecular gas shows irregular non-circular motions in the range $\sim -10$ $\kms$ to 15 $\kms$ in the plane of the disk, in agreement with the simulation provided by \citet{2011ApJ...735....1W}.  Moreover, we observe a region of strong velocity gradient coincident with one of the regions we found to present high SFE (region 11 in Figure \ref{figure_map_sfe}).  In order to have a more detailed view of the velocity field in this region, in the left panel of Figure \ref{figure_pv} we show a position-velocity diagram of region 11 taken parallel to the local velocity gradient.  The molecular gas shows a much steeper velocity gradient compared to the global rotation velocity described by the model.  According to models of spiral galaxies with an active potential, this region may be the result of the convergence of gas into the spiral potential from both sides, as we observe in the residual velocity field regions moving faster and slower than the global rotation model, i.e., a transition of positive to negative velocity values across the region.

The other region with high SFE (region 6 in Figure \ref{figure_ngc6946e_13co}), on the other hand, does not present such a high velocity gradient as the region discussed above.  However, we observe a $\sim$15 $\kms$ difference in velocity between region 6 and the regions located to the north-west and to the south-east.  While those regions surrounding the molecular gas complex in region 6 are moving more or less at $-5$ $\kms$ with respect to the rotating frame, region 6 is moving in the other direction with a velocity of $\sim$10 $\kms$.  This difference in velocity can be observed clearly in the position-velocity diagram shown in the right panel of Figure \ref{figure_pv}.  This may imply that the gas actively forming stars in region 6 could be the result of a recent convergence of flows.  Although we do not observe signatures of a merging or collision process of spiral arms in the gas distribution near region 6, it is possible that the process had already formed a new spiral arm, obscuring the original arm locations.  More detailed analysis of the kinematics of the gas is required to disentangle the actual process involved in the enhancement of star formation efficiency in region 6.

\subsection{Caveats of this work}\label{caveats}
The study presented here can be improved in multiple ways.  Our analysis has been limited to the north-eastern area of the molecular gas disk of NGC 6946 for $\co$ observations, and to some of the brightest regions for $\cotwo$ follow-up observations.  A full census of the GMCs population across the disk is needed in order to perform a detailed study of molecular cloud properties and their relation with the surrounding environment.  Nevertheless, high sensitivity maps with resolutions $\sim 60$ pc remain observationally challenging to obtain, although recent studies have achieved this goal with CARMA+NRO 45m (e.g. DM12).  New facilities like ALMA will provide maps of unprecedented quality, allowing us to perform an unbiased analysis of the molecular gas in nearby galaxies at scales close to GMC sizes.

In our SFR calculations for $\cotwo$ clouds, we have applied a crude global extinction correction A$_\mathrm{H\alpha}=1$ mag to the SFR(H$\alpha$).  Although we showed that such correction is a good approximation to the SFR values found using a combination of FUV+24$\mu$m for $\co$ complexes, a more sophisticated approach to calculate the obscured SFR at resolution close to the GMCs sizes is needed.  Thus, tracers relatively unaffected by dust absorption represent a valuable alternative to estimate the total SFR at smaller scales.  In this direction, observations of Paschen $\alpha$ (P$\alpha$) line could provide unbiased measurements of the number of ionizing photons in these star forming regions (\citealt{2007ApJ...666..870C}). 

Equations \ref{sfr} and \ref{sfr_ha} assume that stars have been formed continuously.  Nevertheless, this assumption may not be longer valid for scales close to GMC sizes, as one approaches the case of  a single stellar population with a discrete age.  Some studies have investigated the uncertainty in using a linear conversion from UV or H$\alpha$ luminosity to estimate the SFR, making use of models assuming an instantaneous burst of star formation (\citealt{2010MNRAS.407.2091G}; \citealt{2012AJ....144....3L}).  \citet{2012AJ....144....3L} showed that while H$\alpha$ emission occurs mostly in the first 10 Myr, FUV presents low, but significant, emission up to 65 Myr after the burst.  They report a factor of 2 uncertainty inferring the SFR from H$\alpha$, and a factor 3-4 uncertainty inferring SFR from FUV.  Resolved star cluster observations in the regions observed in this study could shed light on the ages involved, and provide an independent way to estimate the star formation rate which can be compared to the SFR traced by FUV or H$\alpha$.

\section{Summary}\label{summary}

We have performed one of the most detailed observations of GMCs outside the Milky Way.  Using CARMA, we have observed $\co$ and $\cother$ over the northeastern disk of the non-grand design galaxy NGC 6946, covering a region of 6$\times$6 kpc$^2$ at a resolution of $\sim$110 pc.  Although at this resolution it is possible to resolve some of the largest molecular clouds, we have complemented this map with higher ($\sim$50 pc) resolution maps of $\cotwo$ towards the brightest regions observed in $\co$.  The extended flux has been recovered by merging our interferometric maps from CARMA with single dish maps from NRO 45m and IRAM 30m.  The results are summarized as follows:

\begin{enumerate}

\item We have estimated the properties of the molecular gas structures (size, line width and flux) using the cloud-finding algorithm CPROPS (Section \ref{cl-prop}).  We have resolved the largest molecular complexes into several structures.  At the resolution offered by $\co$ observations, we have identified 45 CO emitting complexes with typical sizes of $\sim$150 pc.  Five of these complexes show significant $\cother$ emission, which appear to be regions of denser gas.  The higher resolution observations of $\cotwo$ towards the brightest regions detected in $\co$ allowed us to find 64 structures which we have identified as GMCs.   

\item In Section \ref{cl-prop-corr} we observed that the properties of the clouds and complexes follow relations similar to those found previously for extragalactic clouds (Figure \ref{figure_line-width}).  In particular, the trends defined by the $\co$ complexes presented in this paper are consistent with those found in the center of NGC 6946 by DM12, despite the differences in resolution and cloud identification algorithms; the exception is the set of GMCs located within 400 pc of the galactic center with large velocity dispersions (> 10 $\kms$).  Additionally, our $M_\mathrm{vir}$-$L_\mathrm{CO}$ relation is located slightly above the relation found by DM12 for the central part of NGC 6946 and is consistent with our choice of the CO to H$_2$ conversion factor $X_\mathrm{CO}$=$2\times10^{20}\mathrm{cm}^{-2}(\mathrm{K}\ \kms)^{-1}$.  

\item We have estimated LTE masses for the regions with significant $\cother$ emission (Figure \ref{figure_13co}).  Although the LTE masses are smaller by a factor of 2 than the luminosity-based masses in our sample of clouds, this discrepancy does not necessarily mean that $\co$ intensities are not suitable to estimate true molecular cloud masses, but may reflect systematic errors arising from our lack of knowledge of the excitation temperature and CO abundance, both needed to infer masses from $\cother$ emission.

\item The star formation rates of the $\co$ complexes were estimated in Section \ref{sfr-result} using multiband {\it Spitzer} imaging from SINGS and FUV GALEX maps of NGC 6946.  In Figure \ref{figure_sfr} we observe a clear correlation between SFR surface density ($\Sigma_\mathrm{SFR}$) and molecular gas surface density ($\Sigma_\mathrm{H2}$).  The bisector linear regression for the relation $\Sigma_\mathrm{SFR}=A\ \Sigma_\mathrm{H2}^N$ yields a slope of $N = 1.43 \pm 0.11$ for the $\co$ emitting complexes, which is steeper than the value found by previous work ($N=0.92$, \citealt{2008AJ....136.2846B}).  In order to detect faint CO complexes, but with active star formation, we have observed with higher sensitivity the $\co$ line over a smaller region located beyond the spiral arms, where atomic gas is comparable (or dominant) to the molecular gas (Figure \ref{figure_mom0_e}).  We interpret these CO-faint star forming regions as isolated structures that could have departed from the main spiral arms, and have been turning their molecular gas into stars since then.  Thus, it is not surprising to find atomic gas peaks in these regions, which could have been produced by photodissociated molecular gas.  Higher sensitivity observations are necessary to detect faint molecular gas associated with such late stage star formation.   

\item For GMCs identified in the $\cotwo$ maps, we estimate their SFR using H$\alpha$ maps.  We have found a steeper slope than the values found for $\co$ complexes, with $N = 1.96 \pm 0.18$.  We observe that Galactic massive dense clumps follow a rough extrapolation of the $\Sigma_\mathrm{SFR}$-$\Sigma_\mathrm{H2}$ relation for these $\cotwo$ GMCs to higher gas surface densities (Figure \ref{figure_sfr_heid}).  ALMA will observe the molecular gas distribution at even higher resolution and sensitivity, allowing a direct comparison between the Galactic and extragalactic $\Sigma_\mathrm{SFR}$-$\Sigma_\mathrm{H2}$ relations at similar scales.

\item In Section \ref{in-on-arm}, on-arm and inter-arm regions have been defined based on stellar mass density enhancements traced by the 3.6 $\mu$m image of NGC 6946 (Figure \ref{figure_four}).  The complexes and GMCs were classified as on-arm and inter-arm structures accordingly.  Although we observe that cloud properties have similar values independent of location in the disk, a small number of clouds are more massive, present broader line widths, and higher star formation rate than the remaining sample of clouds (Figure \ref{figure_arm10}).  We found that these structures appear to be located at two specific regions in the spiral arms (region 6-7 and region 11 in Figure \ref{figure_ngc6946e_13co}).

\item The star formation efficiency (SFE, defined as the ratio between $\Sigma_\mathrm{SFR}$ and $\Sigma_\mathrm{H2}$) is roughly flat over the region observed, with values between $10^{-4}$ and $10^{-3}$ Myr$^{-1}$  (Section \ref{rad-prop}, Figure \ref{figure_rad_sfr}).  However, the SFE is enhanced in the regions where we observed more massive and more luminous GMCs (see Figure \ref{figure_map_sfe} and Table \ref{table-high-cld}).  The velocity field of the molecular gas reveals irregular non-circular motion with values below 15 $\kms$ with respect to the rotating frame (Figure \ref{figure_mom1_e}).  In addition, we found some evidence for convergence of gas flows in one of the regions of high SFE (Figure \ref{figure_pv}).  The other region presenting SFE peaks does not show such a strong velocity gradient.  However, this region shows a 15 $\kms$ difference with respect to the surrounding molecular gas, which could be caused by a previous convergence of gas flow.  Although a more detailed comparison with simulations is needed in order to have a more conclusive picture of the role of gas dynamics in the formation of GMCs and stars in these type of galaxies, these new observations suggest that, in agreement with numerical simulations of non-grand design spiral galaxies, the convergence of gas flows into the spiral arm, and collisions or merging of arms may increase the local gas density to higher values than the mean observed in the disk, creating more massive molecular clouds, and enhancing the rate of star formation.

\end{enumerate}

The authors thank the anonymous referee for comments that have improved the presentation and the discussion of the paper.  We gratefully acknowledge the efforts by the SINGS, {\it GALEX} NGS, and THINGS teams to make their data public, and we want to thank Erwin de Blok for providing the rotation curve of NGC 6946 in tabular form.  We also thank the CARMA staff the support in the observations presented in this work.  The construction of CARMA was supported by the Gordon and Betty Moore Foundation, the Kenneth T. and Eileen L. Norris Foundation, the James S. McDonnell Foundation, the Associates of the California Institute of Technology, the University of Chicago, the states of California, Illinois, and Maryland, and the National Science Foundation.  The current CARMA development and operations are supported by the National Science Foundation and the CARMA partner universities under a cooperative agreement. This work has been partially supported by the Comisi\'{o}n Nacional de Investigaci\'{o}n en Ciencia y Tecnolog\'{i}a of Chile (Conicyt) under the Fulbright-Conicyt fellowship program, and by NSF grant AST 08-38226.  Jin Koda acknowledges support from the NSF through grant AST-1211680.  The National Radio Astronomy Observatory is a facility of the National Science Foundation operated under cooperative agreement by Associated Universities, Inc.


\begin{thebibliography}{}
\bibitem[Akritas \& Bershady(1996)]{1996ApJ...470..706A} Akritas, M.~G. \& Bershady, M.~A. 1996, \apj, 470, 706
\bibitem[Ballesteros-Paredes et al.(2011)]{2011MNRAS.411...65B} Ballesteros-Paredes, J., Hartmann, L.~W., V{\'a}zquez-Semadeni, E., Heitsch, F. \& Zamora-Avil{\'e}s, M.~A. 2011, \mnras, 411, 65-70
\bibitem[Bell et al.(2006)]{2006MNRAS.371.1865B} Bell, T.~A., Roueff, E., Viti, S. \& Williams, D.~A. 2006, \mnras,371, 1865-1872
\bibitem[Bertoldi \& McKee(1992)]{1992ApJ...395..140B} Bertoldi, F. \& McKee, C.~F. 1992, \apj, 395,140-157
\bibitem[Bicay \& Giovanelli(1986)]{1986AJ.....91..705B} Bicay, M.~D. \& Giovanelli, R. 1986, \aj, 91, 705-750
\bibitem[Bigiel et al.(2008)]{2008AJ....136.2846B} Bigiel, F., Leroy, A., Walter, F., et al. 2008, \aj, 136, 2846-2871
\bibitem[Blanc et al.(2009)]{2009ApJ...704..842B} Blanc, G.~A., Heiderman, A., Gebhardt, K., Evans, II, N.~J. \& Adams, J. 2009, \apj, 704, 842-862
\bibitem[Blitz \& Rosolowsky(2004)]{2004ApJ...612L..29B} Blitz, L. \& Rosolowsky, E.  2004, \apjl, 612, L29-L32
\bibitem[Blitz \& Rosolowsky(2006)]{2006ApJ...650..933B} Blitz, L. \& Rosolowsky, E.  2006, \apj, 650, 933-944
\bibitem[Blitz et al.(2007)]{2007prpl.conf...81B} Blitz, L., Fukui, Y., Kawamura, A., et al. 2007, Protostars and Planets V, 81-96
\bibitem[Bolatto et al.(2008)]{2008ApJ...686..948B} Bolatto, A.~D., Leroy, A.~K., Rosolowsky, E., Walter, F. \& Blitz, L. 2008, \apj, 686, 948-965
\bibitem[Boselli et al.(1994)]{1994A&A...285...69B} Boselli, A., Gavazzi, G., Combes, F., Lequeux, J. \& Casoli, F. 1994, \aap, 285, 79-68
\bibitem[Briggs(1990)]{1990ApJ...352...15B} Briggs, F.~H. 1990, \apj, 352, 15-29
\bibitem[Calzetti et al.(2005)]{2005ApJ...633..871C} Calzetti, D., Kennicutt, Jr., R.~C., Bianchi , L., et al. 2005, \apj, 633, 871-893
\bibitem[Calzetti et al.(2007)]{2007ApJ...666..870C} Calzetti, D., Kennicutt, R.~C., Engelbracht, C.~W., et al. 2007, \apj, 666, 870-895
\bibitem[Catinella et al.(2008)]{2008ApJ...685L..13C} Catinella, B., Haynes, M.~P., Giovanelli, R., Gardner, J.~P., \& Connolly, A.~J. 2008, \apjl, 685, L13-L17
\bibitem[Clarke \& Gittins(2006)]{2006MNRAS.371..530C} Clarke, C. \& Gittins, D. 2006, \mnras, 371, 530-536
\bibitem[Consid\`ere \& Athanassoula(1988)]{1988A&AS...76..365C} Consid\`ere, S. \& Athanassoula, E. 1988, \aaps, 76, 365-404
\bibitem[Crosthwaite \& Turner(2007)]{2007AJ....134.1827C} Crosthwaite, L.~P. \& Turner, J.~L., 2007, \aj, 134, 5, 1827
\bibitem[de Blok et al.(2008)]{2008AJ....136.2648D}  de Blok, W.~J.~G., Walter, F., Brinks, E., et al. 2008, \aj, 136, 2648-2719
\bibitem[Dickey \& Kaz\`es(1992)]{1992ApJ...393..530D} Dickey, J.~M. \& Kaz\`es, I., 1992, \apj, 393, 530-543
\bibitem[Dobbs \& Bonnell(2006)]{2006MNRAS.367..873D} Dobbs, C.~L. \& Bonnell, I.~A. 2006, \mnras, 367, 873-878
\bibitem[Dobbs \& Bonnell(2008)]{2008MNRAS.385.1893D} Dobbs, C.~L., \& Bonnell, I.~A., 2008, \mnras. 385, 1893-1902
\bibitem[Donovan Meyer et al.(2012)]{2012ApJ...744...42D} Donovan Meyer, J., Koda, J., Momose, et al. 2012, \apj, 744, 42
\bibitem[Drozdovsky \& Karachentsev(2000)]{2000A&AS..142..425D} Drozdovsky, I.~O. \& Karachentsev, I.~D. 2000, \aaps, 142, 425-432
\bibitem[Durbala et al.(2009)]{2009MNRAS.397.1756D} Durbala, A., Buta, R., Sulentic, J.~W. \& Verdes-Montenegro, L. 2009, \mnras, 397, 1756-1775 
\bibitem[Elmegreen et al.(2011)]{2011ApJ...737...32E} Elmegreen, D.~M., Elmegreen, B.~G., Yau, A., et al.  2011, \apj, 737, 32
\bibitem[Engargiola et al.(2003)]{2003ApJS..149..343E} Engargiola, G., Plambeck, R.~L., Rosolowsky, E. \& Blitz, L. 2003, \apjs, 149, 343-363
\bibitem[Evans et al.(2009)]{2009ApJS..181..321E} Evans, II, N.~J., Dunham, M.~M., J{\o}rgensen, J.~K., et al. 2009, \apjs, 181, 321-350 
\bibitem[Fazio et. al(2004)]{2004ApJS..154...10F} Fazio, G.~G., Hora, J.~L., Allen, L.~E., et al. 2004, \apjs, 154, 10-17
\bibitem[Fouqu\'e et al.(1990)]{1990A&AS...86..473F} Fouqu\'e, P., Durand, N., Bottinelli, L., Gouguenheim, L. \& Paturel, G. 1990, \aaps, 86, 473-502
\bibitem[Foyle et al.(2010)]{2010ApJ...725..534F} Foyle, K., Rix, H.-W., Walter, F. \& Leroy, A.~K. 2010, \apj, 725, 535-541 
\bibitem[Frerking et al.(1982)]{1982ApJ...262..590F} Frerking, M.~A., Langer, W.~D. \& Wilson, R.~W. 1982, \apj,  262, 590-605
\bibitem[Fukui et al.(2008)]{2008ApJS..178...56F} Fukui, Y., Kawamura, A., Minamidani, T., et al. 2008, \apjs, 178, 56-70
\bibitem[Genzel et al.(2010)]{2010MNRAS.407.2091G} Genzel, R., Tacconi, L.~J., Gracia-Carpio, J., et al. 2010, \mnras, 407, 2091-2108
\bibitem[Gil de Paz et al.(2007)]{2007ApJS..173..185G} Gil de Paz, A., Boissier, S., Madore, B.~F., et al. 2007, \apjs, 173, 185-255
\bibitem[Goldsmith et al.(2008)]{2008ApJ...680..428G} Goldsmith, P.~F., Heyer, M., Narayanan, G., et al. 2008, \apj, 680, 428-445 
\bibitem[Gutermuth et al.(2011)]{2011ApJ...739...84G} Gutermuth, R.~A., Pipher, J.~L., Megeath, S.~T., et al. 2011, \apj, 739, 84
\bibitem[Haynes \& Giovanelli(1984)]{1984AJ.....89..758H} Haynes , M.~P. \& Giovanelli, R. 1984, \apj, 89, 758-800
\bibitem[Heiderman et al.(2010)]{2010ApJ...723.1019H} Heiderman, A., Evans, II, N.~J., Allen, L.~E., Huard, T. \& Heyer, M. 2010, \apj, 723, 1019-1037
\bibitem[Helfer et al.(2003)]{2003ApJS..145..259H} Helfer, T.~T., Thornley, M.~D., Regan, M.~W., et al. 2003, \apjs, 145, 259-327
\bibitem[Heyer et al.(2009)]{2009ApJ...699.1092H} Heyer, M., Krawczyk, C., Duval, J. \& Jackson, J.~M. 2009, \apj, 699, 1092-1103
\bibitem[Hirota et al.(2011)]{2010PASJ...62.1261H} Hirota, A., Kuno, N., Sato, N., et al. 2010, \pasj, 62, 1261-1275 
\bibitem[Ho(2007)]{2007ApJ...669..821H}  Ho, L.~C. 2007, \apj, 2007, 669, 821-829
\bibitem[Hughes et al.(2010)]{2010MNRAS.406.2065H}  Hughes, A., Wong, T., Ott, J., et al. 2010, \mnras, 406, 2065-2086
\bibitem[Isobe et al.(1990)]{1990ApJ...364..104I} Isobe, T., Feigelson, E.~D., Akritas, M.~G. \& Babu, G.~J. 1990, \apj, 364, 104-113
\bibitem[Jarrett et al.(2003)]{2003AJ....125..525J} Jarrett, T.~H., Chester, T., Cutri, R., Schneider, S.~E. \& Huchra, J.~P., 2003, \aj, 125, 525-554
\bibitem[Karachentsev et al.(2004)]{2004AJ....127.2031K} Karachentsev, I.~D., Karachentseva, V.~E., Huchtmeier, W.~K. \& Makarov, D.~I. 2004, \aj, 127, 2031-2068
\bibitem[Kennicutt(1998)]{1998ApJ...498..541K} Kennicutt, Jr., R.~C. 1998, \apj, 498, 541
\bibitem[Kennicutt et al.(2003)]{2003PASP..115..928K} Kennicutt, Jr., R.~C., Armus, L., Bendo, G., et al. 2003, \pasp, 115, 928-952
\bibitem[Kennicutt et al.(2007)]{2007ApJ...671..333K} Kennicutt, Jr., R.~C., Calzetti, D., Walter, F., et al. 2007, \apj, 671, 333-348
\bibitem[Kim \& Ostriker(2002)]{2002ApJ...570..132K} Kim, W.-T. \& Ostriker, E.~C. 2002, \apj, 570, 132-151 
\bibitem[Koda et al.(2006)]{2006ApJ...638..191K} Koda, J., Sawada, T., Hasegawa, T., \& Scoville, N.~Z. 2006, \apj, 638, 191-195
\bibitem[Koda et al.(2009)]{2009ApJ...700L.132K} Koda, J., Scoville, N., Sawada, T., et al. 2009, \apjl, 700, L132-L136
\bibitem[Koyama \& Ostriker(2009a)]{2009ApJ...693.1316K} Koyama, H. \& Ostriker, E.~C. 2009, \apj, 693, 1316-1345
\bibitem[Koyama \& Ostriker(2009b)]{2009ApJ...693.1346K} Koyama, H. \& Ostriker, E.~C. 2009, \apj, 693, 1346-1359
\bibitem[Krumholz et al.(2009)]{2009ApJ...699..850K} Krumholz, M.~R., McKee, C.~F. \& Tumlinson, J. 2009, \apj, 699, 850-856
\bibitem[Krumholz et al.(2010)]{2010ApJ...713.1120K} Krumholz, M.~R., Cunningham, A.~J., Klein, R.~I. \& McKee, C.~F. 2010, \apj, 713, 1120-1133
\bibitem[Kuno et al.(2007)]{2007PASJ...59..117K} Kuno, N., Sato, N., Nakanishi, H., et al. 2007, \pasj, 59, 117-166
\bibitem[Larson(1981)]{1981MNRAS.194..809L} Larson, R.~B. 1981, \mnras,194, 809-826
\bibitem[Lavezzi \& Dickey(1997)]{1997AJ....114.2437L} Lavezzi, T.~E. \& Dickey, J.~M. 1997, \aj, 114, 2437
\bibitem[Lavezzi \& Dickey(1998)]{1998AJ....116.2672L} Lavezzi, T.~E. \& Dickey, J.~M. 1998, \aj, 116, 2672-2681 
\bibitem[Leroy et al.(2008)]{2008AJ....136.2782L} Leroy, A.~K., Walter, F., Brinks, E., et al. 2008, \aj, 136, 2782-2845
\bibitem[Leroy et al.(2009)]{2009AJ....137.4670L} Leroy, A.~K., Walter, F., Bigiel, F., et al. 2009, \aj, 137, 4670-4696
\bibitem[Leroy et al.(2011)]{2011ApJ...737...12L} Leroy, A.~K., Bolatto, A., Gordon, K., et al. 2011, \apj, 737, 12
\bibitem[Leroy et al.(2012)]{2012AJ....144....3L} Leroy, A.~K., Bigiel, F., de Blok, W.~J.~G., et al. 2012, \aj, 144, 3
\bibitem[Liu et al.(2011)]{2011ApJ...735...63L} Liu, G., Koda, J., Calzetti, D., Fukuhara, M. \& Momose, R. 2011, \apj, 735, 63 
\bibitem[Markwardt(2009)]{2009ASPC..411..251M} Markwardt, C.~B. 2009, 251
\bibitem[McKee \& Ostriker(2007)]{2007ARA&A..45..565M} McKee, C.~F. \& Ostriker, E.~C. 2007, \araa, 45, 565-687
\bibitem[Milam et al.(2005)]{2005ApJ...634.1126M} Milam, S.~N., Savage, C., Brewster, M.~A., Ziurys, L.~M. \& Wyckoff, S. 2005, \apj, 634, 1126-1132
\bibitem[Miyawaki et al.(2009)]{2009PASJ...61...39M} Miyawaki, R., Hayashi, M. \& Hasegawa, T. 2009, \pasj, 61, 39
\bibitem[Murray et al.(2010)]{2010ApJ...709..191M} Murray, N., Quataert, E. \& Thompson, T.~A. 2010. \apj, 709, 191-209
\bibitem[Nieten et al.(2006)]{2006A&A...453..459N} Nieten, C., Neininger, N., Gu{\'e}lin, M., et al. 2006, \aap, 453, 459-475
\bibitem[Obreschkow et al.(2009)]{2009ApJ...698.1467O}  Obreschkow, D., Croton, D., De Lucia, G., Khochfa, S. \& Rawlings, S. 2009, \apj, 698, 1467-1484
\bibitem[Obreschkow et al.(2009b)]{2009ApJ...703.1890O}  Obreschkow, D., Kl{\"o}ckner, H.-R., Heywood, I., Levrier, F., \& Rawlings, S. 2009b, \apj, 703, 1890-1903
\bibitem[Oka et al.(2001)]{2001ApJ...562..348O} Oka, T., Hasegawa, T., Sato, F., et al. 2001, \apj, 2001, 562, 348-362
\bibitem[Ossenkopf \& Henning(1994)]{1994A&A...291..943O} Ossenkopf, V. \& Henning, T. 1994, \aap, 291, 943-959
\bibitem[Ostriker et al.(2010)]{2010ApJ...721..975O} Ostriker, E.~C., McKee, C.~F. \& Leroy, A.~K. 2010, \apj, 721, 975-994
\bibitem[Rahman et al.(2011)]{2011ApJ...730...72R} Rahman, N., Bolatto, A.~D., Wong, T., et al. 2011, \apj, 730, 72
\bibitem[Rand(1995)]{1995AJ....109.2444R} Rand, R.~J. 1995, \aj, 1995, 109, 2444
\bibitem[Rand \& Kulkarni(1990)]{1990ApJ...349L..43R} Rand, R.~J. \& Kulkarni, S.~R. 1990, \apjl, 349, L43-L46
\bibitem[Rand et al.(1999)]{1999ApJ...513..720R} Rand, R.~J., Lord, S.~D., \& Higdon, J.~L. 1999, \apj, 513, 720-732
\bibitem[Regan \& Vogel(1995)]{1995ApJ...452L..21R} Regan, M.~W. \& Vogel, S.~N. 1995, \apjl, 452, L21  
\bibitem[Rieke et al.(2004)]{2004ApJS..154...25R} Rieke, G., Young, E.~T., Engelbracht, C.~W., et al. 2004, \apjs, 154, 25
\bibitem[Roberts(1969)]{1969ApJ...158..123R} Roberts, W.~W. 1969, \apj, 158, 123
\bibitem[Rohlfs \& Wilson(2004)]{2004tra..book.....R} Rohlfs, K. \& Wilson, T.~L. 2004, Tools of radio astronomy, (4th rev.~and enl.~ed; Berlin: Springer).
\bibitem[Rosolowsky(2007)]{2007ApJ...654..240R} Rosolowsky , E. 2007, \apj, 654, 240-251
\bibitem[Rosolowsky \& Leroy(2006)]{2006PASP..118..590R} Rosolowsky, E. \& Leroy, A. 2006, \pasp, 118, 590-610
\bibitem[Sakamoto et al.(1999)]{1999ApJS..124..403S} Sakamoto, K., Okumura, S.~K., Ishizuki, S. \& Scoville, N.~Z. 1999, \apjs, 124, 403-427
\bibitem[Schlegel et al.(1998)]{1998ApJ...500..525S} Schlegel, D.~J., Finkbeiner, D.~P. \& Davis, M. 1998, \apj, 500, 525
\bibitem[Schmidt(1959)]{1959ApJ...129..243S} Schmidt, M. 1959, \apj, 129, 243
\bibitem[Shetty \& Ostriker(2006)]{2006ApJ...647..997S} Shetty, R. \& Ostriker, E.~C. 2006, \apj, 647, 997
\bibitem[Shetty \& Ostriker(2008)]{2008ApJ...684..978S} Shetty, R. \& Ostriker, E.~C. 2008, \apj, 684, 978
\bibitem[Shetty et al.(2007)]{2007ApJ...665.1138S} Shetty, R., Vogel, S.~N., Ostriker, E.~C. \& Teuben, P.~J. 2007, \apj, 665, 1138-1158
\bibitem[Shetty et al.(2010)]{2010ApJ...712.1049S} Shetty, R., Collins, D.~C., Kauffmann, J., et al. 2010, \apj, 712, 1049-1056
\bibitem[Solomon et al.(1987)]{1987ApJ...319..730S} Solomon, P.~M., Rivolo, A.~R., Barrett, J. \& Yahil, A. 1987, \apj, 319, 730-741
\bibitem[Tamburro et al.(2008)]{2008AJ....136.2872T} Tamburro, D., Rix, H.-W., Walter, F., et al.  2008, \aj, 136, 2872-2885
\bibitem[Tamburro et al.(2009)]{2009AJ....137.4424T} Tamburro, D., Rix, H.-W., Leroy, A.~K., et al. 2009, \aj, 137, 4424-4435
\bibitem[Tasker \& Tan(2009)]{2009ApJ...700..358T} Tasker, E.~J. \& Tan, J.~C. 2009, \apj, 700, 358-375
\bibitem[Tasker(2011)]{2011ApJ...730...11T} Tasker, E.~J. 2011, \apj, 730, 11
\bibitem[Thornley \& Mundy(1997)]{1997ApJ...484..202T} Thornley, M.~D. \& Mundy, L.~G. 1997, \apj, 484, 202
\bibitem[Tifft \& Cocke(1988)]{1988ApJS...67....1T} Tifft, W.~G. \& Cocke, W.~J. 1988, \apjs, 67, 1-75
\bibitem[Tosaki et al.(2003)]{2003PASJ...55..605T} Tosaki, T., Shioya, Y., Kuno, N., Nakanishi, K. \& Hasegawa, T. 2003, \pasj, 55, 605-613
\bibitem[Tully(1988)]{1988JBAA...98..316T} Tully, R.~B., 1988, Journal of the British Astronomical Association, 98, 316
\bibitem[Tully \& Fisher(1977)]{1977A&A....54..661T} Tully, R.~B. \& Fisher, J.~R. 1977, \aap, 54,  661-673
\bibitem[Tutui \& Sofue(1999)]{1999A&A...351..467T} Tutui, Y. \& Sofue, Y. 1999, \aap, 351, 467-471
\bibitem[Vogel et al.(1988)]{1988Natur.334..402V} Vogel, S.~N., Kulkarni, S.~R. \& Scoville, N.~Z. 1988, \nat, 334, 402-406
\bibitem[Wada \& Koda(2004)]{2004MNRAS.349..270W} Wada, K. \& Koda, J. 2004, \mnras, 349, 270-280 
\bibitem[Wada et al.(2011)]{2011ApJ...735....1W} Wada, K., Baba, J. \& Saitoh, T.~R. 2011, \apj, 735, 1
\bibitem[Walter et al.(2004)]{2004ApJ...615L..17W}  Walter, F., Carilli, C., Bertoldi, F., et al. 2004, \apjl, 615, L17-L20
\bibitem[Walter et al.(2008)]{2008AJ....136.2563W} Walter, F., Brinks, E., de Blok, W.~J.~G., et al. 2008, \aj, 136, 2563-2647
\bibitem[Walterbos \& Braun(1996)]{1996ASPC..106....1W} Walterbos, R.~A.~M. \& Braun, R. 1996, The Minnesota Lectures on Extragalactic Neutral Hydrogen, 106, 1-+
\bibitem[Wang et al.(2009)]{2009A&A...507..369W} Wang, K., Wu, Y.~F., Ran, L., Yu, W.~T. \& Miller, M. 2009, \aap, 507, 369-376
\bibitem[Wong \& Blitz(2002)]{2002ApJ...569..157W}  Wong, T. \&  Blitz , L. 2002, \apj, 569, 157-183
\bibitem[Wong et al.(2011)]{2011ApJS..197...16W} Wong, T., Hughes, A., Ott, J., et al. 2011, \apjs, 197, 16.
\bibitem[Wu et al.(2010)]{2010ApJS..188..313W}  Wu, J., Evans, II, N.~J., Shirley, Y.~L. \& Knez, C. 2010, \apj, 188, 313-357
\bibitem[Wyder et al.(2007)]{2007ApJS..173..293W} Wyder, T., Martin, D.~C., Schiminovich, D., et al. 2007, \apjs, 173, 293-314
\bibitem[Yim et al.(2011)]{2011AJ....141...48Y}  Yim, K., Wong, T., Howk, J.~C. \& van der Hulst, J.~M. 2011, \aj, 141, 48-+

\end{thebibliography}
\end{document}